\documentclass[12pt,letterpaper]{article}
\usepackage[T1]{fontenc}
\usepackage[utf8]{inputenc}
\usepackage{newtxtext}
\usepackage{amsmath,amsthm,mathtools}
\usepackage{newtxmath,bm}
\usepackage[margin=1in]{geometry}
\usepackage{microtype}
\usepackage{xcolor,graphicx,booktabs,array,adjustbox}
\usepackage{tikz,pgfplots}
\pgfplotsset{compat=1.18}
\usetikzlibrary{arrows.meta,calc,positioning,fit,backgrounds,decorations.pathreplacing,shapes.geometric}
\usepackage[numbers,sort&compress]{natbib}
\usepackage[font=small,labelfont=bf]{caption}
\usepackage{placeins,needspace,etoolbox}
\usepackage{hyperref}
\newcommand{\doi}[1]{\href{https://doi.org/#1}{\nolinkurl{#1}}}
\hypersetup{colorlinks=true,linkcolor=blue!40!black,citecolor=blue!40!black,urlcolor=blue!40!black,
 pdfauthor={Matthew Can and Mark M. Wilde},
 pdftitle={One-shot entanglement generation and distillation via Petz recovery and binary hypothesis testing}}
\numberwithin{equation}{section}
\newtheorem{theorem}{Theorem}[section]
\newtheorem{lemma}[theorem]{Lemma}
\newtheorem{proposition}[theorem]{Proposition}
\newtheorem{corollary}[theorem]{Corollary}
\theoremstyle{definition}\newtheorem{definition}[theorem]{Definition}
\theoremstyle{remark}
\DeclareMathOperator{\Tr}{Tr}
\DeclareMathOperator{\supp}{supp}

\DeclareMathOperator{\Var}{Var}
\newcommand{\ket}[1]{|#1\rangle}
\newcommand{\bra}[1]{\langle#1|}
\newcommand{\proj}[1]{\ket{#1}\!\bra{#1}}
\newcommand{\norm}[1]{\left\|#1\right\|}
\newcommand{\id}{\operatorname{id}}
\newcommand{\cN}{\mathcal N}
\newcommand{\cD}{\mathcal D}

\newcommand{\ee}{\mathbb E}
\newcommand{\err}{\varepsilon}
\newcommand{\cost}{\mathsf e}
\newcommand{\Frec}{F_{\mathrm P}}
\newcommand{\Ih}{I_H}
\newcommand{\Ic}{I_{\mathrm c}}
\newcommand{\Vc}{V_{\mathrm c}}
\newcommand{\thref}[1]{Theorem~\ref{#1}}
\newcommand{\lemref}[1]{Lemma~\ref{#1}}
\newcommand{\prref}[1]{Proposition~\ref{#1}}
\newcommand{\corref}[1]{Corollary~\ref{#1}}
\definecolor{inkblue}{RGB}{36,69,101}
\definecolor{softblue}{RGB}{232,240,247}
\definecolor{inkgreen}{RGB}{28,103,87}
\definecolor{softgreen}{RGB}{233,244,239}
\tikzset{>=Latex,box/.style={draw=inkblue,fill=softblue,rounded corners=2pt,align=center,inner sep=5pt},
 qwire/.style={line width=.8pt},cwire/.style={line width=.8pt,dashed},flow/.style={->,line width=.8pt}}

\allowdisplaybreaks[1]\widowpenalty=10000\clubpenalty=10000
\title{\textbf{One-shot entanglement generation
and distillation\\[3pt]
via Petz recovery and binary hypothesis testing}}
\author{Matthew Can \qquad Mark M.\ Wilde\\[6pt]
 {\small\itshape School of Electrical and Computer Engineering}\\
 {\small\itshape Cornell University, Ithaca, New York 14850, USA}\\[2pt]
 {\small\href{mailto:mmc347@cornell.edu}{mmc347@cornell.edu}}, \small\href{mailto:wilde@cornell.edu}{wilde@cornell.edu}}
\date{October 21, 2026}
\begin{document}
\maketitle
\begin{abstract}
How much entanglement can two distant laboratories share when their connection is noisy? We relate this question to deciding which of two quantum hypotheses is present. Our one-shot bounds apply to entanglement generation over a quantum channel and to one-way entanglement distillation from a shared state. We use Cheng's trace inequality to bound the infidelity of a Petz decoder in terms of a binary hypothesis testing cost. Applying the inequality before averaging over codes reduces the error analysis to computing a second moment of random subspaces. The proof covers every integer target dimension and treats rank-deficient states explicitly. A corollary expresses the achievable entanglement through hypothesis-testing coherent information, with a single logarithmic error slack and exact integer rounding. We also prove an error-preserving conversion from entanglement generation codes to entanglement transmission codes: the target dimension decreases by only a constant factor at fixed error, and the decoder is the original decoder followed by local processing. Consequently, for both entanglement generation and entanglement transmission, arbitrary pure reference--input states yield the same coherent-information rate, second-order Gaussian correction at the original error tolerance, and explicit achievable logarithmic remainder.  For commuting testing hypotheses, the logarithmic remainder improves to a constant. These results give a direct route from binary distinguishability to explicit recovery guarantees and finite-resource quantum communication rates.
\end{abstract}

\clearpage
\begingroup
\fontsize{10.5}{12.2}\selectfont
\setlength{\parskip}{0pt}
\makeatletter
\patchcmd{\l@section}{1.0em}{0.55em}{}{}
\makeatother
\tableofcontents
\endgroup
\clearpage

\section{Introduction}\label{sec:intro}
\subsection{Background and motivation}
How reliably can two distant laboratories share entanglement when their only connection is noisy? Entanglement is a resource for quantum communication and distributed information processing, but noise can destroy the correlations that make it useful. Three basic tasks address this difficulty. In \emph{entanglement generation}, Alice prepares a joint state, retains one system, and sends the other through a noisy channel for Bob to decode. In \emph{entanglement transmission}, she instead encodes one share of a prescribed maximally entangled state, while leaving its reference share untouched. In \emph{one-way entanglement distillation}, Alice and Bob already share an imperfect state and use local operations and classical communication from Alice to Bob to extract a more nearly ideal entangled state. Figure~\ref{fig:protocols} depicts the distinct resources and operations of all three tasks. At finite blocklength, distinguishing these tasks and their error criteria is essential~\cite[Chapter~14]{KhatriWilde2024}.

A recurring idea in information theory is that reliable communication is closely tied to statistical distinguishability. A classical receiver decides which transmitted message best explains its observation. Binary hypothesis testing isolates the simpler question of distinguishing one explanation from another. Blahut's work develops this relationship systematically for classical information theory~\cite{Blahut1974}. In the quantum setting, one-shot achievability bounds for classical communication can be expressed through hypothesis-testing relative entropy~\cite{WangRenner2012}. Entanglement-assisted classical communication admits analogous bounds in terms of hypothesis-testing mutual information~\cite{AnshuJainWarsi2019,QiWangWilde2018}. These results make a binary statistical quantity useful for understanding communication protocols. Expressing achievable rates through one-shot quantities based on hypothesis-testing relative entropy also connects coding directly to second-order asymptotics: estimates for independent and identically distributed resources give the leading information rate and its second-order Gaussian correction~\cite{TomamichelHayashi2013,Li2014,Cheng2023}. Combined with suitable converse bounds, this approach can sometimes give complete second-order characterizations; an achievability bound alone supplies the corresponding lower bound.

For classical communication over a quantum channel, Cheng's trace inequality bounds the error of the pretty-good measurement, a standard measurement for distinguishing signal states, through binary tests~\cite{Cheng2023}. For each message, the test compares its output state with the sum of the output states associated with the other messages. This separates the choice of analysis tests from the physical decoding measurement.

Quantum communication requires an additional feature: the decoder must preserve superpositions and entanglement, rather than simply identify a classical message.
The Petz recovery channel provides an explicit decoder with a substantial history in quantum error correction~\cite{Petz1986,Petz1988,BarnumKnill2002,NgMandayam2010,MandayamNg2012,Renes2017}. Beigi, Datta, and Leditzky combined Petz decoding and random code spaces to establish one-shot and second-order quantum communication bounds~\cite{BeigiDattaLeditzky2016}. A central difficulty in analyzing random  codes based on Petz recovery is that the inverse output-state operator in the decoder changes with the selected code. The relationship between binary distinguishability and coherent quantum recovery therefore calls for a different analysis from classical message discrimination.

\begin{figure}[!t]
\centering
\captionsetup{skip=12pt}
\begin{tikzpicture}[font=\small,x=1cm,y=1.22cm,
 taskbox/.style={box,minimum height=.8cm,inner sep=6pt}]
 \node[anchor=west,font=\bfseries] at (0,0) {(a) Entanglement generation: Alice chooses the input state};
 \node[taskbox,text width=2.1cm,minimum height=1.25cm] (egprep) at (1.3,-1.1) {Local preparation\\$\zeta_{LA'}$};
 \node[taskbox,minimum width=1.1cm] (egn) at (6.7,-1.45) {$\cN$};
 \node[taskbox,minimum width=1.1cm] (egd) at (9.6,-1.45) {$\cD$};
 \draw[qwire,->] (egprep.east |- 0,-.75)--(12,-.75) node[right=3pt] {$L$};
 \draw[qwire,->] (egprep.east |- 0,-1.45)--node[above=3pt] {$A'$}(egn.west);
 \draw[qwire,->] (egn.east)--node[above=3pt] {$B$}(egd.west);
 \draw[qwire,->] (egd.east)--(12,-1.45) node[right=3pt] {$\widehat L$};
 \node[anchor=west,text=inkblue,font=\footnotesize] at (2.9,-2.12) {Only the lower quantum system is sent; no classical message.};
 \draw[gray!45] (0,-2.5)--(13,-2.5);
 \node[anchor=west,font=\bfseries] at (0,-2.95) {(b) Entanglement transmission: the logical entangled input is prescribed};
 \node[taskbox,text width=2.1cm,minimum height=1.25cm] (etprep) at (1.3,-4.05) {Local input\\$\Phi^m_{L\widetilde L}$};
 \node[taskbox,minimum width=1.0cm] (etenc) at (4.25,-4.4) {$\mathcal E$};
 \node[taskbox,minimum width=1.1cm] (etn) at (6.7,-4.4) {$\cN$};
 \node[taskbox,minimum width=1.1cm] (etd) at (9.6,-4.4) {$\cD$};
 \draw[qwire,->] (etprep.east |- 0,-3.7)--(12,-3.7) node[right=3pt] {$L$};
 \draw[qwire,->] (etprep.east |- 0,-4.4)--node[above=3pt] {$\widetilde L$}(etenc.west);
 \draw[qwire,->] (etenc.east)--node[above=3pt] {$A'$}(etn.west);
 \draw[qwire,->] (etn.east)--node[above=3pt] {$B$}(etd.west);
 \draw[qwire,->] (etd.east)--(12,-4.4) node[right=3pt] {$\widehat L$};
 \node[anchor=west,text=inkblue,font=\footnotesize] at (2.9,-5.07) {The encoder acts on one share; the retained marginal stays $I_L/m$.};
 \draw[gray!45] (0,-5.45)--(13,-5.45);
 \node[anchor=west,font=\bfseries] at (0,-5.9) {(c) One-way entanglement distillation: the noisy state is already shared};
 \node[taskbox,text width=2.1cm,minimum height=1.85cm] (diinput) at (1.3,-7.27) {Shared input\\$\rho_{AB}$};
 \node[taskbox,minimum width=1.3cm] (dienc) at (4.25,-6.72) {$\mathcal E_j$};
 \node[taskbox,minimum width=1.3cm] (didec) at (9.6,-7.82) {$\mathcal D_j$};
 \draw[qwire,->] (diinput.east |- 0,-6.72)--node[above=3pt] {$A$}(dienc.west);
 \draw[qwire,->] (dienc.east)--(12,-6.72) node[right=3pt] {$L$};
 \draw[qwire,->] (diinput.east |- 0,-7.82)--node[pos=.18,above=3pt] {$B$}(didec.west);
 \draw[cwire,->] (dienc.south)--(4.25,-7.22)--node[below=3pt,font=\footnotesize]{classical outcome $j$}(9.6,-7.22)--(didec.north);
 \draw[qwire,->] (didec.east)--(12,-7.82) node[right=3pt] {$\widehat L$};
 \node[anchor=west,text=inkblue,font=\footnotesize] at (2.9,-8.57) {Alice sends the outcome; Bob chooses the corresponding decoder.};
\end{tikzpicture}
\caption{Three tasks with the same final target $\Phi^m_{L\widehat L}$. Time flows from left to right; the upper output belongs to Alice and the lower output to Bob. Solid arrows carry quantum systems and the dashed arrow carries a classical outcome. In entanglement generation and entanglement transmission, the two initial systems are both local to Alice. Only in entanglement distillation is the input initially shared. The definitions and infidelity criteria appear in Section~\ref{subsec:tasks}.}
\label{fig:protocols}
\end{figure}

\subsection{Summary of results}
Here we give a direct binary hypothesis-testing guarantee for the quantum information processing tasks depicted in Figure~\ref{fig:protocols}. Throughout, an \emph{$(m,\err)$ protocol} (or code, for a channel task) produces a maximally entangled target of local dimension $m$ with infidelity at most $\err$. This common convention applies to entanglement generation, entanglement transmission, and entanglement distillation; their different allowed operations are specified in Section~\ref{subsec:tasks}.

Let $\rho_{AB}$ be a state shared by Alice and Bob, with
\begin{equation}
d\coloneqq\dim A\ge2.
\end{equation}
For an integer $m\in\{2,\ldots,d\}$, an $(m,\err)$ entanglement distillation protocol transforms this shared state into an approximation of the maximally entangled target. The target contains $\log_2m$ ebits of entanglement, where one ebit is the entanglement of an ideal pair of qubits. Infidelity is equal to one minus the overlap of the output state with the target state, so $\err=0$ requires perfect recovery.

Here we bound the error $\err$ in terms of a binary testing cost. For positive semidefinite operators $X$ and $Y$ on the same space, define
\begin{equation}\label{eq:intro-cost}
 \cost(X,Y)\coloneqq\min\{\Tr[(I-T)X]+\Tr[TY]:0\le T\le I\}.
\end{equation}
An effect $T$ represents acceptance of the first hypothesis. For arbitrary positive semidefinite $X,Y$, the two traces in \eqref{eq:intro-cost} are nonnegative \emph{error costs}. A probabilistic interpretation of $\cost(X,Y)$ is obtained after normalization: if $s\coloneqq\Tr(X+Y)>0$, use prior weights $\Tr X/s$ and $\Tr Y/s$ and, for each nonzero weight, the states $X/\Tr X$ and $Y/\Tr Y$. Then $\cost(X,Y)/s$ is the optimal Bayesian error probability in discriminating these states. A zero-trace operator contributes zero cost, and $X=Y=0$ gives $\cost(X,Y)=0$. We retain the unnormalized formulation because, in coding applications, $Y$ can be a sum of density operators associated with competing messages. Its trace then counts the terms in that sum and need not equal one. This is the form used in Cheng's classical-communication analysis~\cite{Cheng2023}.

Our principal statement is that, for every such $m$, there exists a one-way $(m,\err)$ entanglement distillation protocol whenever $\err$ satisfies
\begin{align}
 e_{d,m}\left(1-\frac{e_{d,m}}{dm}\right)&\leq \err ,\label{eq:intro-bound}\\
 e_{d,m}&\coloneqq\cost(\rho_{AB},B_{d,m}),\label{eq:intro-e}\\
 B_{d,m}&\coloneqq\frac{dm-1}{d^2-1}
                 (dI_A\otimes\rho_B-\rho_{AB}).\label{eq:intro-alternative}
\end{align}
Thus the protocol produces an approximate maximally entangled state
of two $m$-dimensional systems with infidelity at most
$\err$, for all $\err$ satisfying \eqref{eq:intro-bound}.
Here $\rho_B\coloneqq\Tr_A\rho_{AB}$ is Bob's reduced state. The operator $B_{d,m}$ is positive semidefinite, as proved in Section~\ref{subsec:operators}, and has trace $dm-1$. As we see later, it is the weighted alternative produced by averaging the code construction. Equations~\eqref{eq:intro-bound}--\eqref{eq:intro-alternative} reduce the coding guarantee to distinguishing $\rho_{AB}$ from this one explicitly specified alternative. Smaller testing cost guarantees smaller entanglement infidelity. The factor in parentheses in \eqref{eq:intro-bound} provides an additional improvement; the following bound also holds:
\begin{equation}
e_{d,m}\left(1-\frac{e_{d,m}}{dm}\right)\leq e_{d,m},
\end{equation}
implying that there exists a one-way $(m,\err)$ entanglement distillation protocol whenever $\err$ satisfies the simpler guarantee $e_{d,m} \leq \err $.

Theorem~\ref{thm:main} proves this statement for one-way entanglement distillation, in which Alice sends a classical outcome to Bob, and for unassisted entanglement generation. In the channel setting, choose a normalized pure reference--input state $\phi_{AA'}$ and set $\rho_{AB}=(\id_A\otimes\cN)(\phi_{AA'})$ for a channel $\cN\colon A'\to B$. The same expression in \eqref{eq:intro-bound} certifies an $(m,\err)$ entanglement generation code consisting of a selected pure input and its Petz decoder. If the reference marginal is maximally mixed, $\phi_A=I_A/d$, Corollary~\ref{cor:flat} gives an $(m,\err)$ \emph{entanglement transmission} code with exactly the same dimension and error bound. The selected state then has a maximally mixed retained marginal and is prepared by an isometric encoder acting on one share of the prescribed target. 

The key analytical step in the proof of Theorem~\ref{thm:main} is to apply Cheng's trace inequality (Lemma~\ref{lem:cheng}) \emph{before} averaging over codes. It bounds the recovery error by a binary testing cost whose dependence on a comparison operator is linear. Only a second moment then needs to be averaged, while coherence inside each code space is retained. For comparison, the one-shot Petz bound of Beigi, Datta, and Leditzky~\cite[Theorem~3.1]{BeigiDattaLeditzky2016} involves the minimum of two information-spectrum expressions, with linked error parameters and an input-purity term. Their proof controls a collision expression containing the code-dependent inverse from the Petz recovery channel while averaging, using weak dephasing and joint convexity~\cite[Sections~3.2--3.4]{BeigiDattaLeditzky2016}. By applying Cheng's trace inequality first, we avoid that additional averaging analysis. The binary-testing bound in
\eqref{eq:intro-bound}--\eqref{eq:intro-alternative}
and its single-slack consequence in \eqref{eq:intro-rate}
give a more compact statement and proof for the tasks
and input assumptions considered here. 

A second formulation states how much entanglement can be obtained at a chosen error. For a testing tolerance $0\le\delta<1$, define the hypothesis-testing coherent information as
\begin{equation}\label{eq:intro-IH}
 \Ih^\delta(A\rangle B)_\rho
 \coloneqq D_H^\delta(\rho_{AB}\Vert I_A\otimes\rho_B),
\end{equation}
where the hypothesis-testing relative entropy $D_H^\delta$ is defined in \eqref{eq:DH}. It is equal to the negative logarithm of the smallest false-acceptance weight attainable by a binary test that accepts the true state with probability at least $1-\delta$. A larger value means stronger statistical separation and permits a larger entangled target. For a desired final infidelity $\err$, Corollary~\ref{cor:testing} gives the sufficient condition
\begin{equation}\label{eq:intro-rate}
 \log_2m\le\Ih^{\err-\nu}(A\rangle B)_\rho-\log_2\frac1\nu,
 \qquad 0<\nu\le\err<1.
\end{equation}
The left side is the number of ebits to be produced. Equation~\eqref{eq:intro-rate} has the same single-slack structure as Cheng's one-shot bounds for unassisted classical communication and entanglement-assisted classical communication~\cite[Propositions~2 and~4]{Cheng2023}: a hypothesis-testing information quantity at tolerance $\err-\nu$, followed by the logarithmic deduction $-\log_2(1/\nu)$. The information quantity here is coherent information rather than the Holevo or mutual information quantity. For fixed $\nu$, we may choose $m$ to be the largest integer in $\{2,\ldots,d\}$ satisfying \eqref{eq:intro-rate}; if there is no such integer, the trivial target $m=1$ is always achievable with zero error. Corollary~\ref{cor:testing} keeps the integer rounding exactly. Corollary~\ref{cor:profile} gives the stronger choice that retains the full finite-dimensional bound.

The appearance of $\Ih^\delta$ gives a common information-theoretic formulation to three central communication tasks. Unassisted classical communication uses hypothesis-testing Holevo information for a classical--quantum state~\cite{WangRenner2012}; entanglement-assisted classical communication uses the hypothesis-testing mutual information for a reference--output state~\cite{AnshuJainWarsi2019,QiWangWilde2018}. Here, entanglement generation and the flat-reference entanglement transmission specialization are controlled directly by the hypothesis-testing coherent information in~\eqref{eq:intro-IH}. For general reference--input states, an error-preserving conversion gives an entanglement transmission guarantee with only a constant loss in log-dimension at fixed error, as described next. The alternatives are different---a product of marginals for mutual information, and $I_A\otimes\rho_B$ for coherent information---but each bound turns distinguishability measured by $D_H$ into an achievable communication guarantee. One-way entanglement distillation shares the same guarantee. 

The conversion in Lemma~\ref{lem:eg-et-conversion} takes an $(M,\err)$ entanglement generation code to a $(\max\{1,\lfloor M\sqrt{1-\err}\rfloor\},\err)$ entanglement transmission code. Thus a general entanglement generation input state need not have a maximally mixed marginal: reducing the target dimension, rather than increasing the error, suffices to produce an entanglement transmission encoder. At fixed $\err<1$, the loss in log-dimension is bounded independently of the number of channel uses. The decoder is the entanglement generation decoder followed by a local channel. The exact flat-reference statement in Corollary~\ref{cor:flat} remains stronger at the one-shot level because it needs neither modification.

For $n$ independent copies and fixed target infidelity $\err$, the entanglement generation and entanglement distillation bounds yield
\begin{equation}\label{eq:intro-second}
 \log_2m_n\ge
 \left[n\Ic(A\rangle B)_\rho+
       \sqrt{n\Vc(A\rangle B)_\rho}\,\Phi^{-1}(\err)
       -\tfrac12\log_2n-c\right]_+,
\end{equation}
when $\Vc(A\rangle B)_\rho>0$, for a constant $c$ independent of $n$. Here $m_n$ is the integer dimension of each share of the maximally entangled target produced from $n$ copies of $\rho_{AB}$, or by an entanglement generation code using the channel $n$ times. It depends on the fixed state and target infidelity as well as $n$; an explicit choice is given in \eqref{eq:second-mn}. Theorem~\ref{thm:second-transmission} gives the same bound for an entanglement transmission code, with $m_n$ replaced by its reduced target dimension $k_n$ and a possibly larger constant.

The quantity $\Ic(A\rangle B)_\rho=H(B)_\rho-H(AB)_\rho$ is the coherent information, introduced by Schumacher and Nielsen~\cite{SchumacherNielsen1996}: it is the leading rate supplied by this construction. The nonnegative quantity $\Vc(A\rangle B)_\rho$, defined in \eqref{eq:Vc}, is its information variance. This is the conditional version of the quantum relative entropy variance developed in second-order quantum information theory~\cite{TomamichelHayashi2013,Li2014}, and used for coherent-information coding bounds in~\cite{BeigiDattaLeditzky2016,TomamichelBertaRenes2016}. It describes the fluctuations responsible for the order-$\sqrt n$ correction. The function $\Phi^{-1}$ is the standard normal quantile, and $[x]_+\coloneqq\max\{0,x\}$ prevents a negative claimed yield. In particular, $\Phi^{-1}(\err)<0$ for $\err<1/2$, reflecting the finite-blocklength cost of requiring small error.

The remainder in \eqref{eq:intro-second} follows the same mechanism as Cheng's second-order analysis for classical communication~\cite[Section~3.1]{Cheng2023}. Choosing $\nu_n=n^{-1/2}$ in the single-slack bound produces the achievable term $-\tfrac12\log_2n-O(1)$ while retaining the Gaussian quantile at the prescribed final error. The conversion from entanglement generation to entanglement transmission costs only $O_{\err}(1)$ additional ebits, so it preserves this logarithmic coefficient as well.

Asymptotic achievability of coherent information was established for quantum channel coding by Devetak~\cite{Devetak2005} and for one-way entanglement distillation by Devetak and Winter~\cite{DevetakWinter2005}. Non-asymptotic bounds and second-order achievability were subsequently developed for quantum communication by Beigi, Datta, and Leditzky~\cite{BeigiDattaLeditzky2016} and Tomamichel, Berta, and Renes~\cite{TomamichelBertaRenes2016}, and for one-way entanglement distillation by~\cite[Proposition~21, Section~6.2, and Appendix~C]{WildeTomamichelBerta2017}. Our contribution is a direct proof through the one-shot hypothesis testing bound, with explicit control of its remainder and an error-preserving conversion from entanglement generation to entanglement transmission. Section~\ref{subsec:second-transmission} explains how the conversion supplies an omitted task-conversion step in the general second-order entanglement transmission argument of Tomamichel, Berta, and Renes, while distinguishing that issue from the independently established entanglement transmission result of Beigi, Datta, and Leditzky. The logarithmic term in \eqref{eq:intro-second} is a guarantee, not an optimal third-order coefficient. Appendix~\ref{app:third-order} explains its relation to classical testing asymptotics and proves a constant-remainder improvement when $\rho_{AB}$ commutes with $I_A\otimes\rho_B$. Zero information variance is handled separately. 

The argument builds on established recovery, pinching, and random-subspace tools. We distinguish our tasks and error conventions from the smooth-entropy statements in~\cite{BuscemiDattaDistillation2010,MorganWinter2014,KhatriWilde2024}. Cheng's trace inequality is invoked as a critical part of the analysis. 

\begin{figure}[!t]
\centering
\captionsetup{skip=10pt}
\begin{tikzpicture}[font=\footnotesize,x=1cm,y=1cm]
 \foreach \x in {0,3.8,7.6,11.4}{
  \node[draw=inkblue,fill=softblue,rounded corners=2pt,minimum width=3.45cm,minimum height=5.4cm] at (\x,0) {};
 }
 \node[font=\footnotesize\bfseries,align=center,text width=3.15cm] at (0,1.95) {1. Keep a\\quantum subspace};
 \node[font=\footnotesize\bfseries,align=center,text width=3.15cm] at (3.8,1.95) {2. Decode\\with Petz};
 \node[font=\footnotesize\bfseries,align=center,text width=3.15cm] at (7.6,1.95) {3. Compare by\\Cheng's inequality};
 \node[font=\footnotesize\bfseries,align=center,text width=3.15cm] at (11.4,1.95) {4. Average\\and select};
 \draw[gray!65,fill=white,line width=.5pt] (0,.26) ellipse (1.42cm and .98cm);
 \node[font=\scriptsize,fill=white,inner sep=1pt] at (1.16,.91) {$A$};
 \filldraw[fill=softgreen,draw=inkgreen,line width=.7pt]
   (-1.20,-.25)--(-.36,.25)--(1.20,-.13)--(.36,-.63)--cycle;
 \coordinate (origin) at (-.70,.05);
 \coordinate (input) at (-.08,.86);
 \coordinate (projected) at (.55,-.12);
 \draw[->,inkblue,line width=.8pt] (origin)--(input);
 \node[font=\scriptsize,fill=white,inner sep=1pt,text=inkblue] at (-.13,1.06) {$\ket\psi$};
 \draw[gray!70,dashed,line width=.6pt] (input)--(projected);
 \draw[->,inkgreen,line width=.9pt] (origin)--(projected);
 \node[font=\scriptsize,fill=softgreen,inner sep=1pt,text=inkgreen] at (.69,.10) {$Q_j\ket\psi$};
 \node[font=\scriptsize,text=inkgreen] at (.39,-.49) {$\operatorname{ran}Q_j$};
 \node[font=\scriptsize,align=center] at (0,-1.00) {$\ket{\psi_j}\propto Q_j\ket\psi$};
 \node[align=center,text width=3.05cm] at (0,-1.70)
   {Keep the projected\\quantum state.};
 \node[box,minimum width=.85cm,inner sep=3pt] (dec) at (3.8,.3) {$\cD_\tau$};
 \draw[qwire,->] (2.48,.3) node[above] {$B$}--(dec.west);
 \draw[qwire,->] (dec.east)--(5.12,.3) node[above] {$\widehat L$};
 \node[align=center,text width=3.0cm] at (3.8,-1.70) {Write its fidelity\\as a joint-state trace.};
 \node[draw=inkgreen,fill=softgreen,ellipse,inner sep=5pt] at (6.65,.32) {$\rho$};
 \node[draw=inkblue,fill=white,ellipse,inner sep=5pt] at (8.20,.32) {$\Gamma^G-\rho$};
 \node[align=center,text width=3.1cm] at (7.6,-1.70) {Bound the fixed-code\\error by a binary\\testing cost.};
 \node[align=center] at (11.4,.28) {$\mathbb E_G\Gamma^G-\rho=B_{d,m}$};
 \node[align=center,text width=3.05cm] at (11.4,-1.70) {Compute one second\\moment; then choose\\a fixed good code.};
 \foreach \x in {1.76,5.56,9.36}{\draw[->,thick,inkblue] (\x,0)--(\x+.27,0);}
 \draw[decorate,decoration={brace,mirror,amplitude=5pt},inkblue] (-1.7,-3.00)--(5.5,-3.00);
 \node[align=center,text=inkblue] at (1.9,-3.49) {Physical construction};
 \draw[decorate,decoration={brace,mirror,amplitude=5pt},inkgreen] (5.9,-3.00)--(13.1,-3.00);
 \node[align=center,text=inkgreen] at (9.5,-3.49) {Analysis and existence of a good code};
\end{tikzpicture}
\caption{The main idea of the proof of Theorem~\ref{thm:main}. In the first panel, the plane represents a selected subspace $\operatorname{ran}Q_j$ of Alice's space $A$. Conditioned on this outcome, the retained state is the normalized, nonzero projection $\ket{\psi_j}\propto Q_j\ket\psi$, not a measured basis label. Petz decoding gives an explicit physical recovery. Cheng's trace inequality is applied \emph{before} the random code is averaged, so the averaged upper bound on the error depends linearly on a comparison operator instead of an inverse operator. The final binary test need not be implemented by Alice and Bob. Here $\rho=\rho_{AB}$ and $\Gamma^G$ is the code-dependent comparison operator defined in \eqref{eq:fixed-fidelity-comparison}. Figure~\ref{fig:proof-route} gives a detailed proof guide.}
\label{fig:intro-overview}
\end{figure}
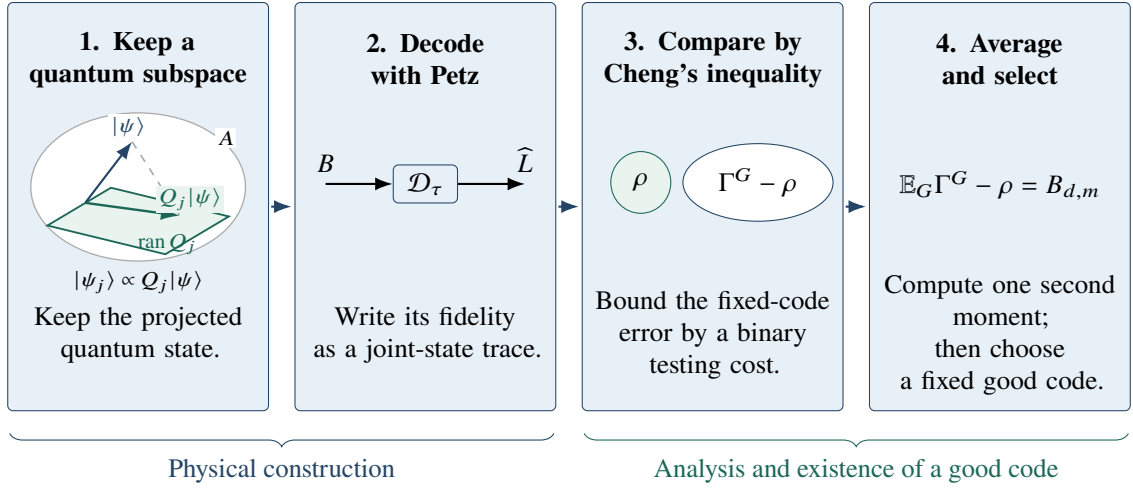

Figure~\ref{fig:intro-overview} summarizes the central proof idea: select a quantum subspace, write the Petz fidelity as a trace, apply Cheng's trace inequality while the code is fixed, and only then take the Haar average. The detailed proof in Section~\ref{sec:theorem} follows this same order.

\subsection{Paper organization}
Section~\ref{sec:tools} defines the three tasks and develops the testing, recovery, and phase-averaging tools. Section~\ref{sec:theorem} states Theorem~\ref{thm:main} and immediately proves it: a fixed-code lemma isolates Cheng's trace inequality, and a Haar-averaging lemma determines the alternative. Section~\ref{sec:consequences} gives the hypothesis-testing corollaries, the direct entanglement transmission specialization, and the error-preserving code conversion. Section~\ref{sec:examples} examines noiseless and erasure channels. Section~\ref{sec:second} treats independent repetitions, including the Gaussian correction, entanglement transmission at the same error for arbitrary reference--input states, zero variance, and an erasure-channel illustration. Section~\ref{sec:conclusion} summarizes the implications and remaining questions. Appendix~\ref{app:embedding} proves the isometric-embedding inequality, Appendix~\ref{app:haar} gives the elementary two-copy averaging proof, Appendix~\ref{app:integer} evaluates the refined integer threshold, Appendix~\ref{app:conversion} proves the code conversion, and Appendix~\ref{app:third-order} analyzes logarithmic terms and commuting hypotheses.
\FloatBarrier

\section{Operational setting and tools}\label{sec:tools}
The introduction described how a binary testing problem can certify an entanglement protocol. To make that connection precise, we now specify the physical tasks and develop the two ingredients of the proof: Cheng's trace inequality and an explicit Petz decoder. We define entanglement generation, entanglement transmission, and one-way entanglement distillation with a common infidelity convention. We then specialize the Petz formula to a received joint state and derive the trace expression for its fidelity. Finally, a phase-averaging identity explains how to record a subspace label without measuring the quantum state inside that subspace. These tools lead directly to the proof of Theorem~\ref{thm:main}. One additional operator inequality is introduced within that proof, at the point where a more general choice of encoding subspaces requires it.

Throughout this development, the physical decoder and the analysis test play different roles. Bob implements a channel on $B$. A joint binary effect on $AB$ certifies that channel's performance but is not an operation that the laboratories must perform.

\subsection{Notation}
Throughout we take all Hilbert spaces to be finite dimensional. A state is a positive semidefinite operator of trace one, and a subnormalized state is a positive semidefinite operator of trace at most one. A quantum channel is a completely positive, trace-preserving superoperator. An instrument is a family of completely positive, trace-nonincreasing superoperators whose sum is a channel. For Hermitian operators, $X\ge Y$ means that $X-Y$ is positive semidefinite. The identity operator is denoted by $I$, a projector satisfies $Q=Q^\dagger=Q^2$, and an effect satisfies $0\le T\le I$. We use $\norm X$ for the operator norm, $\norm X_1\coloneqq\Tr\sqrt{X^\dagger X}$ for the trace norm, and $\norm{\ket v}\coloneqq\sqrt{\langle v|v\rangle}$ for the Hilbert-space norm. We denote the support projector of a positive semidefinite operator $X$ by $\Pi_X$. An inverse is evaluated on the support and set to zero on the kernel unless a regularization is stated explicitly.

System labels specify where an operator acts. Identities on other systems may be implicit once those systems have been specified. For example, $Y_B\tau_{LB}Y_B$ means $(I_L\otimes Y_B)\tau_{LB}(I_L\otimes Y_B)$. We retain system labels in the coding calculations, where Alice's input and output spaces are different.

We take all information logarithms to have base two. The von Neumann entropy is defined as
\begin{equation}\label{eq:entropy}
 H(\xi)\coloneqq-\Tr[\xi\log_2\xi],\qquad H(B)_\rho\coloneqq H(\rho_B).
\end{equation}
The second notation specifies the reduced state of a multipartite system. 

The fidelity of states $\rho$ and $\sigma$ is defined as~\cite{Uhlmann1976}
\begin{equation}\label{eq:fidelity}
 F(\rho,\sigma)\coloneqq\norm{\sqrt\rho\sqrt\sigma}_1^2.
\end{equation}
Let $L$ and $\widehat L$ be $m$-dimensional quantum systems held at the end by Alice and Bob, respectively. Fix matching orthonormal bases and define the target
\begin{equation}\label{eq:target}
 \ket{\Phi^m}_{L\widehat L}\coloneqq
 \frac1{\sqrt m}\sum_{i=0}^{m-1}\ket i_L\ket i_{\widehat L},
 \qquad \Phi^m_{L\widehat L}\coloneqq\proj{\Phi^m}_{L\widehat L}.
\end{equation}
This state contains $\log_2m$ ebits of entanglement. 

\subsection{Coding tasks}\label{subsec:tasks}
The three tasks in Figure~\ref{fig:protocols} have the same target but different initial resources. In all of them, a normalized final state $\omega_{L\widehat L}$ has \emph{entanglement infidelity}
$1-F(\Phi^m_{L\widehat L},\omega_{L\widehat L})
=1-\Tr[\Phi^m_{L\widehat L}\omega_{L\widehat L}]$.
The maximally entangled target is fixed by \eqref{eq:target}. For an integer $m\ge1$ and $0\le\err\le1$, an \emph{$(m,\err)$ protocol} has output systems of dimension $m$ and infidelity at most $\err$. In the channel tasks we also call it an $(m,\err)$ code. The next definitions specify the operations allowed in each case.

\emph{Entanglement generation.}
An entanglement generation code for a channel $\cN\colon A'\to B$ consists of a normalized pure state $\zeta_{LA'}$ prepared by Alice and a decoding channel $\cD\colon B\to\widehat L$. Alice retains $L$ and transmits only $A'$. Its error is
\begin{equation}\label{eq:eg-error}
 \err_{\mathrm{eg}}\coloneqq
 1-\Tr\!\left[\Phi^m_{L\widehat L}
 (\id_L\otimes\cD\circ\cN)(\zeta_{LA'})\right].
\end{equation}
This is an $(m,\err)$ entanglement generation code when $\err_{\mathrm{eg}}\le\err$.  The code is selected in advance, and its state is prepared directly. 

\emph{Entanglement transmission.}
Alice starts with the prescribed maximally entangled state $\Phi^m_{L\widetilde L}$ in her laboratory, where $\widetilde L$ is an $m$-dimensional logical input system. An encoding channel $\mathcal E\colon\widetilde L\to A'$ acts only on $\widetilde L$. After sending the encoded system through $\cN$, Bob applies $\cD\colon B\to\widehat L$. The error is
\begin{equation}\label{eq:et-error}
 \err_{\mathrm{et}}\coloneqq
 1-\Tr\!\left[\Phi^m_{L\widehat L}
 (\id_L\otimes\cD\circ\cN\circ\mathcal E)(\Phi^m_{L\widetilde L})\right].
\end{equation}
This is an $(m,\err)$ entanglement transmission code when $\err_{\mathrm{et}}\le\err$. The reference marginal remains $I_L/m$, because every operation acts on the other share. We call the encoder \emph{isometric} when $\mathcal E(X)=VXV^\dagger$ for an isometry $V\colon\widetilde L\to A'$. Corollary~\ref{cor:flat} states the additional input condition under which our construction gives an encoder of this form. 

A relation between these two channel tasks is developed in Section~\ref{subsec:transmission-consequences}. In particular, Lemma~\ref{lem:eg-et-conversion} gives a conversion that preserves the error tolerance while reducing the target dimension.

\emph{One-way entanglement distillation.}
Alice and Bob instead start with a shared state $\rho_{AB}$. Alice applies an encoding instrument $(\mathcal E_j\colon A\to L)_j$ and sends its outcome $j$ to Bob, who applies the conditional decoding channel $\mathcal D_j\colon B\to\widehat L$. Define the outcome probabilities and, for positive-probability outcomes, the normalized conditional states by
\begin{align}
 p_j&\coloneqq\Tr[(\mathcal E_j\otimes\id_B)(\rho_{AB})],\label{eq:distillation-probability}\\
 \tau^j_{LB}&\coloneqq\frac{(\mathcal E_j\otimes\id_B)(\rho_{AB})}{p_j},
 \qquad p_j>0.\label{eq:distillation-instrument}
\end{align}
The final state and its infidelity are given by
\begin{align}
 \omega_{L\widehat L}
 &\coloneqq\sum_{j:p_j>0}p_j(\id_L\otimes\mathcal D_j)(\tau^j_{LB}),
 \label{eq:distillation-output}\\
 \err_{\to}
 &\coloneqq1-\Tr[\Phi^m_{L\widehat L}\omega_{L\widehat L}]\label{eq:distillation-error-definition}\\
 &=\sum_{j:p_j>0}p_j\left(1-\Tr\!\left[
 \Phi^m_{L\widehat L}(\id_L\otimes\mathcal D_j)(\tau^j_{LB})\right]\right).
 \label{eq:distillation-error}
\end{align}
This is an $(m,\err)$ one-way entanglement distillation protocol when $\err_{\to}\le\err$. The sum of Alice's outcome operations is trace preserving, so $\sum_jp_j=1$ and the final state is normalized. Outcomes with $p_j=0$ do not contribute. No limit is imposed on forward classical communication; however, no message is allowed to be sent from Bob to Alice.

These definitions specify the physical operations allowed in the three protocols of Figure~\ref{fig:protocols}. The joint binary effects introduced next are used only to analyze their performance.

\subsection{Binary testing and Cheng's trace inequality}\label{subsec:cheng}
A binary effect $T$ accepts the hypothesis represented by $X$ and rejects the hypothesis represented by $Y$. The two nonnegative error costs add to the objective in \eqref{eq:intro-cost}; they have the normalized Bayesian interpretation specified after that equation, but neither operator is required to have unit trace. The Helstrom--Holevo theorem states that~\cite{Helstrom1967,Holevo1973}
\begin{equation}\label{eq:helstrom}
 \cost(X,Y)=\frac{\Tr(X+Y)-\norm{X-Y}_1}{2}.
\end{equation}
Thus our coding bound ultimately involves a directly computable trace norm. The next inequality is the step that makes this binary cost useful for analyzing a recovery channel.

\begin{lemma}[Cheng's trace inequality]\label{lem:cheng}
Let $X$ and $Y$ be positive semidefinite operators on the same finite-dimensional space. Set $S\coloneqq X+Y$, $e\coloneqq\cost(X,Y)$ as in \eqref{eq:intro-cost}, and $s\coloneqq\Tr S$. If $s>0$, then
\begin{equation}\label{eq:cheng}
 \Tr[XS^{-1/2}YS^{-1/2}]
 \le e\left(1-\frac es\right)\le e
 \le\Tr[(I-T)X]+\Tr[TY]
\end{equation}
for every effect $T$. If $s=0$, all these quantities vanish.
\end{lemma}
This is \cite[Appendix~A, Lemma~1]{Cheng2023}; an alternative proof appears in \cite[Lemma~2.1]{Wilde2026Tilted}. The quadratic improvement is related to the binary near-optimality bound of Renes~\cite{Renes2017}. In our proof, we construct a positive semidefinite \emph{comparison operator} $\Gamma_{AB}\ge\rho_{AB}$ and apply the lemma with $X=\rho_{AB}$ and $Y=\Gamma_{AB}-\rho_{AB}$. For a fixed effect $T$, the last expression in \eqref{eq:cheng} is linear in $\Gamma_{AB}$. We can therefore average that expression over code choices without averaging an inverse operator.

For a state $\rho$, a positive semidefinite operator $\sigma$ on the same space, and $0\le\delta<1$, define the hypothesis-testing relative entropy by
\begin{equation}\label{eq:DH}
 D_H^\delta(\rho\Vert\sigma)
 \coloneqq-\log_2\min\{\Tr[T\sigma]:0\le T\le I,\ \Tr[T\rho]\ge1-\delta\}.
\end{equation}
The constraint bounds the probability of rejecting the normalized state $\rho$ by $\delta$. The minimized trace is a false-acceptance \emph{weight}. When $\sigma$ is a state, this weight is the usual type-II error probability. The minimum is attained by compactness. This quantity and closely related conventions occur in~\cite{BuscemiDatta2010,BrandaoDatta2011,WangRenner2012,DupuisEtAl2013}. The hypothesis-testing coherent information in \eqref{eq:intro-IH} uses the actual marginal $\rho_B$ and the alternative $I_A\otimes\rho_B$, whose trace is $d$.

\Needspace{9\baselineskip}
\subsection{Petz recovery and its coding representation}\label{subsec:petz}\label{subsec:petz-identification}
Section~\ref{subsec:cheng} introduced binary testing as a way to bound a trace expression through Cheng's trace inequality \eqref{eq:cheng}. To use that inequality for communication, we must first specify a physical decoder and relate its fidelity to such a trace. The Petz recovery channel provides the decoder: it is constructed from a reference input state and the channel describing the noise, and it acts only on Bob's received system.

This subsection explains that construction and identifies the reference state and effective channel associated with a pure coding input. We include the action outside the support of the reference output so that the decoder is a channel on Bob's entire space. Lemma~\ref{lem:petz-identification} then expresses the decoder through the operator-valued blocks of the received joint state. That block formula is the link to Section~\ref{subsec:state-decoder}, where we define the same decoder for an arbitrary shared state and prove the fidelity identity \eqref{eq:Frec} used in the coding analysis.

\begin{definition}[Petz recovery channel]\label{def:petz}
Let $\mathcal M$ be a quantum channel, and let $\sigma$ and $\xi$ be states on its input space, with $\xi$ fixed as the completion state. The completed Petz recovery channel, from the output space of $\mathcal M$ to its input space, is defined as
\begin{equation}\label{eq:petz-general}
 \mathcal P_{\sigma,\mathcal M}(Y)
 \coloneqq\sigma^{1/2}\mathcal M^\dagger
       \!\left(\mathcal M(\sigma)^{-1/2}Y\mathcal M(\sigma)^{-1/2}\right)\sigma^{1/2}
 +\Tr[(I-\Pi_{\mathcal M(\sigma)})Y]\xi.
\end{equation}
The inverse is the support inverse specified in Section~\ref{sec:tools}. The adjoint is defined by trace duality,
$\Tr[X\mathcal M(Y)]=\Tr[\mathcal M^\dagger(X)Y]$.
We denote the first term of \eqref{eq:petz-general} by $\mathcal P_{\sigma,\mathcal M,0}(Y)$ and call it the \emph{support part}. The choice of $\xi$ is fixed whenever this notation is used.
\end{definition}
Both terms in \eqref{eq:petz-general} are completely positive: the adjoint of a completely positive map is completely positive, and the second term is a measurement followed by state preparation. Trace duality applied to \eqref{eq:petz-general} gives
\begin{align}
 \Tr[\mathcal P_{\sigma,\mathcal M}(Y)]
 &=\Tr[\mathcal M(\sigma)^{-1/2}\mathcal M(\sigma)\mathcal M(\sigma)^{-1/2}Y]
       +\Tr[(I-\Pi_{\mathcal M(\sigma)})Y]
 \label{eq:petz-completion-trace-step-1}\\
 &=\Tr[\Pi_{\mathcal M(\sigma)} Y]+\Tr[(I-\Pi_{\mathcal M(\sigma)})Y]
 \label{eq:petz-completion-trace-step-2}\\
 &=\Tr Y.
 \label{eq:petz-completion-trace}
\end{align}
Thus \eqref{eq:petz-general} defines a channel on the full output space, not just on $\supp\mathcal M(\sigma)$. The completion term in \eqref{eq:petz-general} prepares a fixed state when the input lies outside the support accessible from the reference state. This construction does not assert that arbitrary noise can be perfectly reversed; its value here is an explicit decoder whose error can be bounded~\cite{BarnumKnill2002,BeigiDattaLeditzky2016,Renes2017}.

For the coding setting, fix a channel $\cN\colon A'\to B$ and a normalized pure reference--input state $\psi_{LA'}$, with $\dim L=m$. Set $\tau_{LB}\coloneqq(\id_L\otimes\cN)(\psi_{LA'})$. The operator-valued matrix blocks of this joint output in an orthonormal basis of $L$ are
\begin{equation}\label{eq:tau-blocks}
 \tau_B^{ij}\coloneqq\bra i_L\tau_{LB}\ket j_L,
 \qquad 0\le i,j<m.
\end{equation}
They are operators on $B$, and reconstruct the state and its marginal as
\begin{equation}\label{eq:tau-expansion}
 \tau_{LB}=\sum_{i,j=0}^{m-1}\ket i\bra j_L\otimes\tau_B^{ij},
 \qquad \tau_B=\sum_i\tau_B^{ii}.
\end{equation}
To identify the reference state and the effective noise in the Petz formula, suppose that the normalized pure input $\psi_{LA'}$ on $LA'$, with $\dim L=m$, has Schmidt decomposition
\begin{equation}\label{eq:schmidt-code}
 \ket\psi_{LA'}=\sum_{i:\lambda_i>0}\sqrt{\lambda_i}\ket i_L\ket{v_i}_{A'},
 \qquad \sigma_L\coloneqq\Tr_{A'}\psi_{LA'}=\sum_i\lambda_i\proj i_L.
\end{equation}
Introduce an $m$-dimensional logical input system $\widetilde L$ with the matching basis, the state $\sigma\coloneqq\sum_i\lambda_i\proj i_{\widetilde L}$, and support projector $P\coloneqq\Pi_\sigma$. Define $V\colon\widetilde L\to A'$ by $V\ket i=\ket{v_i}$ for $\lambda_i>0$ and by zero otherwise. Thus $V^\dagger V=P$. For an arbitrary fixed state $\chi_B$, define
\begin{equation}\label{eq:logical-extension}
 \mathcal M(X)\coloneqq
 \cN(VXV^\dagger)+\Tr[(I_{\widetilde L}-P)X]\chi_B.
\end{equation}
On the occupied subspace, this map sends the logical input through $V$ and the noise $\cN$; on the unused subspace it prepares $\chi_B$. The channel property of $\mathcal M$ is verified in Lemma~\ref{lem:petz-identification} below. We identify the matching bases of $\widetilde L$ and $\widehat L$ when writing the recovered output below.

\begin{lemma}[Petz decoder for a pure channel input]\label{lem:petz-identification}
For the input, output, and map in \eqref{eq:schmidt-code}--\eqref{eq:logical-extension}, $\mathcal M$ is a channel and $\mathcal M(\sigma)=\tau_B$, with $\tau_B$ given in \eqref{eq:tau-expansion}. For a fixed state $\xi_{\widehat L}$, its completed Petz decoder is
\begin{equation}\label{eq:petz}
 \cD_\tau(Y_B)
 \coloneqq\sigma^{1/2}\mathcal M^\dagger
       (\tau_B^{-1/2}Y_B\tau_B^{-1/2})\sigma^{1/2}
       +\Tr[(I_B-\Pi_{\tau_B})Y_B]\xi_{\widehat L},
\end{equation}
where the first term is written on $\widehat L$ using the matching bases. Writing $\cD_{\tau,0}$ for that term, its adjoint satisfies
\begin{equation}\label{eq:petz-block-adjoint}
 \cD_{\tau,0}^\dagger(\ket i\bra j_{\widehat L})
 =\tau_B^{-1/2}\tau_B^{ij}\tau_B^{-1/2},
 \qquad 0\le i,j<m,
\end{equation}
with the blocks defined in \eqref{eq:tau-blocks}. Thus the decoder is specified by the joint output, its matching bases, and the completion state.
\end{lemma}
\begin{proof}
The map \eqref{eq:logical-extension} first measures whether its input belongs to $\supp\sigma$. On that subspace it applies $V$ and $\cN$; on the orthogonal subspace it prepares $\chi_B$. This also removes cross terms between the two subspaces. Both terms in \eqref{eq:logical-extension} are completely positive. Since $V^\dagger V=P$ and $\cN$ is trace preserving, their traces are $\Tr[PX]$ and $\Tr[(I-P)X]$, respectively, so that $\mathcal M$ is a channel.

Expanding the input in \eqref{eq:schmidt-code} and applying $\cN$ gives
\begin{align}
 \tau_{LB}
 &=\sum_{a,b:\lambda_a\lambda_b>0}\sqrt{\lambda_a\lambda_b}\ket a\bra b_L
                       \otimes\cN(\ket{v_a}\bra{v_b})
 \label{eq:schmidt-output-expansion-step-1}\\
 &=\sum_{a,b}\sqrt{\lambda_a\lambda_b}\ket a\bra b_L
                       \otimes\mathcal M(\ket a\bra b_{\widetilde L}).
 \label{eq:schmidt-output-expansion}
\end{align}
The terms with a zero Schmidt coefficient vanish. Taking a partial matrix element yields
\begin{equation}\label{eq:schmidt-blocks}
 \tau_B^{ij}=\sqrt{\lambda_i\lambda_j}\mathcal M(\ket i\bra j),
\end{equation}
and then taking the partial trace of $ \tau_{LB}$ over $L$ yields
\begin{equation}\label{eq:petz-output-marginal}
  \tau_B=\sum_i\lambda_i\mathcal M(\proj i)
             =\mathcal M(\sigma).
\end{equation}
Using \eqref{eq:petz-output-marginal}, the decoder in \eqref{eq:petz} is exactly the channel in \eqref{eq:petz-general}, with the logical and recovered output bases identified and with completion state $\xi_{\widehat L}$.
To determine its adjoint on matrix units, use
$\sigma\ket i=\lambda_i\ket i$ and hence
$\sigma^{1/2}\ket i=\sqrt{\lambda_i}\ket i$:
\begin{align}
 \cD_{\tau,0}^\dagger(\ket i\bra j_{\widehat L})
 &=\tau_B^{-1/2}\mathcal M(\sigma^{1/2}
          \ket i\bra j_{\widetilde L}\sigma^{1/2})\tau_B^{-1/2}
 \label{eq:petz-adjoint3-step-1}\\
 &=\sqrt{\lambda_i\lambda_j}\tau_B^{-1/2}
           \mathcal M(\ket i\bra j_{\widetilde L})\tau_B^{-1/2}
 \label{eq:petz-adjoint3-step-2}\\
 &=\tau_B^{-1/2}\tau_B^{ij}\tau_B^{-1/2}.
 \label{eq:petz-adjoint3}
\end{align}
This proves \eqref{eq:petz-block-adjoint}, including indices with zero Schmidt coefficients. The choice of $\chi_B$ on the unused input subspace has disappeared from the formula.
\end{proof}
For $\sigma=I_{\widetilde L}/m$, \eqref{eq:petz} is the transpose decoder for a maximally mixed code input studied in~\cite{BarnumKnill2002,NgMandayam2010,BeigiDattaLeditzky2016}. Ng and Mandayam established near-optimality of the transpose channel for approximate subspace error correction under a worst-case pure-state fidelity criterion~\cite{NgMandayam2010}; Mandayam and Ng extended this approach to approximate subsystem codes, with near-optimality guarantees for specified classes of codes and noise~\cite{MandayamNg2012}. Those results provide context for this explicit recovery construction; the entanglement-infidelity bound proved here follows directly from \eqref{eq:Frec} and Cheng's trace inequality \eqref{eq:cheng}, rather than from a conversion of their worst-case criterion. The logical register in the lemma is used to describe the reference state and the effective noise; no additional system is sent during the protocol.

\subsection{A decoder specified by the received joint state}\label{subsec:state-decoder}
The block formula \eqref{eq:petz-block-adjoint} no longer requires a channel description or a Schmidt decomposition. We therefore use it to specify Bob's decoder for an arbitrary conditional state in entanglement distillation, as well as for a channel output in entanglement generation. We first verify that it is a channel, and then derive the single trace expression that will be used throughout the coding proof.

\begin{definition}[State-dependent decoding channel]\label{def:recovery}
Let $\tau_{LB}$ be a state with $\dim L=m$, and let its blocks in the matching target basis be as in \eqref{eq:tau-blocks}. Define the support part $\cD_{\tau,0}$ by its adjoint,
\begin{equation}\label{eq:recovery}
 \cD_{\tau,0}^\dagger(\ket i\bra j_{\widehat L})
 \coloneqq\tau_B^{-1/2}\tau_B^{ij}\tau_B^{-1/2}.
\end{equation}
For a fixed state $\xi_{\widehat L}$, define its completion by
\begin{equation}\label{eq:recovery-action}
 \cD_\tau(Y_B)
 \coloneqq\Tr[(I_B-\Pi_{\tau_B})Y_B]\xi_{\widehat L} + \sum_{i,j=0}^{m-1}
 \Tr[\tau_B^{-1/2}\tau_B^{ji}\tau_B^{-1/2}Y_B]\ket i\bra j_{\widehat L}.
\end{equation}
For a conditional state $\tau^j_{LB}$, we write $\mathcal D_j\coloneqq\cD_{\tau^j}$.
\end{definition}
The reversed block indices $(j,i)$ in \eqref{eq:recovery-action} follow from trace duality. The next proposition proves this explicitly and justifies the word ``channel'' in the definition.

\begin{proposition}[Complete positivity and trace preservation]\label{prop:recovery-channel}
For every state $\tau_{LB}$ and fixed completion state $\xi_{\widehat L}$, the map $\cD_\tau$ in \eqref{eq:recovery-action} is completely positive and trace preserving. Its support part has the adjoint \eqref{eq:recovery}. The completion term has no effect on $(\id_L\otimes\cD_\tau)(\tau_{LB})$.
\end{proposition}
\begin{proof}
For an operator $Y_B$ supported on $\supp\tau_B$, trace duality gives
\begin{align}
 \bra i\cD_{\tau,0}(Y_B)\ket j
 &=\Tr[\ket j\bra i_{\widehat L}\cD_{\tau,0}(Y_B)]
 \label{eq:recovery-entry3-step-1}\\
 &=\Tr[\cD_{\tau,0}^\dagger(\ket j\bra i_{\widehat L})Y_B]
 \label{eq:recovery-entry3-step-2}\\
 &=\Tr[\tau_B^{-1/2}\tau_B^{ji}\tau_B^{-1/2}Y_B].
 \label{eq:recovery-entry3}
\end{align}
The matrix entries in \eqref{eq:recovery-entry3} give the sum in \eqref{eq:recovery-action}. Complete positivity follows from the Choi criterion~\cite{Choi1975}: the Choi operator of its adjoint is
\begin{equation}\label{eq:recovery-choi}
 \sum_{i,j}\ket i\bra j_L\otimes
 \cD_{\tau,0}^\dagger(\ket i\bra j_{\widehat L})
 =(I_L\otimes\tau_B^{-1/2})\tau_{LB}(I_L\otimes\tau_B^{-1/2})\ge0.
\end{equation}
The preparation term in \eqref{eq:recovery-action} is also completely positive. Summing the diagonal entries specified by \eqref{eq:recovery-entry3} gives
\begin{align}
 \Tr[\cD_\tau(Y_B)]
 &=\Tr[\tau_B^{-1/2}\tau_B\tau_B^{-1/2}Y_B]
   +\Tr[(I_B-\Pi_{\tau_B})Y_B]
 \label{eq:recovery-trace2-step-1}\\
 &=\Tr[\Pi_{\tau_B}Y_B]+\Tr[(I_B-\Pi_{\tau_B})Y_B]
 \label{eq:recovery-trace2-step-2}\\
 &=\Tr Y_B.
 \label{eq:recovery-trace2}
\end{align}
Finally, $\tau_{LB}$ is supported on $L\otimes\supp\tau_B$. To see this, take $\ket b\in\ker\tau_B$. The nonnegative numbers
$\langle i,b|\tau_{LB}|i,b\rangle$ sum to $\langle b|\tau_B|b\rangle=0$, so they all vanish. Positivity then implies $\tau_{LB}\ket{i,b}=0$ for every $i$. The completion term therefore vanishes on the state being decoded.
\end{proof}

\begin{lemma}[Trace formula for the decoding fidelity]\label{lem:recovery-fidelity}
Let $\tau_{LB}$ be a state with $\dim L=m$, let $\cD_\tau$ be the channel in \eqref{eq:recovery-action}, and let $\Phi^m_{L\widehat L}$ be the target in \eqref{eq:target}. Then
\begin{align}
 \Frec(\tau_{LB})
 &\coloneqq\Tr[\Phi^m_{L\widehat L}(\id_L\otimes\cD_\tau)(\tau_{LB})]
 \label{eq:Frec-step-1}\\
 &=\frac1m\Tr[\tau_{LB}(I_L\otimes\tau_B^{-1/2})
                    \tau_{LB}(I_L\otimes\tau_B^{-1/2})].
 \label{eq:Frec}
\end{align}
The formula includes singular $\tau_B$ and is independent of $\xi_{\widehat L}$.
\end{lemma}
Equation~\eqref{eq:Frec-step-1} defines the operational fidelity, and \eqref{eq:Frec} gives the trace expression used in the proof of Theorem~\ref{thm:main}. It contains only the joint state and the inverse square root of Bob's marginal. Cheng's trace inequality will eventually remove this inverse from the error bound before we average over code choices.
\begin{proof}[Proof of Lemma~\ref{lem:recovery-fidelity}]
For an operator $X_{\widehat L}$, expansion of the target gives
\begin{equation}\label{eq:entangled-overlap-entry}
 \bra{\Phi^m}(\ket i\bra j_L\otimes X_{\widehat L})\ket{\Phi^m}
 =\frac1m\bra iX_{\widehat L}\ket j.
\end{equation}
Insert \eqref{eq:tau-expansion} and use \eqref{eq:recovery-entry3}:
\begin{align}
 \Frec(\tau_{LB})
 &=\frac1m\sum_{i,j}\bra i\cD_\tau(\tau_B^{ij})\ket j
 \label{eq:Frec-blocks-step-1}\\
 &=\frac1m\sum_{i,j}
 \Tr[\tau_B^{ij}\tau_B^{-1/2}\tau_B^{ji}\tau_B^{-1/2}]
 \label{eq:Frec-blocks-step-2}\\
 &=\frac1m\Tr[\tau_{LB}(I_L\otimes\tau_B^{-1/2})
                    \tau_{LB}(I_L\otimes\tau_B^{-1/2})].
 \label{eq:Frec-blocks}
\end{align}
For \eqref{eq:Frec-blocks}, multiply the block expansions in \eqref{eq:tau-expansion} and use
$\Tr[\ket i\bra j\ket a\bra b]=\delta_{ja}\delta_{ib}$. The surviving indices are $(i,j)$ and $(j,i)$, with no commutativity or reality assumption. Proposition~\ref{prop:recovery-channel} makes the completion term zero.
\end{proof}

For completeness, the Petz interpretation also holds for a mixed joint state in an arbitrary fixed basis. On a logical register $\widetilde L$, set $\sigma_{\widetilde L}\coloneqq\tau_L^{\top}$, where $\top$ is matrix transpose in that basis, and define
\begin{equation}\label{eq:canonical-petz-reference}
 \ket{\psi_\sigma}_{L\widetilde L}
 \coloneqq(I_L\otimes\sigma_{\widetilde L}^{1/2})
              \sum_{i=0}^{m-1}\ket i_L\ket i_{\widetilde L}.
\end{equation}
This vector is normalized and has marginals $\tau_L$ and $\sigma_{\widetilde L}$. Purify $\tau_{LB}$ to $\psi_{LBE}$ and compare the Schmidt decompositions of this purification and \eqref{eq:canonical-petz-reference} across $L$. An isometry on the occupied logical subspace maps the latter purification to the former. Follow it by $\Tr_E$ and extend the unused input as in \eqref{eq:logical-extension}. This gives a channel $\mathcal M\colon\widetilde L\to B$ satisfying
\begin{equation}\label{eq:canonical-petz-blocks}
 \tau_{LB}=(\id_L\otimes\mathcal M)(\psi_\sigma),
 \qquad \tau_B^{ij}=\mathcal M(\sigma^{1/2}\ket i\bra j\sigma^{1/2}).
\end{equation}
The adjoint of its Petz support part is therefore exactly \eqref{eq:recovery}. The purification is only a mathematical description, not a system supplied to Bob. This observation justifies the notation $F_{\mathrm P}$ for all the conditional states used below. A basis change $W$ on $L$ together with $\overline W$ on $\widehat L$ leaves $\ket{\Phi^m}$ invariant, in agreement with the basis-independent trace in \eqref{eq:Frec}.

\subsection{Removing coherence between subspaces}\label{subsec:operators}
The fidelity formula \eqref{eq:Frec} tells us what must be controlled after Alice's operation. To organize those conditional states, we next describe the effect of recording a subspace label. Alice's local operation will retain an $m$-dimensional subspace and identify it with her output system $L$. We call this a \emph{subspace selection}; it preserves superpositions inside the selected space. Recording which subspace was selected removes coherence between different outcome labels, but it need not reveal the state within the selected space. The operator tool describing exactly this loss of coherence is a pinching channel.

For mutually orthogonal projectors $(P_j)_{j=0}^{r-1}$ summing to the identity, define
\begin{equation}\label{eq:pinch}
 \overline{\Delta}(X)\coloneqq\sum_{j=0}^{r-1}P_jXP_j
 =\frac1r\sum_{k=0}^{r-1}W_kXW_k^\dagger,
 \qquad W_k\coloneqq\sum_{j=0}^{r-1}e^{2\pi\mathrm i kj/r}P_j.
\end{equation}
Each $W_k$ is unitary, and $W_0=I$. On a block $P_jXP_\ell$, the phase average is equal to one for $j=\ell$ and equal to zero otherwise, which proves the equality. If $X\ge0$, every conjugated term is positive semidefinite and $W_0=I$. Retaining just this identity term gives the \emph{phase-average order bound}
\begin{equation}\label{eq:pinching-order}
 X\le r\overline\Delta(X),\qquad X\ge0.
\end{equation}
This is the inequality used later to compare a state with its block-diagonal version. The same unitary-averaging idea also gives a useful dimension bound.

\begin{lemma}[Dimension bound from phase and shift averaging]\label{lem:dimension}
Let $X_{LB}$ be a positive semidefinite operator, with $\dim L=m$, and set $X_B\coloneqq\Tr_LX_{LB}$. Then
\begin{equation}\label{eq:dimension-bound}
 X_{LB}\le m I_L\otimes X_B.
\end{equation}
\end{lemma}
\begin{proof}
Use the rank-one projectors $P_i=\proj i_L$ in \eqref{eq:pinch}, and define the cyclic shift $\mathsf X_L\ket i\coloneqq\ket{i+1\bmod m}$. First averaging the phases removes off-diagonal blocks, and then averaging the shifts places the sum of the diagonal blocks in every diagonal position. Consequently,
\begin{equation}\label{eq:weyl-average}
 \frac1{m^2}\sum_{a,k=0}^{m-1}
 (\mathsf X_L^aW_k\otimes I_B)X_{LB}
 (\mathsf X_L^aW_k\otimes I_B)^\dagger
 =\frac{I_L}{m}\otimes X_B.
\end{equation}
The $(a,k)=(0,0)$ summand is $X_{LB}$. Discarding all the other positive semidefinite summands gives $X_{LB}/m^2\le I_L\otimes X_B/m$, which is \eqref{eq:dimension-bound}. This unitary-averaging proof is used in \cite[proof of Proposition~3.5]{DupuisEtAl2013}.
\end{proof}
In particular, $dI_A\otimes\rho_B-\rho_{AB}\ge0$, which verifies positivity of the alternative in \eqref{eq:intro-alternative}. The same inequality gives the support inclusions needed for the inverses in the proof.

\Needspace{24\baselineskip}
\section{One-shot coding theorem and proof}\label{sec:theorem}\label{sec:proof}
The preceding tools now connect an explicit decoder to a binary test. We first state the achievable error for every integer target dimension, then prove it before drawing its information-theoretic consequences. 

\begin{theorem}[Direct binary-testing achievability]\label{thm:main}
Let $\rho_{AB}$ be a state on finite-dimensional systems, with $d\coloneqq\dim A\ge2$, and let $m\in\{2,\ldots,d\}$ be an integer. Define
\begin{equation}\label{eq:alternative}
 B_{d,m}\coloneqq\frac{dm-1}{d^2-1}(dI_A\otimes\rho_B-\rho_{AB}),
 \qquad e_{d,m}\coloneqq\cost(\rho_{AB},B_{d,m}),
\end{equation}
using \eqref{eq:intro-cost}. There is a one-way entanglement distillation protocol from $\rho_{AB}$, with at most $d$ classical outcomes sent by Alice, whose error \eqref{eq:distillation-error}, measured against \eqref{eq:target}, obeys
\begin{equation}\label{eq:main-bound}
 \err_{\to}\le f_{d,m}(e_{d,m}),
 \qquad f_{d,m}(x)\coloneqq x\left(1-\frac{x}{dm}\right).
\end{equation}
Bob uses the decoding channel of Proposition~\ref{prop:recovery-channel} for each normalized conditional state; Section~\ref{subsec:state-decoder} gives its Petz representation, including mixed conditional states. If $\rho_{AB}=(\id_A\otimes\cN)(\phi_{AA'})$ for a channel $\cN\colon A'\to B$ and a normalized pure reference--input state, there is also an unassisted entanglement generation code whose error \eqref{eq:eg-error} satisfies \eqref{eq:main-bound}. The sender directly prepares a fixed normalized pure state selected in the code design. No classical side channel or runtime postselection is used. The target $m=1$ has error zero and is treated separately.
\end{theorem}

The alternative $ B_{d,m}$ has trace $dm-1$, so its sum with $\rho$ has trace $dm$. Also $0\le e_{d,m}\le1$ by choosing $T=0$ in the testing problem. Therefore $f_{d,m}$ is increasing and concave on $[0,1]$, and the right-hand side of \eqref{eq:main-bound} takes values between zero and one. In particular, the weaker bound $\err_{\to}\le e_{d,m}$ also holds for the constructed entanglement distillation protocol, and the same relaxation applies to its entanglement generation code counterpart.

\subsection{Proof strategy and the encoding instruments}\label{subsec:construction}
\begin{figure}[!p]
\centering
\captionsetup{skip=11pt}
\begin{tikzpicture}[font=\small,x=1cm,y=1cm,
 routebox/.style={box,text width=5.4cm,minimum height=1.05cm,inner sep=6pt},
 connectornote/.style={font=\footnotesize,align=center,fill=white,inner sep=3pt}]
 \node[anchor=west,font=\bfseries,text=inkblue] at (-.1,1.18)
 {Choose $G$ and keep it fixed during the recovery analysis};
 \node[routebox] (orth) at (3,0)
 {Orthogonal spaces: $m$ divides $d$\\$A=\bigoplus_j\operatorname{ran}Q_j$};
 \node[routebox] (cyc) at (10,0)
 {Cyclic windows: every $m\le d$\\$\sum_jQ_j=mI_A$};
 \foreach \j in {0,1,2}{
  \draw[fill=softgreen,draw=inkgreen,rounded corners=1pt]
     (.14+1.94*\j,-2.10) rectangle (1.90+1.94*\j,-1.23);
  \node[font=\footnotesize] at (1.02+1.94*\j,-1.67)
     {$\ket{g_{\the\numexpr2*\j\relax}},\ket{g_{\the\numexpr2*\j+1\relax}}$};
 }
 \node[font=\footnotesize,align=center,text width=5.8cm] (orthiconnote) at (3,-2.70)
 {Each block retains two coherent amplitudes.};
 \foreach \a in {0,...,4}{\node[font=\footnotesize] at (7.60+1.20*\a,-1.42) {$\ket{g_\a}$};}
 \draw[inkgreen,line width=1pt] (7.16,-1.86)--(10.44,-1.86);
 \draw[inkgreen,line width=1pt] (7.16,-1.78)--(7.16,-1.94);
 \draw[inkgreen,line width=1pt] (10.44,-1.78)--(10.44,-1.94);
 \node[font=\scriptsize,text=inkgreen,anchor=east] at (6.99,-1.86) {$Q_0$};
 \draw[inkblue,line width=1pt] (8.36,-2.22)--(11.64,-2.22);
 \draw[inkblue,line width=1pt] (8.36,-2.14)--(8.36,-2.30);
 \draw[inkblue,line width=1pt] (11.64,-2.14)--(11.64,-2.30);
 \node[font=\scriptsize,text=inkblue,anchor=east] at (8.19,-2.22) {$Q_1$};
 \node[font=\footnotesize,align=center,text width=5.8cm] (cyciconnote) at (10,-2.85)
 {Shared basis directions; distinct outcome labels.};
 \draw[flow] (orth.south)--(3,-1.08);
 \draw[flow] (cyc.south)--(10,-1.00);
 \node[routebox] (orthstate) at (3,-4.35)
 {Record a block label\\$\rho_{AB}\longmapsto R_{AB}$};
 \node[routebox] (cycstate) at (10,-4.35)
 {Keep a coherent outcome register\\$J_G\colon A\to CL$, then dephase $C$};
 \draw[flow,shorten <=3pt] (orthiconnote.south)--(orthstate.north);
 \draw[flow,shorten <=3pt] (cyciconnote.south)--(cycstate.north);
 \node[routebox] (orthgam) at (3,-6.65)
 {Phase average on $AB$\\$\Gamma^G_{AB}=dZ_{AB}$};
 \node[routebox] (cycgam) at (10,-6.65)
 {Return the trace to $AB$\\$\Gamma^G_{AB}=J_G^\dagger(mdZ_{CLB})J_G$};
 \draw[flow] (orthstate.south)--(orthgam.north);
 \draw[flow] (cycstate.south)--(cycgam.north);
 \node[connectornote] at (3,-5.50) {Keep the identity phase};
 \node[connectornote] at (10,-5.50) {Lemma~\ref{lem:isometric-recovery-trace}};
 \node[box,fill=softgreen,draw=inkgreen,text width=11.55cm,
       minimum height=1.22cm,inner sep=7pt] (cheng) at (6.5,-9.00)
 {\textbf{Fixed-code bound: Lemma~\ref{lem:fixed-code}}\\[3pt]
 $\Gamma^G_{AB}\ge\rho_{AB}$,\quad $\Tr\Gamma^G_{AB}=dm$,\quad
 $1-F_G\le f_{d,m}(e_G)$};
 \draw[flow] (orthgam.south)--(cheng.north -| 3,0);
 \draw[flow] (cycgam.south)--(cheng.north -| 10,0);
 \node[connectornote] at (6.5,-7.78) {Cheng's trace inequality};
 \node[box,text width=11.55cm,minimum height=1.5cm,inner sep=7pt] (avg) at (6.5,-11.60)
 {\textbf{Now average $G$: Lemma~\ref{lem:haar-comparison}}\\[3pt]
 $\mathbb E_G\Gamma^G_{AB}-\rho_{AB}=B_{d,m}$; a fixed test and concavity give\\[3pt]
 $1-\mathbb E_GF_G\le f_{d,m}(e_{d,m})$.};
 \draw[flow] (cheng.south)--(avg.north);
 \node[connectornote] at (6.5,-10.10) {Average a linear testing expression};
 \node[routebox,minimum height=1.03cm] (dist) at (3,-14.20)
 {\textbf{Entanglement distillation}\\Fix $G$; send the actual outcome $j$.};
 \node[routebox,minimum height=1.03cm] (gen) at (10,-14.20)
 {\textbf{Entanglement generation}\\Fix $G,j$; prepare the selected state.};
 \draw[flow] (avg.south)--(6.5,-13.08)-|(dist.north);
 \draw[flow] (6.5,-13.08)-|(gen.north);
\end{tikzpicture}
\caption{A guide to the proof, with the two subspace constructions shown separately. The small diagrams illustrate $d=6,m=2$ on the left and two cyclic windows for $d=5,m=3$ on the right. Each green block on the left is a retained two-dimensional subspace; the bars on the right mark two overlapping three-dimensional subspaces without crossing the basis labels. No individual basis state is measured. Only the right-hand route requires the isometric-embedding inequality. Both routes reach the same fixed-code guarantee and the same Haar average. In the right-hand comparison-operator formula, $J_G$ is tensored with $I_B$. Here $e_G=\mathsf e(\rho_{AB},\Gamma^G_{AB}-\rho_{AB})$.}
\label{fig:proof-route}
\end{figure}
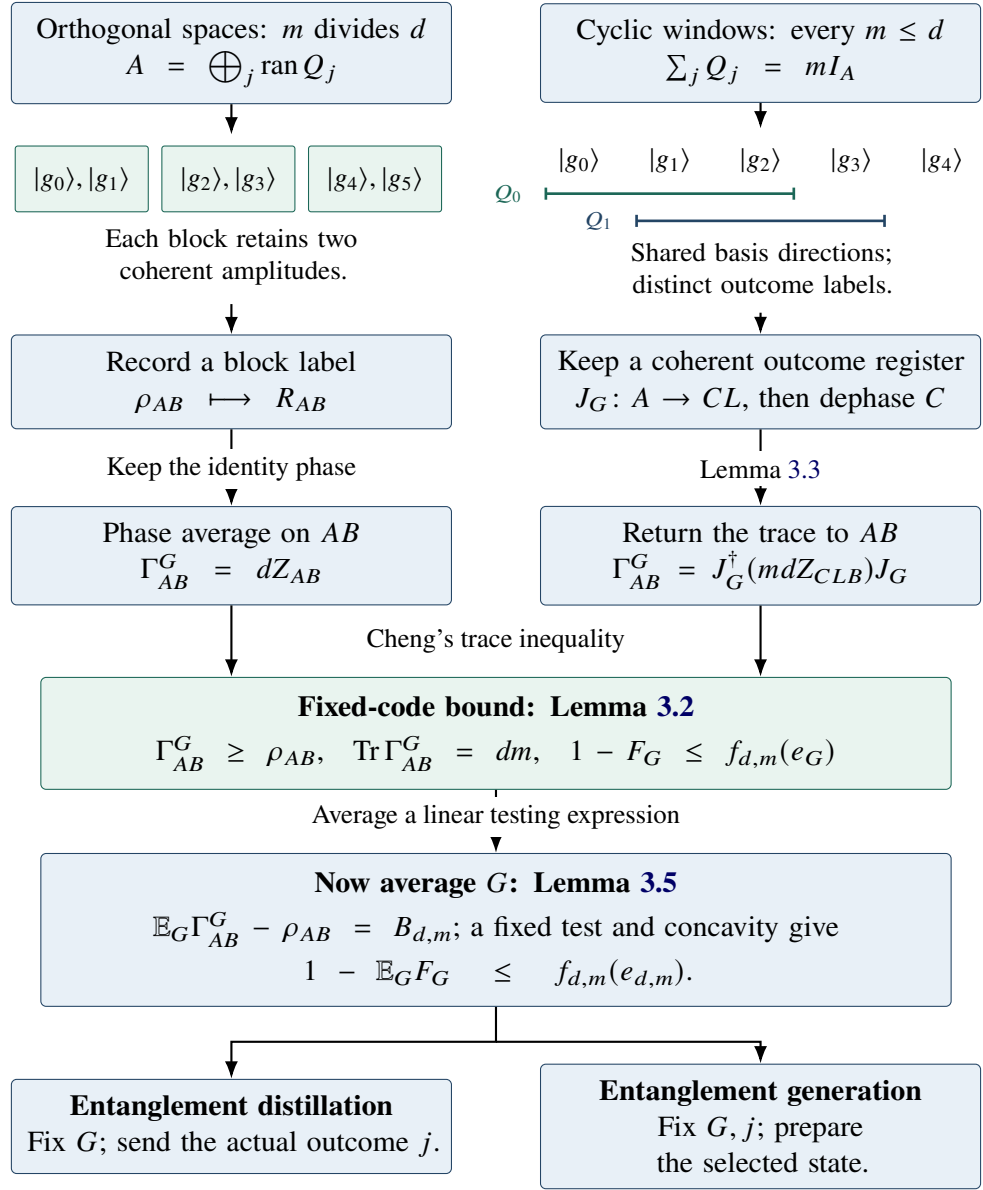

We prove Theorem~\ref{thm:main} in two stages. First, with the code fixed, Lemma~\ref{lem:fixed-code} bounds its decoding error by a binary cost. Second, Lemma~\ref{lem:haar-comparison} evaluates the average comparison operator; a fixed test and a concavity argument then give the theorem. Figure~\ref{fig:proof-route} distinguishes the physical construction from these two analytical stages.

The code is specified by a unitary $G_A$, which is fixed for now. Alice selects an $m$-dimensional subspace without measuring the individual amplitudes within it. We give two choices of subspaces. The first makes the proof transparent when $m$ divides $d$; the second covers every integer $m$ without enlarging Alice's input space.

\emph{Orthogonal blocks when $m$ divides $d$.}
Write $r=d/m$, fix mutually orthogonal rank-$m$ projectors $(P_j)_{j=0}^{r-1}$ with sum $I_A$, and set $Q_j\coloneqq G_AP_jG_A^\dagger$. Choose isometries $V_j\colon L\to A$ satisfying $V_jV_j^\dagger=Q_j$. Alice applies the quantum instrument $(\mathcal E_j\colon A\to L)_{j=0}^{r-1}$, whose outcome superoperators are $\mathcal E_j(X)\coloneqq K_jXK_j^\dagger$ with $K_j\coloneqq V_j^\dagger$. Thus $K_j$ is a Kraus operator representing the operation for the classical outcome $j$, rather than a channel by itself. Since $K_j^\dagger K_j=Q_j$ and $\sum_jQ_j=I_A$, the sum of these superoperators is trace preserving. The instrument records which block is occupied and identifies that block with $L$, while preserving every superposition within it.

\emph{Cyclic windows for an arbitrary target dimension, when $m$ need not divide $d$.}
Set $r=d$ and $\ket{g_a}_A\coloneqq G_A\ket a_A$, with basis indices taken modulo $d$. Define
\begin{equation}\label{eq:cyclic-projectors}
 Q_j\coloneqq\sum_{a=0}^{m-1}\proj{g_{j+a}}_A,
 \qquad V_j\coloneqq\sum_{a=0}^{m-1}\ket{g_{j+a}}_A\bra a_L,
 \quad 0\le j<d,
\end{equation}
\begin{equation}\label{eq:cyclic-K}
 K_j\coloneqq\frac1{\sqrt m}V_j^\dagger
 =\frac1{\sqrt m}\sum_{a=0}^{m-1}\ket a_L\bra{g_{j+a}}_A.
\end{equation}
Every basis vector occurs in exactly $m$ windows. Consequently,
\begin{equation}\label{eq:cyclic-identities}
 \sum_jQ_j=mI_A,\qquad K_j^\dagger K_j=Q_j/m,\qquad
 K_jK_j^\dagger=I_L/m,\qquad \sum_jK_j^\dagger K_j=I_A.
\end{equation}
An outcome identifies a window, not a basis vector inside it. Figure~\ref{fig:cyclic} shows a nondivisible example. When $m=d$, each outcome selects the whole input space, so no quantum dimension is lost.

\begin{figure}[!t]
\centering
\begin{tikzpicture}[font=\small,x=1.05cm,y=.61cm]
 \node[anchor=west,font=\bfseries] at (-2,5.4) {$d=5$, $m=3$: five overlapping code spaces};
 \foreach \a in {0,...,4}{\node at (\a,4.55) {$\ket{g_\a}$};}
 \foreach \j in {0,...,4}{
  \node[anchor=east] at (-.8,3.7-\j) {$Q_\j$};
  \foreach \a in {0,...,4}{
   \pgfmathtruncatemacro{\rel}{mod(\a-\j+5,5)}
   \ifnum\rel<3
    \filldraw[fill=softblue,draw=inkblue] (\a-.33,3.38-\j) rectangle (\a+.33,4.02-\j);
    \node at (\a,3.7-\j) {$\checkmark$};
   \else
    \draw[gray!45] (\a-.33,3.38-\j) rectangle (\a+.33,4.02-\j);
   \fi
  }
 }
 \node[align=left,text width=4.3cm,anchor=west] at (5.0,2.0)
 {Every column occurs three times.\\[5pt]Therefore $\sum_jQ_j=3I_A$.\\[5pt]The factor $1/\sqrt3$ in $K_j$ makes the instrument trace preserving.};
 \node[anchor=west,align=left] at (-1,-1.65) {An outcome retains three coherent amplitudes; it does not reveal which basis vector was present.};
\end{tikzpicture}
\caption{Cyclic windows remove the divisibility restriction. A marked entry belongs to the indicated subspace. Wrapping the window around the last column gives equal coverage of every basis vector. The measurement outcome identifies a selected subspace, while its output remains a three-dimensional quantum system.}
\label{fig:cyclic}
\end{figure}

The two constructions have a common normalization:
\begin{equation}\label{eq:frame-identities}
 \sum_{j=0}^{r-1}Q_j=\frac{rm}{d}I_A,
 \qquad K_j=\sqrt{\frac d{rm}}\,V_j^\dagger,
 \qquad \sum_jK_j^\dagger K_j=I_A.
\end{equation}
Thus the family $(\mathcal E_j)_{j=0}^{r-1}$, with $\mathcal E_j(X)\coloneqq K_jXK_j^\dagger$, is a quantum encoding instrument in either case. Equation~\eqref{eq:frame-identities} proves trace preservation after summing its $r\le d$ outcomes. Define the subnormalized conditional states, their marginals, and their probabilities by
\begin{equation}\label{eq:conditional-r}
 \theta^j_{LB}\coloneqq(K_j\otimes I_B)\rho_{AB}(K_j^\dagger\otimes I_B),
 \qquad \eta_B^j\coloneqq\Tr_L\theta^j_{LB},
\end{equation}
\begin{equation}\label{eq:conditional-probability}
 p_j\coloneqq\Tr\theta^j_{LB},\qquad
 \tau^j_{LB}\coloneqq\theta^j_{LB}/p_j\quad(p_j>0).
\end{equation}
Completeness gives $\sum_jp_j=1$. Bob uses $\mathcal D_j\coloneqq\cD_{\tau^j}$ from Definition~\ref{def:recovery}; an outcome with $p_j=0$ can have an arbitrary decoder. For this fixed $G_A$, set
\begin{equation}\label{eq:fixed-fidelity-comparison}
 F_G\coloneqq\sum_{j:p_j>0}p_j\Frec(\tau^j_{LB}),
 \qquad \Gamma^G_{AB}\coloneqq d\sum_{j=0}^{r-1}Q_j\otimes\eta_B^j.
\end{equation}
Here $\Frec$ is the fidelity in \eqref{eq:Frec}, so $1-F_G$ is the physical entanglement distillation error in \eqref{eq:distillation-error}. The operator $\Gamma^G_{AB}$ will connect that error to a binary test. The following lemma states this connection before its proof, so the purpose of the ensuing trace calculations is explicit.

\subsection{The fixed-code bound from Cheng's trace inequality}\label{subsec:direct}
\begin{lemma}[Fixed-code recovery bound]\label{lem:fixed-code}
Let $\rho_{AB}$ be a state with $d=\dim A\ge2$, let $2\le m\le d$, and use either encoding construction in Section~\ref{subsec:construction} for a fixed unitary $G_A$. Let $F_G$ and $\Gamma^G_{AB}$ be as in \eqref{eq:fixed-fidelity-comparison}. Then
\begin{equation}\label{eq:fixed-code-order}
 \Gamma^G_{AB}\ge\rho_{AB},\qquad \Tr\Gamma^G_{AB}=dm.
\end{equation}
Set $e_G\coloneqq\cost(\rho_{AB},\Gamma^G_{AB}-\rho_{AB})$, using \eqref{eq:intro-cost}, and use $f_{d,m}$ from \eqref{eq:main-bound}. For every effect $T_{AB}$,
\begin{align}
 1-F_G&\le f_{d,m}(e_G)\le e_G
 \label{eq:fixed-code-error-step-1}\\
 &\le\Tr[(I_{AB}-T_{AB})\rho_{AB}]
       +\Tr[T_{AB}(\Gamma^G_{AB}-\rho_{AB})].
 \label{eq:fixed-code-error}
\end{align}
\end{lemma}
Equation~\eqref{eq:fixed-code-error} is the decisive simplification of the error analysis. The decoding fidelity in \eqref{eq:Frec} contains inverse square roots of conditional output states, each of which changes with the code. The error bound in \eqref{eq:fixed-code-error} replaces that nonlinear dependence by an expression that is affine in the single comparison operator $\Gamma^G_{AB}$. Because the inequality holds for every effect $T_{AB}$, we may fix one test independently of $G$ and then average: only $\mathbb E_G\Gamma^G_{AB}$ remains to be calculated. There is no need to average a recovery channel or an inverse operator, or to optimize a different decoder for each analysis test. This is precisely where Cheng's trace inequality \eqref{eq:cheng} turns the recovery calculation into a tractable coding bound.

Equation~\eqref{eq:fixed-code-error-step-1} also retains the quadratic improvement $f_{d,m}(e_G)$; its concavity will preserve that improvement in the averaging step \eqref{eq:mean-error}. We now prove Lemma~\ref{lem:fixed-code}, first for orthogonal blocks and then for the overlapping subspaces of the cyclic construction.

\subsubsection{Proof for orthogonal blocks}\label{subsec:orthogonal}
\begin{proof}[Proof of Lemma~\ref{lem:fixed-code} in the orthogonal case]
\renewcommand{\qedsymbol}{}
Here $d=rm$ and $K_j=V_j^\dagger$. Write the unnormalized state in the $j$th block, still on the original space, as
\begin{equation}
\rho^j_{AB}\coloneqq(Q_j\otimes I_B)\rho_{AB}(Q_j\otimes I_B).
\end{equation}
Then, from \eqref{eq:conditional-r},
\begin{align}
\theta^j_{LB}& =(V_j^\dagger\otimes I_B)\rho^j_{AB}(V_j\otimes I_B),\\
\eta_B^j & =\Tr_A\rho^j_{AB}.
\end{align}
Define
\begin{equation}\label{eq:orthogonal-RZ}
 R_{AB}\coloneqq\sum_j\rho^j_{AB},\qquad
 Z_{AB}\coloneqq\sum_jQ_j\otimes\eta_B^j.
\end{equation}
Both operators in \eqref{eq:orthogonal-RZ} are block diagonal in the mutually orthogonal subspaces. In particular, functional calculus gives
$Z_{AB}^{-1/2}=\sum_jQ_j\otimes(\eta_B^j)^{-1/2}$, with the support convention throughout.

We now calculate each outcome contribution to the fidelity in \eqref{eq:fixed-fidelity-comparison}, displaying the normalization factors. For $p_j>0$, the marginal of $\tau^j$ is $\eta_B^j/p_j$, and
$(\eta_B^j/p_j)^{-1/2}=\sqrt{p_j}(\eta_B^j)^{-1/2}$. Lemma~\ref{lem:recovery-fidelity} therefore gives
\begin{align}
 p_j\Frec(\tau^j_{LB})
 &=\frac{p_j}{m}\Tr\!\left[
 \frac{\theta^j_{LB}}{p_j}
 (I_L\otimes\sqrt{p_j}(\eta_B^j)^{-1/2})
 \frac{\theta^j_{LB}}{p_j}
 (I_L\otimes\sqrt{p_j}(\eta_B^j)^{-1/2})\right]
 \label{eq:weighted-fidelity}\\
 &=\frac1m\Tr[\theta^j_{LB}(I_L\otimes(\eta_B^j)^{-1/2})
                \theta^j_{LB}(I_L\otimes(\eta_B^j)^{-1/2})]
 \label{eq:weighted-fidelity-conditional}\\
 &=\frac1m\Tr[\rho^j_{AB}(Q_j\otimes(\eta_B^j)^{-1/2})
                \rho^j_{AB}(Q_j\otimes(\eta_B^j)^{-1/2})].
 \label{eq:weighted-fidelity-block}
\end{align}
The scalar factors in \eqref{eq:weighted-fidelity} multiply to $p_jp_j/p_j^2=1$. Equation~\eqref{eq:weighted-fidelity-block} follows by inserting $V_j^\dagger V_j=I_L$ and $V_jV_j^\dagger=Q_j$ and cycling the trace. If $p_j=0$, positivity implies $\theta^j=\rho^j=\eta_B^j=0$, so the contribution is zero. Products of distinct blocks vanish. Summing \eqref{eq:weighted-fidelity-block} consequently yields
\begin{equation}\label{eq:orthogonal-F}
 F_G=\frac1m\Tr[R_{AB}Z_{AB}^{-1/2}R_{AB}Z_{AB}^{-1/2}].
\end{equation}
Equation~\eqref{eq:weighted-fidelity-conditional} holds for an arbitrary instrument outcome, including the cyclic construction; only \eqref{eq:weighted-fidelity-block} uses the orthogonal isometry.

Let $\overline\Delta$ be pinching in the projectors $Q_j\otimes I_B$, and let $W_k$ be its phase unitaries from \eqref{eq:pinch}. Then $R_{AB}=\overline\Delta(\rho_{AB})$. Pinching is self-adjoint for the trace inner product: for arbitrary operators $X,Y$,
\begin{align}
 \Tr[\overline\Delta(X)Y]
 &=\sum_j\Tr[(Q_j\otimes I_B)X(Q_j\otimes I_B)Y]
 \label{eq:pinching-trace-duality-step-1}\\
 &=\Tr[X\overline\Delta(Y)].
 \label{eq:pinching-trace-duality}
\end{align}
Since $Z_{AB}^{-1/2}R_{AB}Z_{AB}^{-1/2}$ is block diagonal, it is fixed by $\overline\Delta$. Applying \eqref{eq:pinching-trace-duality} to the first factor in \eqref{eq:orthogonal-F}, and then using \eqref{eq:pinch} for the other factor, gives
\begin{align}
 F_G
 &=\frac1m\Tr[\rho_{AB}Z_{AB}^{-1/2}R_{AB}Z_{AB}^{-1/2}]
 \label{eq:orthogonal-phase3}\\
 &=\frac1{mr}\sum_{k=0}^{r-1}
  \Tr[\rho_{AB}Z_{AB}^{-1/2}W_k\rho_{AB}W_k^\dagger Z_{AB}^{-1/2}]
 \label{eq:orthogonal-phase-average}\\
 &\ge\frac1d\Tr[\rho_{AB}Z_{AB}^{-1/2}\rho_{AB}Z_{AB}^{-1/2}]
 \label{eq:orthogonal-identity-phase}\\
 &=\Tr[\rho_{AB}(\Gamma^G_{AB})^{-1/2}
                        \rho_{AB}(\Gamma^G_{AB})^{-1/2}].
 \label{eq:orthogonal-comparison-fidelity}
\end{align}
Every summand in \eqref{eq:orthogonal-phase-average} is nonnegative: it is the trace of a product of two positive semidefinite operators. Equation~\eqref{eq:orthogonal-identity-phase} keeps only $k=0$ and uses $W_0=I_{AB}$ and $mr=d$. Equation~\eqref{eq:orthogonal-comparison-fidelity} then uses $\Gamma^G_{AB}=dZ_{AB}$ from \eqref{eq:fixed-fidelity-comparison}. Scaling by $d$ scales the support inverse square root by $d^{-1/2}$. Figure~\ref{fig:blocks} shows why this argument does not destroy coherence inside a selected space. 

\begin{figure}[!t]
\centering
\begin{tikzpicture}[font=\small,x=.66cm,y=.66cm]
 \node[anchor=west,font=\bfseries] at (0,7.3) {Before recording the outcome};
 \node[anchor=west,font=\bfseries] at (10.6,7.3) {After recording the outcome};
 \fill[softblue] (0,0) rectangle (6,6);
 \foreach \j in {0,1,2}{
  \fill[softgreen] (10.6+2*\j,4-2*\j) rectangle (12.6+2*\j,6-2*\j);
 }
 \foreach \shift in {0,10.6}{
  \draw[step=1,gray!35,thin] (\shift,0) grid (\shift+6,6);
  \draw[thick] (\shift,0) rectangle (\shift+6,6);
  \foreach \j in {2,4}{
   \draw[gray!70] (\shift+\j,0)--(\shift+\j,6);
   \draw[gray!70] (\shift,\j)--(\shift+6,\j);
  }
 }
 \foreach \row in {0,1,2}{
  \foreach \col in {0,1,2}{
   \node[fill=softblue,inner sep=2pt] at (2*\col+1,5-2*\row) {$X_{\row\col}$};
   \ifnum\row=\col
    \node[fill=softgreen,inner sep=2pt] at (11.6+2*\col,5-2*\row) {$X_{\row\col}$};
   \else
    \node[fill=white,inner sep=2pt,text=gray] at (11.6+2*\col,5-2*\row) {$0$};
   \fi
  }
 }
 \draw[->,thick] (6.55,3)--node[above,align=center]{dephase only\\the outcome label}(10,3);
 \node[align=center,text width=5cm] at (3,-1.2) {Each block acts on a whole quantum code space and on $B$.};
 \node[align=center,text width=5cm] at (13.6,-1.2) {Within each retained block, all quantum coherences remain.};
\end{tikzpicture}
\caption{The pinching used in the proof. The small squares are schematic coordinates, not separately measured outcomes. An orthogonal code-space decomposition, or the outcome register of its coherent instrument representation, determines the large blocks. Removing coherence only between these blocks preserves the within-block recovery fidelity. This is why no conversion from classical guessing error is needed.}
\label{fig:blocks}
\end{figure}
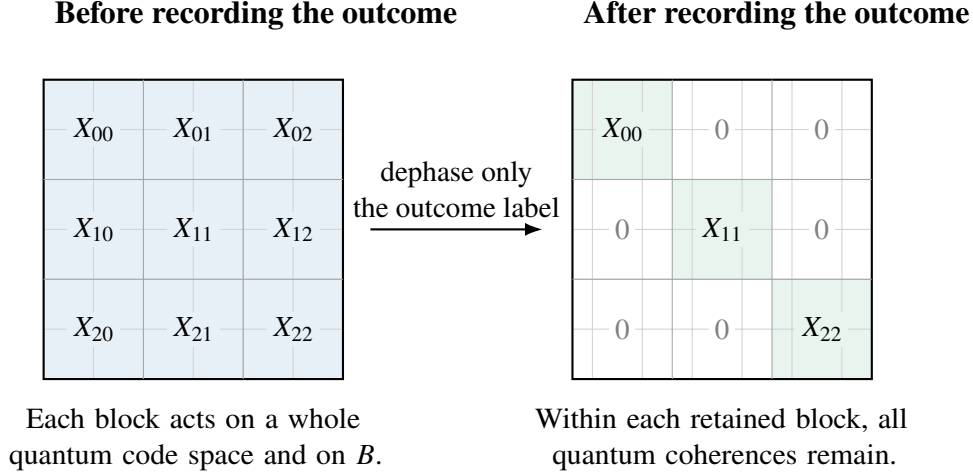

It remains to verify the order and trace needed for Cheng's trace inequality. The phase-average bound \eqref{eq:pinching-order} gives $\rho_{AB}\le rR_{AB}$. Apply the dimension bound \eqref{eq:dimension-bound} to each $m$-dimensional block to get $\rho^j_{AB}\le mQ_j\otimes\eta_B^j$. After summing over $j$,
\begin{equation}\label{eq:orthogonal-order}
 \rho_{AB}\le rR_{AB}\le rmZ_{AB}=\Gamma^G_{AB},
 \qquad \Tr\Gamma^G_{AB}=d\sum_jm\Tr\eta_B^j=dm.
\end{equation}
In particular, $\supp\rho_{AB}\subseteq\supp\Gamma^G_{AB}$, and hence
$\Tr[\rho_{AB}(\Gamma^G_{AB})^{-1/2}\Gamma^G_{AB}(\Gamma^G_{AB})^{-1/2}]=1$.
Subtracting the lower bound in \eqref{eq:orthogonal-comparison-fidelity} from one and applying \eqref{eq:cheng} with $X=\rho_{AB}$ and $Y=\Gamma^G_{AB}-\rho_{AB}$ gives
\begin{align}
 1-F_G
 &\le\Tr[\rho_{AB}(\Gamma^G_{AB})^{-1/2}
       (\Gamma^G_{AB}-\rho_{AB})(\Gamma^G_{AB})^{-1/2}]
 \label{eq:direct-Cheng4-step-1}\\
 &\le e_G\left(1-\frac{e_G}{dm}\right)\le e_G
 \label{eq:direct-Cheng4-step-2}\\
 &\le\Tr[(I_{AB}-T_{AB})\rho_{AB}]
                 +\Tr[T_{AB}(\Gamma^G_{AB}-\rho_{AB})].
 \label{eq:direct-Cheng4}
\end{align}
This proves \eqref{eq:fixed-code-order}--\eqref{eq:fixed-code-error} in the orthogonal case, including the quadratic correction.
\end{proof}

\subsubsection{Proof for cyclic windows}\label{subsec:cyclic}\label{subsec:embedding}
Cyclic windows are not mutually orthogonal subspaces of $A$, so pinching directly in the $Q_j$ would be incorrect. Instead, we represent the instrument coherently in an outcome register $C$, whose labels \emph{are} orthogonal. After applying the same phase argument there, we must return its trace expression to the original $AB$ space. The following technical lemma supplies exactly that step. It is introduced here because the orthogonal proof did not need it; its elementary proof, including singular operators, is in Appendix~\ref{app:embedding}. It is also a consequence of operator Jensen's inequality~\cite{HansenPedersen2003}.

\begin{lemma}[Recovery trace under an isometric embedding]\label{lem:isometric-recovery-trace}
Let $J\colon\mathcal H\to\mathcal K$ be an isometry between finite-dimensional Hilbert spaces, let $\rho$ be a state on $\mathcal H$, and let $S\ge\Omega\coloneqq J\rho J^\dagger$ be a positive semidefinite operator on $\mathcal K$. Set $\Gamma\coloneqq J^\dagger SJ$, so $\Gamma\ge\rho$. Then
\begin{equation}\label{eq:isometric-recovery-trace}
 \Tr[\Omega S^{-1/2}\Omega S^{-1/2}]
 \ge\Tr[\rho\Gamma^{-1/2}\rho\Gamma^{-1/2}],
\end{equation}
where inverses are evaluated on their supports.
\end{lemma}
This is a statement about traces, not an unrestricted order relation between generalized inverses. That distinction matters when $S$ has a kernel.

\begin{proof}[Proof of Lemma~\ref{lem:fixed-code} in the cyclic case]
Completeness in \eqref{eq:cyclic-identities} makes the coherent instrument map an isometry:
\begin{equation}\label{eq:isometry}
 J_G\coloneqq\sum_{j=0}^{d-1}\ket j_C\otimes K_j\colon A\to CL,
 \qquad J_G^\dagger J_G=\sum_jK_j^\dagger K_j=I_A.
\end{equation}
Define operators on $CLB$ by
\begin{align}
 \Omega_{CLB}&\coloneqq(J_G\otimes I_B)\rho_{AB}(J_G^\dagger\otimes I_B),
 \label{eq:coherent-operators}\\
 R_{CLB}&\coloneqq\sum_j\proj j_C\otimes\theta^j_{LB},
 \label{eq:coherent-dephased-state}\\
 Z_{CLB}&\coloneqq\sum_j\proj j_C\otimes I_L\otimes\eta_B^j.
 \label{eq:coherent-marginal-operator}
\end{align}
Dephasing $C$ in $\Omega_{CLB}$ produces $R_{CLB}$: its $j$th diagonal block is $(K_j\otimes I_B)\rho_{AB}(K_j^\dagger\otimes I_B)=\theta^j_{LB}$. The states in \eqref{eq:coherent-operators} and \eqref{eq:coherent-dephased-state} are normalized. By \eqref{eq:weighted-fidelity-conditional}, the block sum of the outcome fidelities is
\begin{equation}\label{eq:general-F}
 F_G=\frac1m\Tr[R_{CLB}Z_{CLB}^{-1/2}R_{CLB}Z_{CLB}^{-1/2}].
\end{equation}
Equation~\eqref{eq:general-F} retains the actual probabilities $p_j$; it does not assume a maximally mixed source marginal.

Now use the $d$ phase unitaries
$W_k\coloneqq\sum_j e^{2\pi\mathrm i kj/d}\proj j_C\otimes I_{LB}$.
Using the marginal operator in \eqref{eq:coherent-marginal-operator}, define
\begin{equation}\label{eq:general-comparison-operator}
 S_{CLB}\coloneqq mdZ_{CLB}.
\end{equation}
The trace-duality argument in \eqref{eq:pinching-trace-duality}, applied to $C$, and the average in \eqref{eq:pinch} give
\begin{align}
 F_G
 &=\frac1m\Tr[\Omega_{CLB}Z_{CLB}^{-1/2}R_{CLB}Z_{CLB}^{-1/2}]
 \label{eq:general-phase3}\\
 &=\frac1{md}\sum_{k=0}^{d-1}\Tr[\Omega_{CLB}Z_{CLB}^{-1/2}
                  W_k\Omega_{CLB}W_k^\dagger Z_{CLB}^{-1/2}]
 \label{eq:general-phase-average}\\
 &\ge\frac1{md}\Tr[\Omega_{CLB}Z_{CLB}^{-1/2}
                              \Omega_{CLB}Z_{CLB}^{-1/2}]
 \label{eq:general-identity-phase}\\
 &=\Tr[\Omega_{CLB}S_{CLB}^{-1/2}\Omega_{CLB}S_{CLB}^{-1/2}].
 \label{eq:general-comparison-fidelity}
\end{align}
Equation~\eqref{eq:general-identity-phase} retains the nonnegative identity-phase term from \eqref{eq:general-phase-average}. Equation~\eqref{eq:general-comparison-fidelity} uses $S_{CLB}^{-1/2}=(md)^{-1/2}Z_{CLB}^{-1/2}$ on the support, by \eqref{eq:general-comparison-operator}. Equations~\eqref{eq:pinching-order} and \eqref{eq:dimension-bound}, applied respectively to the outcome register and to each $L$ block, imply
\begin{equation}\label{eq:general-order}
 \Omega_{CLB}\le dR_{CLB}\le mdZ_{CLB}=S_{CLB}.
\end{equation}
We may therefore apply Lemma~\ref{lem:isometric-recovery-trace} with the isometry $J_G\otimes I_B$. Its compressed operator is exactly the $\Gamma^G_{AB}$ already defined in \eqref{eq:fixed-fidelity-comparison}:
\begin{align}
 (J_G^\dagger\otimes I_B)S_{CLB}(J_G\otimes I_B)
 &=md\sum_j(K_j^\dagger K_j)\otimes\eta_B^j
 \label{eq:comparison-operator-step-1}\\
 &=d\sum_jQ_j\otimes\eta_B^j
 \label{eq:comparison-operator-step-2}\\
 &=\Gamma^G_{AB}.
 \label{eq:comparison-operator}
\end{align}
Equation~\eqref{eq:comparison-operator-step-1} uses the block diagonal form of $S_{CLB}$ from \eqref{eq:general-comparison-operator} and \eqref{eq:coherent-marginal-operator}, so only matching outcome labels survive. Equation~\eqref{eq:comparison-operator-step-2} uses $K_j^\dagger K_j=Q_j/m$. Combining \eqref{eq:general-comparison-fidelity} with \eqref{eq:isometric-recovery-trace} gives
\begin{equation}\label{eq:F-lower}
 F_G\ge\Tr[\rho_{AB}(\Gamma^G_{AB})^{-1/2}
                       \rho_{AB}(\Gamma^G_{AB})^{-1/2}],
 \qquad \Gamma^G_{AB}\ge\rho_{AB}.
\end{equation}
The order follows by compressing \eqref{eq:general-order}, since $J_G$ is an isometry. Furthermore,
\begin{equation}\label{eq:comparison-trace}
 \Tr\Gamma^G_{AB}=d\sum_j(\Tr Q_j)(\Tr\eta_B^j)
                 =dm\sum_jp_j=dm.
\end{equation}
We have now established the same trace lower bound, operator order, and trace normalization as in the orthogonal proof. Applying \eqref{eq:direct-Cheng4-step-1}--\eqref{eq:direct-Cheng4} proves \eqref{eq:fixed-code-error-step-1}--\eqref{eq:fixed-code-error} for cyclic windows as well.
\end{proof}

\subsection{A single Haar moment evaluates the comparison operator}\label{subsec:haar}
The fixed-code lemma (Lemma~\ref{lem:fixed-code}) has removed the inverse operators from the testing expression. We can now average the unitary $G_A$. Haar measure is the probability measure on unitaries invariant under multiplication by a fixed unitary. Under this measure each $Q_j$ is a uniformly distributed rank-$m$ projector, although different $Q_j$ are generally correlated. Independence is not needed: the comparison operator is a sum of terms involving one projector at a time.

Let us begin by stating a standard two-copy Haar averaging identity in a form that applies to an arbitrary operator. Its symmetric/antisymmetric expression explains the averaging, while its identity/swap expression will make the partial trace particularly simple.

Let $\mathbb F$ be the swap operator on two $d$-dimensional registers, defined by $\mathbb F\ket u\ket v=\ket v\ket u$, and define
\begin{equation}\label{eq:symmetry-projectors}
 \Pi_{\mathcal S}\coloneqq\frac{I+\mathbb F}{2},\qquad
 \Pi_{\mathcal A}\coloneqq\frac{I-\mathbb F}{2}.
\end{equation}
These project onto the symmetric and antisymmetric subspaces. Since $\Tr I=d^2$ and $\Tr\mathbb F=d$, their traces are $d(d+1)/2$ and $d(d-1)/2$.

\begin{lemma}[Two-copy unitary average]\label{lem:two-copy-average}
Let $X$ be an arbitrary operator on the tensor product of two $d$-dimensional spaces, with $d\ge2$, and let $G$ be Haar distributed on the unitary group of one such space. With $\mathbb F$ the swap operator and $\Pi_{\mathcal S},\Pi_{\mathcal A}$ as in \eqref{eq:symmetry-projectors}, the following identity holds~\cite[Eqs.~(7.175) and~(7.179)]{Watrous2018}:
\begin{align}
 \mathbb E_G[(G\otimes G)X(G\otimes G)^\dagger]
 &=\frac{2\Tr[\Pi_{\mathcal S}X]}{d(d+1)}\Pi_{\mathcal S}
   +\frac{2\Tr[\Pi_{\mathcal A}X]}{d(d-1)}\Pi_{\mathcal A}
 \label{eq:werner-twirl-step-1}\\
 &=\frac{d\Tr X-\Tr[\mathbb F X]}{d(d^2-1)}I
   +\frac{d\Tr[\mathbb F X]-\Tr X}{d(d^2-1)}\mathbb F.
 \label{eq:werner-twirl}
\end{align}
\end{lemma}
The average preserves the trace against each projector in \eqref{eq:symmetry-projectors} and produces a scalar multiple of that projector. For positive semidefinite $X$, this distributes the trace weight uniformly within each subspace; the identity itself holds for an arbitrary complex $X$. Appendix~\ref{app:haar} proves \eqref{eq:werner-twirl-step-1}--\eqref{eq:werner-twirl} using only phases, permutations, and one superposition, without representation theory. We next use this identity to evaluate the comparison operator from the fixed-code bound.

\begin{lemma}[Haar average of the comparison operator]\label{lem:haar-comparison}
Let $\rho_{AB}$ be a state with $d=\dim A\ge2$, let $m \in \{2, \ldots, d\}$, and let $G_A$ be Haar distributed. For either construction in Section~\ref{subsec:construction}, the operator $\Gamma^G_{AB}$ in \eqref{eq:fixed-fidelity-comparison} satisfies
\begin{equation}\label{eq:averaged-comparison}
 \mathbb E_G\Gamma^G_{AB}-\rho_{AB}
 =\frac{dm-1}{d^2-1}(dI_A\otimes\rho_B-\rho_{AB})=B_{d,m},
\end{equation}
where $B_{d,m}$ is the alternative in \eqref{eq:alternative}.
\end{lemma}
\begin{proof}
We apply Lemma~\ref{lem:two-copy-average} to the sum of the tensor squares of Alice's projectors, then take a partial trace. Keeping the two copies of Alice's system explicit will show why the identity term produces $I_A\otimes\rho_B$ and the swap term produces $\rho_{AB}$.

Let $\bar A$ be a copy of $A$, and write $\rho_{\bar AB}$ for the same state with $A$ relabeled. For either construction, $Q_j=G_AP_jG_A^\dagger$ for fixed rank-$m$ projectors $P_j$. The normalization \eqref{eq:frame-identities} implies
\begin{align}
 \eta_B^j
 &=\Tr_L[(K_j\otimes I_B)\rho_{AB}(K_j^\dagger\otimes I_B)]
 \label{eq:outcome-marginal-step-1}\\
 &=\Tr_A[(K_j^\dagger K_j\otimes I_B)\rho_{AB}]
 \label{eq:outcome-marginal-step-2}\\
 &=\frac d{rm}\Tr_A[(Q_j\otimes I_B)\rho_{AB}].
 \label{eq:outcome-marginal}
\end{align}
To verify the change of traced system in \eqref{eq:outcome-marginal-step-2}, test both sides against an arbitrary operator $O_B$:
\begin{align}
 \Tr[O_B\eta_B^j]
 &=\Tr[(I_L\otimes O_B)(K_j\otimes I_B)\rho_{AB}(K_j^\dagger\otimes I_B)]
 \label{eq:outcome-marginal-duality-step-1}\\
 &=\Tr[(K_j^\dagger K_j\otimes O_B)\rho_{AB}]
 \label{eq:outcome-marginal-duality-step-2}\\
 &=\Tr\!\left[O_B\Tr_A[(K_j^\dagger K_j\otimes I_B)\rho_{AB}]\right].
 \label{eq:outcome-marginal-duality}
\end{align}
Equation~\eqref{eq:outcome-marginal} then uses $K_j^\dagger K_j=dQ_j/(rm)$ from \eqref{eq:frame-identities}. Set
\begin{equation}\label{eq:projector-sum}
 X_0\coloneqq\sum_{j=0}^{r-1}P_j\otimes P_j,\qquad
 X_G\coloneqq(G\otimes G)X_0(G\otimes G)^\dagger
            =\sum_{j=0}^{r-1}Q_j\otimes Q_j.
\end{equation}
The two copies of each projector act on $A$ and $\bar A$, respectively. For a single summand, pulling an operator on the untraced system $A$ outside the partial trace gives
\begin{align}
 \Tr_{\bar A}[(Q_{j,A}\otimes Q_{j,\bar A}\otimes I_B)
                    (I_A\otimes\rho_{\bar AB})]
 &=\Tr_{\bar A}[Q_{j,A}\otimes
                    ((Q_{j,\bar A}\otimes I_B)\rho_{\bar AB})]
 \label{eq:projector-partial-trace-step-1}\\
 &=Q_{j,A}\otimes
                \Tr_{\bar A}[(Q_{j,\bar A}\otimes I_B)\rho_{\bar AB}].
 \label{eq:projector-partial-trace}
\end{align}
Here $Q_{j,A}$ and $Q_{j,\bar A}$ are matching copies of $Q_j$. Substituting \eqref{eq:outcome-marginal} into the definition of $\Gamma^G_{AB}$ in \eqref{eq:fixed-fidelity-comparison}, then using \eqref{eq:projector-partial-trace} and summing, gives
\begin{align}
 \Gamma^G_{AB}
 &=d\sum_jQ_{j,A}\otimes\eta_B^j
 \label{eq:comparison-expansion-step-1}\\
 &=\frac{d^2}{rm}\sum_jQ_{j,A}\otimes
             \Tr_{\bar A}[(Q_{j,\bar A}\otimes I_B)\rho_{\bar AB}]
 \label{eq:comparison-expansion-step-2}\\
 &=\frac{d^2}{rm}\Tr_{\bar A}\!\left[
       \left(\sum_jQ_{j,A}\otimes Q_{j,\bar A}\otimes I_B\right)
       (I_A\otimes\rho_{\bar AB})\right]
 \label{eq:comparison-expansion-step-3}\\
 &=\frac{d^2}{rm}\Tr_{\bar A}
       [(X_G\otimes I_B)(I_A\otimes\rho_{\bar AB})].
 \label{eq:comparison-expansion}
\end{align}
Equation~\eqref{eq:comparison-expansion} uses precisely $X_G=\sum_jQ_{j,A}\otimes Q_{j,\bar A}$ from \eqref{eq:projector-sum}. Relabeling the traced copy of $A$ does not introduce a transpose.

According to Lemma~\ref{lem:two-copy-average}, only two scalar traces are needed to evaluate the expectation of $\Gamma^G_{AB}$ over Haar random $G$. The identity
$\Tr[\mathbb F(M\otimes N)]=\Tr[MN]$ gives
\begin{equation}\label{eq:twirl-weights}
 \Tr X_0=\sum_j(\Tr P_j)^2=rm^2,\qquad
 \Tr[\mathbb F X_0]=\sum_j\Tr[P_j^2]=rm.
\end{equation}
Applying Lemma~\ref{lem:two-copy-average}, specifically the identity/swap expression in \eqref{eq:werner-twirl}, to the two traces in \eqref{eq:twirl-weights} yields
\begin{equation}\label{eq:haar-projector}
 \mathbb E_G X_G
 =\frac{rm(dm-1)}{d(d^2-1)}I_{A\bar A}
  +\frac{rm(d-m)}{d(d^2-1)}\mathbb F_{A\bar A}.
\end{equation}
The partial traces of the identity and swap terms are
\begin{align}
 \Tr_{\bar A}[I_A\otimes\rho_{\bar AB}]&=I_A\otimes\rho_B,
 \label{eq:haar-contraction-step-1}\\
 \Tr_{\bar A}[(\mathbb F_{A\bar A}\otimes I_B)
                   (I_A\otimes\rho_{\bar AB})]&=\rho_{AB}.
 \label{eq:haar-contraction}
\end{align}
For the swap identity in \eqref{eq:haar-contraction}, first consider a tensor product $M_{\bar A}\otimes N_B$ in place of $\rho_{\bar AB}$. Testing its partial trace against an arbitrary $O_A$ gives
\begin{align}
 \Tr\!\left[O_A\Tr_{\bar A}[\mathbb F_{A\bar A}(I_A\otimes M_{\bar A})]\right]
 &=\Tr[(O_A\otimes M_{\bar A})\mathbb F_{A\bar A}]
 \label{eq:swap-partial-trace-check-step-1}\\
 &=\Tr[O_AM_A].
 \label{eq:swap-partial-trace-check}
\end{align}
Equation~\eqref{eq:swap-partial-trace-check-step-1} uses trace duality and cyclicity; \eqref{eq:swap-partial-trace-check} is the swap trace identity used in \eqref{eq:twirl-weights}. Hence the partial trace is $M_A$; the untraced factor $N_B$ is unchanged. Tensor products of operators span the full operator space on $\bar AB$, so linearity proves \eqref{eq:haar-contraction} for $\rho_{\bar AB}$. In particular, there is no transpose or reality assumption.

Finally, average \eqref{eq:comparison-expansion}, insert \eqref{eq:haar-projector}, and use \eqref{eq:haar-contraction-step-1}--\eqref{eq:haar-contraction}:
\begin{align}
 \mathbb E_G\Gamma^G_{AB}-\rho_{AB}
 &=\frac{d(dm-1)}{d^2-1}I_A\otimes\rho_B
   +\left(\frac{d(d-m)}{d^2-1}-1\right)\rho_{AB}
 \label{eq:averaged-comparison-calculation-step-1}\\
 &=\frac{dm-1}{d^2-1}(dI_A\otimes\rho_B-\rho_{AB}).
 \label{eq:averaged-comparison-calculation}
\end{align}
The factor $r$ cancels in \eqref{eq:averaged-comparison-calculation-step-1}, and \eqref{eq:averaged-comparison-calculation} uses $d(d-m)-(d^2-1)=-(dm-1)$. Thus orthogonal blocks and cyclic windows give exactly the same average alternative.
\end{proof}

\subsection{Completing the theorem and choosing physical protocols}\label{subsec:realization}
The two lemmas now supply the promised proof of the theorem: the fixed-code error is controlled by a binary cost, and the average comparison operator is known. It remains to choose one test for the averaging argument and then select a deterministic protocol.

\begin{proof}[Completion of the proof of Theorem~\ref{thm:main}]
Use the cyclic construction, which works for every $2\le m\le d$ and has $d$ outcomes. Let $T_*$ attain $e_{d,m}=\cost(\rho_{AB},B_{d,m})$. Keep this effect fixed as $G$ varies. Applying \eqref{eq:fixed-code-error} to $e_G$, and then Lemma~\ref{lem:haar-comparison}, gives
\begin{align}
 \mathbb E_G e_G
 &\le\Tr[(I_{AB}-T_*)\rho_{AB}]
       +\Tr[T_*(\mathbb E_G\Gamma^G_{AB}-\rho_{AB})]
 \label{eq:mean-cost-step-1}\\
 &=\Tr[(I_{AB}-T_*)\rho_{AB}]+\Tr[T_*B_{d,m}]
 \label{eq:mean-cost-step-2}\\
 &=e_{d,m}.
 \label{eq:mean-cost}
\end{align}
The choice $T=0$ implies $0\le e_G\le1$. The function $f_{d,m}$ in \eqref{eq:main-bound} is increasing and concave on this interval, since
$f'_{d,m}(x)=1-2x/(dm)\ge0$ and $f''_{d,m}(x)=-2/(dm)\le0$.
Equation~\eqref{eq:fixed-code-error-step-1}, concavity, and then monotonicity therefore give
\begin{align}
 1-\mathbb E_GF_G
 &\le\mathbb E_G f_{d,m}(e_G)
 \label{eq:mean-error-step-1}\\
 &\le f_{d,m}(\mathbb E_G e_G)
 \label{eq:mean-error-step-2}\\
 &\le f_{d,m}(e_{d,m}).
 \label{eq:mean-error}
\end{align}
Only a linear testing expression has been averaged. In particular, no inverse operator has been averaged, and no maximization over decoders has been interchanged with an expectation.

For entanglement distillation, choose a fixed unitary $G_A$ such that $F_G\ge\mathbb E_GF_G$. Such a choice exists because a real-valued measurable function cannot be strictly below its mean everywhere. Alice uses its complete instrument $(\mathcal E_j)_j$, sends the outcome, and Bob applies $\mathcal D_j=\cD_{\tau^j}$. By \eqref{eq:fixed-fidelity-comparison}, this protocol has fidelity $F_G$ and satisfies \eqref{eq:main-bound}. All $d$ outcomes, including any of zero probability, belong to the instrument. No shared randomness is needed during its operation.

For entanglement generation, let $\rho_{AB}=(\id_A\otimes\cN)(\phi_{AA'})$ with $\phi_{AA'}$ pure. The convex average $F_G=\sum_jp_j\Frec(\tau^j_{LB})$ has a positive-probability outcome with $\Frec(\tau^j_{LB})\ge F_G$. Fix that outcome together with $G_A$ in the code design. Alice directly prepares
\begin{equation}\label{eq:fixed-input}
 \ket{\phi_{G,j}}_{LA'}\coloneqq
 \frac{(K_j\otimes I_{A'})\ket\phi_{AA'}}{\sqrt{p_j}}.
\end{equation}
Trace preservation of $\cN$ gives
$p_j=\Tr[(K_j^\dagger K_j\otimes I_{A'})\phi_{AA'}]$, so this vector is normalized. Its channel output is exactly $\tau^j_{LB}$, which Bob decodes with the fixed channel $\mathcal D_j$. The resulting infidelity is at most $f_{d,m}(e_{d,m})$. Equation~\eqref{eq:fixed-input} specifies a state to prepare, not a measurement to repeat until successful: no runtime postselection or classical message is used. These two realizations are the distinct tasks drawn in Figure~\ref{fig:protocols}.
\end{proof}

\Needspace{10\baselineskip}
\section{Hypothesis-testing and entanglement transmission consequences}\label{sec:consequences}
With Theorem~\ref{thm:main} proved, we can now translate its binary cost into more familiar coding statements. The first corollary gives a short, explicit integer dimension at a chosen infidelity. The second retains every finite-dimensional correction and shows that optimizing the testing tolerance recovers the theorem exactly. We then distinguish the exact flat-reference entanglement transmission specialization from the general conversion, which trades a constant factor in target dimension for the same error.

\subsection{A simple integer dimension from one hypothesis test}
The hypothesis-testing coherent information in \eqref{eq:intro-IH} asks how small the false-acceptance weight can be when rejection of $\rho_{AB}$ is bounded by a prescribed tolerance. For coding, we reserve a small part of the final error for the remaining term in the theorem. 

\begin{corollary}[Hypothesis-testing achievability with exact rounding]\label{cor:testing}
Let $\rho_{AB}$ be a state with $d=\dim A\ge2$, and let $\Ih^\delta(A\rangle B)_\rho$ be defined by \eqref{eq:intro-IH} and \eqref{eq:DH}. Fix $0<\err<1$ and $0<\nu\le\err$, and set
\begin{equation}\label{eq:integer-x}
 x\coloneqq\nu\,2^{\Ih^{\err-\nu}(A\rangle B)_\rho}.
\end{equation}
In each operational setting of Theorem~\ref{thm:main}, an $(m,\err)$ protocol is attainable with the integer target dimension
\begin{equation}\label{eq:integer-choice}
 m\coloneqq\max\{1,\lfloor x\rfloor\}\le d.
\end{equation}
The resulting entanglement yield is exactly
\begin{equation}\label{eq:integer-yield}
 \log_2m=\log_2\max\!\left\{1,
 \left\lfloor\nu\,2^{\Ih^{\err-\nu}(A\rangle B)_\rho}\right\rfloor\right\}.
\end{equation}
\end{corollary}
\begin{proof}
We evaluate the theorem's binary cost using an optimal test from \eqref{eq:DH}, and then choose the largest admissible integer. For $0\le\delta<1$, let $T$ attain that test, so
$\Tr[(I-T)\rho_{AB}]\le\delta$ and
$\Tr[T(I_A\otimes\rho_B)]=2^{-\Ih^\delta(A\rangle B)_\rho}$.
For an integer $2\le m\le d$, evaluate the minimum in \eqref{eq:intro-cost} at this particular effect $T$. Expanding \eqref{eq:alternative} and then using the two properties of the optimal test gives
\begin{align}
 e_{d,m}
 &\le\Tr[(I-T)\rho_{AB}]+\Tr[T B_{d,m}]
 \label{eq:relaxed-test-evaluate}\\
 &=\Tr[(I-T)\rho_{AB}]
   +\frac{dm-1}{d^2-1}\left(d\Tr[T(I_A\otimes\rho_B)]-\Tr[T\rho_{AB}]\right)
 \label{eq:relaxed-test-expand}\\
 &\le\delta+\frac{d(dm-1)}{d^2-1}2^{-\Ih^\delta(A\rangle B)_\rho}
       -\frac{dm-1}{d^2-1}\Tr[T\rho_{AB}]
 \label{eq:relaxed-test-substitute}\\
 &\le\delta+\frac{d(dm-1)}{d^2-1}2^{-\Ih^\delta(A\rangle B)_\rho}
 \label{eq:relaxed-test-drop}\\
 &\le\delta+m\,2^{-\Ih^\delta(A\rangle B)_\rho}.
 \label{eq:relaxed-test-bound}
\end{align}
The step \eqref{eq:relaxed-test-drop} discards a nonpositive term because $T\ge0$, $\rho_{AB}\ge0$, and $(dm-1)/(d^2-1)>0$. For \eqref{eq:relaxed-test-bound}, the required coefficient comparison is
\begin{align}
 m-\frac{d(dm-1)}{d^2-1}
 &=\frac{m(d^2-1)-d(dm-1)}{d^2-1}
 \label{eq:relaxed-coefficient-step-1}\\
 &=\frac{d-m}{d^2-1}\ge0.
 \label{eq:relaxed-coefficient}
\end{align}
Theorem~\ref{thm:main} gives error at most $f_{d,m}(e_{d,m})\le e_{d,m}$. With $\delta=\err-\nu$, every integer $2\le m\le\min\{x,d\}$ therefore has error at most $\delta+\nu=\err$. If $\lfloor x\rfloor<2$, take the zero-error target $m=1$.

It remains to check that this rule never exceeds the input dimension. Apply \eqref{eq:dimension-bound} to a feasible hypothesis test to obtain
\begin{equation}\label{eq:dimension-IH}
 2^{-\Ih^\delta(A\rangle B)_\rho}\ge\frac{1-\delta}{d},
 \qquad \Ih^\delta(A\rangle B)_\rho\le\log_2d-\log_2(1-\delta).
\end{equation}
Consequently,
\begin{equation}\label{eq:integer-dimension}
 x\le d\,\frac{\nu}{1-\err+\nu}<d.
\end{equation}
This proves the dimension assertion and the exact yield.
\end{proof}
For $x\ge1$ and $m=\lfloor x\rfloor$, the loss from integer rounding is
\begin{equation}\label{eq:rounding-loss}
 0\le\log_2x-\log_2m=\log_2\frac{x}{\lfloor x\rfloor}
 <\log_2\!\left(1+\frac1m\right).
\end{equation}
It vanishes when $x$ is an integer and tends to zero as $m$ grows. A useful consequence, including $x<1$, is
\begin{equation}\label{eq:rounding-yield}
 \log_2m\ge\left[
 \Ih^{\err-\nu}(A\rangle B)_\rho-\log_2\frac1\nu
 -\log_2\!\left(1+\frac1m\right)\right]_+.
\end{equation}
When $x<1$, the bracket is negative and the trivial target applies. This keeps the distinction between a sufficient real-valued rate condition and an achievable integer dimension. The exact formula \eqref{eq:integer-yield} will also be used in the numerical examples.

\subsection{Keeping the full finite-dimensional bound}\label{subsec:testing-profile}
The proof of Corollary~\ref{cor:testing}, specifically \eqref{eq:relaxed-test-drop} and the replacement $f_{d,m}(e_{d,m})\le e_{d,m}$, discarded two potentially useful terms: the subtraction of a multiple of $\rho_{AB}$ from the alternative and the quadratic improvement in Cheng's trace inequality (Lemma~\ref{lem:cheng}). We now retain both. For $2\le m\le d$, define
\begin{equation}\label{eq:testing-coefficients}
 c_{d,m}\coloneqq\frac{dm-1}{d^2-1},\qquad
 \lambda_{d,m}\coloneqq d c_{d,m}.
\end{equation}
These are the coefficients of $-\rho_{AB}$ and $I_A\otimes\rho_B$ in $B_{d,m}$. The same hypothesis-testing curve therefore suffices to evaluate the full guarantee.

\begin{corollary}[The full bound from the hypothesis-testing profile]\label{cor:profile}
Let $\rho_{AB}$, $d$, and an integer $2\le m\le d$ satisfy the hypotheses of Theorem~\ref{thm:main}. Use $\Ih^\delta$ from \eqref{eq:intro-IH}, $e_{d,m}$ from \eqref{eq:alternative}, $f_{d,m}$ from \eqref{eq:main-bound}, and $c_{d,m}$ from \eqref{eq:testing-coefficients}. For $0\le\delta<1$, set
\begin{equation}\label{eq:test-refinement}
 v_{d,m}^{\delta}\coloneqq\min\!\left\{1,\,
 \delta+c_{d,m}\left(d\,2^{-\Ih^\delta(A\rangle B)_\rho}-1+\delta\right)\right\}.
\end{equation}
In each operational setting of the theorem, an achievable infidelity is at most
\begin{equation}\label{eq:test-bound}
 f_{d,m}(v_{d,m}^{\delta})
 \le\min\!\left\{1-\frac1{dm},\,
       \delta+m\,2^{-\Ih^\delta(A\rangle B)_\rho}\right\}.
\end{equation}
Moreover, optimizing the tolerance recovers the full theorem exactly:
\begin{equation}\label{eq:profile-exact}
 \inf_{0\le\delta<1}f_{d,m}(v_{d,m}^{\delta})
 =f_{d,m}(e_{d,m}).
\end{equation}
\end{corollary}
\begin{proof}
First use an optimal test $T$ for $\Ih^\delta$ and let
$\delta_T\coloneqq\Tr[(I-T)\rho_{AB}]\le\delta$. Retaining all terms of the alternative gives
\begin{align}
 e_{d,m}
 &\le\delta_T+\lambda_{d,m}2^{-\Ih^\delta(A\rangle B)_\rho}
                    -c_{d,m}(1-\delta_T)
 \label{eq:test-cost-end-step-1}\\
 &=(1+c_{d,m})\delta_T+
       \lambda_{d,m}2^{-\Ih^\delta(A\rangle B)_\rho}-c_{d,m}
 \label{eq:test-cost-end-step-2}\\
 &\le\delta+c_{d,m}
             \left(d\,2^{-\Ih^\delta(A\rangle B)_\rho}-1+\delta\right).
 \label{eq:test-cost-end}
\end{align}
Equation~\eqref{eq:dimension-IH} makes the quantity in parentheses nonnegative. Since $e_{d,m}\le1$, it follows that $e_{d,m}\le v_{d,m}^{\delta}\le1$. Monotonicity of $f_{d,m}$ proves achievability of $f_{d,m}(v_{d,m}^{\delta})$. For its upper bounds, use $f_{d,m}(x)\le x$, $f_{d,m}(1)=1-1/(dm)$, $\lambda_{d,m}\le m$, and $c_{d,m}(\delta-1)\le0$.

To prove exactness after optimization, take an optimal effect $T_*$ for $\cost(\rho_{AB},B_{d,m})$. If it accepts $\rho_{AB}$ with positive probability, set $\delta_*\coloneqq1-\Tr[T_*\rho_{AB}]<1$. This effect is feasible for $\Ih^{\delta_*}$, so
\begin{align}
 v_{d,m}^{\delta_*}
 &\le(1+c_{d,m})\delta_*+
       \lambda_{d,m}2^{-\Ih^{\delta_*}(A\rangle B)_\rho}-c_{d,m}
 \label{eq:profile-reverse-step-1}\\
 &\le(1+c_{d,m})\delta_*+
       \lambda_{d,m}\Tr[T_*(I_A\otimes\rho_B)]-c_{d,m}
 \label{eq:profile-reverse-step-2}\\
 &=e_{d,m}.
 \label{eq:profile-reverse}
\end{align}
Together with the opposite inequality already proved, this gives equality. If an optimal effect never accepts $\rho_{AB}$, its cost is at least one, so $e_{d,m}=1$ because $T=0$ has cost one. The inequalities $1=e_{d,m}\le v_{d,m}^{\delta}\le1$ proved using \eqref{eq:test-cost-end} then force $v_{d,m}^{\delta}=1$ for every $\delta$. In either case $\inf_{\delta<1}v_{d,m}^{\delta}=e_{d,m}$, and continuity and monotonicity of $f_{d,m}$ establish \eqref{eq:profile-exact}.
\end{proof}

For a desired error $0<\err<1$, the largest integer certified by this bound is therefore
\begin{equation}\label{eq:best-integer}
 m_{\mathrm{cert}}\coloneqq\max\!\left(
 \{1\}\cup\left\{m\in\{2,\ldots,d\}:
 \inf_{0\le\delta<1}f_{d,m}(v_{d,m}^{\delta})\le\err\right\}\right).
\end{equation}
Equality at the threshold is allowed: \eqref{eq:profile-exact} equals the attained binary cost in the theorem. This dimension is no smaller than the choice in Corollary~\ref{cor:testing}. It is the strongest dimension certified by this particular bound, not a claim of optimality over all codes. Appendix~\ref{app:integer} supplies an explicit fixed-tolerance threshold, including capped and zero-denominator cases.

Alternatively, the trace norm in \eqref{eq:helstrom} evaluates the same certificate directly. As $m$ increases, $B_{d,m}$ increases in operator order, so $e_{d,m}$ increases. The function $f_{d,m}(x)$ increases with both $m$ and $x\in[0,1]$. Thus the certificate is nondecreasing in the integer dimension, and a binary search finds \eqref{eq:best-integer}.

Both corollaries use a joint effect only in the analysis. The decoder remains $\cD_{\tau^j}$, not an implementation of that effect. Replacing the information quantity in the simpler sufficient condition by $\inf_{\sigma_B}D_H^\delta(\rho_{AB}\Vert I_A\otimes\sigma_B)$, with the infimum over normalized states, gives a no-larger quantity and preserves that sufficient condition. Such an optimization is unnecessary for the actual-marginal bound proved here.

\Needspace{10\baselineskip}
\subsection{Entanglement transmission and code conversion}\label{subsec:transmission-consequences}
The bounds in Sections~\ref{sec:theorem} and~\ref{subsec:testing-profile} directly certify entanglement generation. To obtain entanglement transmission, the input must be prepared by a channel acting on one share of a prescribed maximally entangled state, as required by \eqref{eq:et-error}. We first show that a maximally mixed reference marginal gives this property with no change in target dimension or error. We then give an error-preserving conversion that applies to arbitrary entanglement generation codes, at the cost of a smaller target dimension.

\begin{corollary}[Entanglement transmission from a maximally entangled input]\label{cor:flat}
Let $\cN\colon A'\to B$ be a channel, let $\phi_{AA'}$ be a normalized pure state with $d=\dim A\ge2$ and $\phi_A=I_A/d$, and set $\rho_{AB}=(\id_A\otimes\cN)(\phi_{AA'})$. For every integer $2\le m\le d$, there is an isometric $(m,\err)$ entanglement transmission code in the sense of \eqref{eq:et-error} with $\err=f_{d,m}(e_{d,m})$, where $e_{d,m}$ and $f_{d,m}$ are defined in \eqref{eq:alternative}--\eqref{eq:main-bound}. Corollaries~\ref{cor:testing} and~\ref{cor:profile} therefore also apply to this task under the stated input assumption. The target $m=1$ is trivial.
\end{corollary}
\begin{proof}
Use the cyclic operators in \eqref{eq:cyclic-K}. Their normalization and the flat marginal give $p_j=\Tr[K_j(I_A/d)K_j^\dagger]=1/d$. The retained marginal of the selected pure state \eqref{eq:fixed-input} is therefore
\begin{equation}\label{eq:flat-marginal}
 \frac{K_j(I_A/d)K_j^\dagger}{p_j}=I_L/m.
\end{equation}
Let $\ket\zeta_{LA'}=\ket{\phi_{G,j}}_{LA'}$ denote the selected state. Define a linear map on the prescribed logical basis by
\begin{equation}\label{eq:flat-encoder}
 V_{G,j}\ket a_{\widetilde L}
 \coloneqq\sqrt m\,(\bra a_L\otimes I_{A'})\ket\zeta_{LA'},
 \qquad 0\le a<m.
\end{equation}
Its matrix entries obey
$\bra aV_{G,j}^\dagger V_{G,j}\ket b
=m\bra b\zeta_L\ket a=\delta_{ab}$ by \eqref{eq:flat-marginal}. Hence it is an isometry, and
$(I_L\otimes V_{G,j})\ket{\Phi^m}_{L\widetilde L}=\ket\zeta_{LA'}$ by \eqref{eq:flat-encoder}. The encoding channel $\mathcal E(X)=V_{G,j}XV_{G,j}^\dagger$ followed by $\cN$ and the selected decoder therefore produces exactly the entanglement generation code output already proved to satisfy \eqref{eq:main-bound}. Lemma~\ref{lem:petz-identification} gives its completed Petz decoder for a maximally mixed logical input.
\end{proof}

For an arbitrary reference marginal, we instead use a conversion from entanglement generation to entanglement transmission. Keeping the target dimension fixed can increase the error. The following lemma instead preserves the error tolerance and reduces the target dimension by only a constant factor. Its proof, given in Appendix~\ref{app:conversion}, uses a change of Kraus representation, singular vectors, and an elementary scalar inequality.

\begin{lemma}[Error-preserving conversion from entanglement generation to entanglement transmission]\label{lem:eg-et-conversion}
Let $\cN\colon A'\to B$ admit an $(M,\err)$ entanglement generation code as in \eqref{eq:eg-error}, with decoding channel $\cD\colon B\to B_0$, where $M\ge1$ is an integer, $\dim B_0=M$, and $0\le\err<1$. Then $\cN$ admits a $(k,\err)$ entanglement transmission code as in \eqref{eq:et-error}, with
\begin{equation}\label{eq:conversion-dimension}
 k\coloneqq\max\!\left\{1,\left\lfloor M\sqrt{1-\err}\right\rfloor\right\}.
\end{equation}
Its decoder is $\mathcal C_k\circ\cD$ for a channel $\mathcal C_k\colon B_0\to\widehat L$, with $\dim\widehat L=k$. The encoder can be chosen isometric if $k\le\dim A'$. Without that dimensional condition, a general encoding channel achieves \eqref{eq:conversion-dimension}, and an isometric encoder achieves the target dimension $k_{\mathrm{iso}}\coloneqq\min\{k,\dim A'\}$. 
\end{lemma}
In particular, the target in \eqref{eq:conversion-dimension} satisfies
\begin{equation}\label{eq:conversion-log-loss}
 \log_2 k\ge\log_2 M-1-\frac12\log_2\frac1{1-\err}.
\end{equation}
The constant one in \eqref{eq:conversion-log-loss} is only a convenient upper bound on the rounding loss; the exact choice \eqref{eq:conversion-dimension} does not deduct one bit. At fixed $\err$, the conversion therefore preserves the first- and second-order terms, and even an explicit logarithmic term, in a sequence of achievable log-dimensions. Section~\ref{subsec:second-transmission} applies this observation to our entanglement generation codes. The proof is related to the Kraus-operator and encoding arguments of Barnum, Knill, and Nielsen~\cite[Theorem~4 and Lemmas~10--11]{BarnumKnillNielsen2000}; here the dimension is allowed to decrease so that the original error tolerance is retained.

For a general pure reference--input state, Theorem~\ref{thm:main} and Lemma~\ref{lem:eg-et-conversion} give another entanglement transmission consequence. For an integer $2\le m\le d$, set $\eta_m\coloneqq f_{d,m}(e_{d,m})<1$, with the quantities in \eqref{eq:alternative}--\eqref{eq:main-bound}. An achievable entanglement transmission dimension at infidelity at most $\eta_m$ is
\begin{equation}\label{eq:general-one-shot-transmission}
 k_m\coloneqq\max\!\left\{1,\left\lfloor m\sqrt{1-\eta_m}\right\rfloor\right\}.
\end{equation}
Its decoder is the entanglement generation code Petz decoder followed by the channel supplied by the conversion. The same rule applies to an entanglement generation dimension from Corollary~\ref{cor:testing} or~\ref{cor:profile} at its chosen error tolerance. The encoder is isometric under the dimension condition in Lemma~\ref{lem:eg-et-conversion}; otherwise one may either use a general encoder or reduce to the dimension specified there. Unlike Corollary~\ref{cor:flat}, this conclusion requires no maximally mixed reference marginal. Section~\ref{subsec:second-transmission} shows why the bounded loss in log-dimension leaves the second-order and logarithmic terms intact.

\section{Examples and quantitative scope}\label{sec:examples}
Two elementary channels help distinguish the proved bound from the exact performance of a particular recovery. The first checks that no unavoidable error was introduced by the proof. The second exhibits a remaining gap, so that coefficient one is not mistaken for global optimality.

\subsection{A noiseless channel}
A noiseless channel checks whether the proof can certify perfect recovery when the task allows it. Let $\rho_{AB}=\Phi^d_{AB}$, the output of a noiseless $d$-dimensional channel on a maximally entangled input. Since $\rho_B=I_B/d$,
\begin{equation}\label{eq:noiseless}
 B_{d,m}=\frac{dm-1}{d^2-1}(I_{AB}-\Phi^d),
 \qquad e_{d,m}=0.
\end{equation}
The test $T=\Phi^d$ separates the two operators perfectly. The theorem consequently certifies zero error for every $m\le d$. Retaining the subtraction of $\rho$ in \eqref{eq:alternative} is essential for this exact conclusion; the relaxed bound $\delta+m2^{-\Ih^\delta(A\rangle B)_\rho}$ discards that information.

\subsection{An erasure channel}
An erasure flag makes both the testing cost and the actual selected-code error explicit, allowing a direct comparison. Let a $d$-dimensional erasure channel transmit its input with probability $1-p$ and output an orthogonal flag with probability $p$, for $0\le p\le1$. With a maximally entangled input,
\begin{equation}\label{eq:erasure-state}
 \rho_{AB}=(1-p)\Phi^d_{AB_0}\oplus p\,\pi_A\otimes\proj e_B,
 \qquad \pi_A\coloneqq I_A/d.
\end{equation}
The transmitted component of $B_{d,m}$ is orthogonal to $\Phi^d$. On the flag block, the alternative is $(dm-1)p\pi_A\otimes\proj e$, which dominates the true flag component for $m\ge2$. Thus
\begin{equation}\label{eq:erasure-bound}
 e_{d,m}=p,
 \qquad \err_{\mathrm{eg}}\le p\left(1-\frac{p}{dm}\right).
\end{equation}
For every selected flat code, the actual Petz decoder transmits perfectly when no erasure occurs and outputs the maximally mixed logical state on erasure. Its exact infidelity is
\begin{equation}\label{eq:erasure-exact}
 \err_{\mathrm P}=p(1-m^{-2}).
\end{equation}
Indeed, the erased joint output is $\pi_L\otimes\pi_{\widehat L}$, whose target overlap is $1/m^2$. Since $p/(dm)\le1/m^2$ for $m\le d$, \eqref{eq:erasure-exact} is no larger than the general guarantee. This example leaves room for a more channel-specific analysis even after the conversion prefactor is removed.

The theorem concerns infidelity, arbitrary joint decoding, and finite dimensions. Its simple testing corollary is not claimed to uniformly improve every smooth-entropy bound in~\cite{MorganWinter2014,KhatriWilde2024}, since their task and error parameters must also be matched. Nor does the code-existence proof provide an efficient circuit for the Petz channel. Its value is that both operations and the origin of each error term are explicit.

\section{Achievable second-order asymptotics}\label{sec:second}
The one-shot bound is particularly convenient for independent repetitions because its information quantity is already a hypothesis-testing divergence. Only two choices are needed: apply the bound to $\rho^{\otimes n}$, and let the unused error slack decrease slowly with $n$. We give the complete passage, including the unnormalized alternative, moving testing tolerance, and integer dimension. The result is a fixed-source or fixed-input achievable rate, not an optimized capacity formula. After deriving the entanglement generation and entanglement distillation bounds, Section~\ref{subsec:second-transmission} applies the error-preserving conversion to obtain entanglement transmission for arbitrary reference--input states with the same Gaussian and logarithmic terms.

\subsection{Coherent information, variance, and a testing estimate}\label{subsec:variance}
Fix a state $\rho_{AB}$ with $d=\dim A\ge2$. The relative entropy of a state and a positive semidefinite operator is defined by~\cite{Umegaki1962}
\begin{equation}\label{eq:relative-entropy}
 D(\rho\Vert\sigma)\coloneqq\Tr[\rho(\log_2\rho-\log_2\sigma)]
 \quad\text{when }\supp\rho\subseteq\supp\sigma.
\end{equation}
For the alternative $\sigma\coloneqq I_A\otimes\rho_B$, the support condition always holds by Lemma~\ref{lem:dimension}. Coherent information~\cite{SchumacherNielsen1996} and coherent information variance~\cite{TomamichelHayashi2013,Li2014,BeigiDattaLeditzky2016,TomamichelBertaRenes2016} are defined as
\begin{align}
 \Ic(A\rangle B)_\rho
 &\coloneqq D(\rho_{AB}\Vert I_A\otimes\rho_B)
 =H(B)_\rho-H(AB)_\rho,\label{eq:Ic}\\
 \Vc(A\rangle B)_\rho
 &\coloneqq\Tr\!\left[\rho_{AB}
   \bigl(\log_2\rho_{AB}-\log_2(I_A\otimes\rho_B)-\Ic(A\rangle B)_\rho I_{AB}\bigr)^2\right],\label{eq:Vc}
\end{align}
where the von Neumann entropy $H$ is defined in \eqref{eq:entropy}. We abbreviate the two quantities by $\Ic$ and $\Vc$ when the state is fixed. We work on $\supp\sigma$, and extend $\log_2\rho$ by zero on $\ker\rho$ when forming the Hermitian log-likelihood difference. The expectations in \eqref{eq:Ic}--\eqref{eq:Vc} are finite and do not depend on this unused extension. The variance is nonnegative and measures fluctuations of the quantum log-likelihood difference in the state being tested. Its additivity on product states follows because the log-likelihood difference is a sum of single-copy terms and each centered cross term has expectation zero.

Let
\begin{equation}\label{eq:normal}
 \Phi(x)\coloneqq\frac1{\sqrt{2\pi}}\int_{-\infty}^x e^{-t^2/2}\,\mathrm dt,
 \qquad \varphi(x)\coloneqq\frac1{\sqrt{2\pi}}e^{-x^2/2}.
\end{equation}
The normal approximation to quantum hypothesis testing was established independently by Li and by Tomamichel and Hayashi~\cite{Li2014,TomamichelHayashi2013}. For our proof, a one-sided finite-sample form is enough. The following version makes its applicability to a varying tolerance explicit.

\begin{lemma}[Finite-sample testing estimate]\label{lem:normal-testing}
For the fixed state in \eqref{eq:Ic}--\eqref{eq:Vc}, suppose $\Vc>0$. There is a finite constant $b_\rho\ge0$ such that, for $0<\delta<1$ and an integer $n\ge1$ satisfying $\delta-b_\rho/\sqrt n\in(0,1)$,
\begin{equation}\label{eq:normal-testing}
 \Ih^\delta(A^n\rangle B^n)_{\rho^{\otimes n}}
 \ge n\Ic+\sqrt{n\Vc}\,
                \Phi^{-1}\!\left(\delta-\frac{b_\rho}{\sqrt n}\right).
\end{equation}
The constant depends on $\rho$, not on $\delta$ or $n$.
\end{lemma}
\begin{proof}
This is the lower bound in \cite[Theorem~5]{Li2014}, expressed in bits. To check the normalization, set $\sigma\coloneqq I_A\otimes\rho_B$ and introduce the state $\bar\sigma\coloneqq d^{-1}I_A\otimes\rho_B$. Then
\begin{align}
 D_H^\delta(\rho^{\otimes n}\Vert\sigma^{\otimes n})
 &=D_H^\delta(\rho^{\otimes n}\Vert\bar\sigma^{\otimes n})-n\log_2d,\label{eq:testing-scaling}\\
 D(\rho\Vert\bar\sigma)&=\Ic+\log_2d,
 \qquad V(\rho\Vert\bar\sigma)=\Vc.\label{eq:variance-scaling}
\end{align}
Here $V(\rho\Vert\zeta)$ denotes the relative entropy variance
$\Tr[\rho(\log_2\rho-\log_2\zeta-D(\rho\Vert\zeta)I)^2]$.
Multiplying the alternative by a positive scalar shifts the log-likelihood by a constant, leaving this centered variance unchanged. Apply Li's normalized-state estimate and cancel the $n\log_2d$ term. More concretely, if $(\lambda_x,\ket{e_x})$ and $(\mu_y,\ket{f_y})$ are spectral data for $\rho$ and $\bar\sigma$, use probabilities $p(x,y)\coloneqq\lambda_x|\langle e_x|f_y\rangle|^2$. One may take $b_\rho$ to be a universal Berry--Esseen constant times the third absolute centered moment of $\log_2(\lambda_x/\mu_y)$ under $p$, divided by $\Vc^{3/2}$. Only pairs with $p(x,y)>0$ enter this moment, and the support condition makes it finite. Thus \eqref{eq:normal-testing} is uniform over the tolerances used below, rather than a pointwise expansion with an uncontrolled remainder.
\end{proof}

\subsection{Second-order achievability at fixed infidelity}\label{subsec:second-theorem}
The slack in \eqref{eq:intro-rate} should vanish so that the Gaussian quantile is evaluated at the desired final error. Choosing a slack of order $n^{-1/2}$ changes the Gaussian term by only a constant and costs $\tfrac12\log_2n$ in the yield.

\Needspace{13\baselineskip}
\begin{theorem}[Achievable second-order rates]\label{thm:second}
Let $\rho_{AB}$ be a fixed finite-dimensional state with $d=\dim A\ge2$ and quantities \eqref{eq:Ic}--\eqref{eq:Vc}. Fix $0<\err<1$. If $\Vc>0$, there is a finite constant $c_{\rho,\err}$ such that, for every sufficiently large $n$, a one-way entanglement distillation protocol on $\rho_{AB}^{\otimes n}$ has infidelity at most $\err$ and target dimension $m_n$ satisfying
\begin{equation}\label{eq:second-distill}
 \log_2m_n\ge
 \left[n\Ic+\sqrt{n\Vc}\,\Phi^{-1}(\err)
                 -\tfrac12\log_2n-c_{\rho,\err}\right]_+.
\end{equation}
If $\rho_{AB}=(\id_A\otimes\cN)(\phi_{AA'})$ for a fixed channel and pure reference--input state, the same dimension and infidelity are achievable for unassisted entanglement generation over the parallel channel $\cN^{\otimes n}$. If the reference--input state satisfies $\phi_A=I_A/d$, the statement also holds for isometric entanglement transmission. In each case the decoder is the completed channel \eqref{eq:recovery-action}; Section~\ref{subsec:state-decoder} gives its Petz representation, including mixed states. If $\Vc=0$, a constant-remainder version is given in \prref{prop:zero-variance}.
\end{theorem}
\begin{proof}
Set $\nu_n\coloneqq n^{-1/2}$ and $\delta_n\coloneqq\err-\nu_n$ for sufficiently large $n$ that $0<\nu_n<\err$. Apply \corref{cor:testing} to the block state, whose left dimension is $d^n$. It gives error at most $\err$ for
\begin{equation}\label{eq:second-mn}
 m_n\coloneqq\max\left\{1,\left\lfloor
       \nu_n\,2^{\Ih^{\delta_n}(A^n\rangle B^n)_{\rho^{\otimes n}}}
                                 \right\rfloor\right\}.
\end{equation}
The dimension check \eqref{eq:integer-dimension} applies with $d^n$; therefore $m_n\le d^n$ automatically. It is not necessary to assume a gap between $\Ic$ and $\log_2d$.

For large $n$, \lemref{lem:normal-testing} with the moving tolerance $\delta_n$ gives
\begin{equation}\label{eq:moving-tolerance}
 \Ih^{\delta_n}(A^n\rangle B^n)_{\rho^{\otimes n}}
 \ge n\Ic+\sqrt{n\Vc}\,
        \Phi^{-1}\!\left(\err-\frac{1+b_\rho}{\sqrt n}\right).
\end{equation}
For $n$ large enough, the argument lies in $[\err/2,\err]$. The inverse normal function has bounded derivative on this interval. In particular, set
\begin{equation}\label{eq:quantile-constant}
 L_\err\coloneqq\max_{u\in[\err/2,\err]}
                  \frac1{\varphi(\Phi^{-1}(u))}<\infty.
\end{equation}
The mean value theorem yields
\begin{equation}\label{eq:quantile-shift}
 \sqrt{n\Vc}\,\Phi^{-1}\!\left(\err-\frac{1+b_\rho}{\sqrt n}\right)
 \ge\sqrt{n\Vc}\,\Phi^{-1}(\err)
       -\sqrt{\Vc}(1+b_\rho)L_\err.
\end{equation}
Finally, use the exact rounding estimate \eqref{eq:rounding-yield}, $\log_2\nu_n=-\tfrac12\log_2n$, and \eqref{eq:moving-tolerance}--\eqref{eq:quantile-shift}. The rounding term is $\log_2(1+1/m_n)\le1$, so it can be absorbed into the constant in the asymptotic display. This gives \eqref{eq:second-distill} with
$c_{\rho,\err}=1+\sqrt{\Vc}(1+b_\rho)L_\err$. The one-shot formula \eqref{eq:integer-yield} still keeps the exact integer rather than deducting one bit. When $\Ic>0$, $m_n$ grows exponentially and the actual rounding loss tends to zero exponentially.

For a channel, apply the same theorem to the reference--input state $\phi_{AA'}^{\otimes n}$ and $\cN^{\otimes n}$. The selected state is prepared directly as in \eqref{eq:fixed-input}. Although the initial reference--input state is a product, the subspace selected on its reference can correlate the channel inputs across uses; this is an allowed block code. No forward classical message is added. The statement for a maximally entangled input follows from \corref{cor:flat} on the block systems.
\end{proof}

For positive $\Ic$, dividing \eqref{eq:second-distill} by $n$ displays the familiar coherent-information rate and its $n^{-1/2}$ correction. More generally, the positive part also covers zero or negative coherent information: it does not assert a negative entanglement yield. The normal quantile has the original target infidelity $\err$ because the testing tolerance approaches $\err$. This is why the coefficient-one one-shot statement is convenient for the asymptotic argument.

The Gaussian term for one-way entanglement distillation agrees with the lower bound obtained from the one-shot entanglement distillation result and its second-order analysis in~\cite[Proposition~21, Section~6.2, and Appendix~C]{WildeTomamichelBerta2017}; Petz-based second-order quantum coding is studied in~\cite{BeigiDattaLeditzky2016}, and general finite-resource quantum coding and dispersion bounds are developed in~\cite{TomamichelBertaRenes2016}. Equation~\eqref{eq:second-distill} gives a direct proof with an explicit achievable logarithmic remainder. It does not determine the optimal dispersion of an arbitrary channel or the optimal third-order coefficient. Optimizing a fixed reference--input state or applying the result to a fixed multi-use reference--input state gives further lower bounds; exchanging such optimizations with asymptotic limits is not needed here.

The $-\tfrac12\log_2n$ remainder in \eqref{eq:second-distill} is an achievable guarantee, not an asserted optimal third-order term. Appendix~\ref{app:third-order} gives the more specialized analysis: Proposition~\ref{prop:commuting-second} improves the remainder to a constant for commuting hypotheses, and the surrounding discussion explains why a logarithmic term from binary testing cannot simply be transferred to a coding yield. Before treating zero variance and the erasure example, we establish the corresponding entanglement transmission statement for arbitrary reference--input states.

\subsection{Entanglement transmission at the same error for arbitrary reference--input states}\label{subsec:second-transmission}
Theorem~\ref{thm:second} allows an arbitrary pure reference--input state for entanglement generation. Its direct entanglement transmission specialization assumes a maximally mixed reference marginal. Lemma~\ref{lem:eg-et-conversion} removes that restriction at the level of asymptotic rates: it preserves the prescribed error and changes the log-dimension by only a constant. In particular, it does not replace the desired Gaussian quantile by one evaluated at a smaller error.

\begin{theorem}[Arbitrary-input entanglement transmission with an explicit logarithmic remainder]\label{thm:second-transmission}
Let $\cN\colon A'\to B$ be a finite-dimensional channel, let $\phi_{AA'}$ be a normalized pure state with $d\coloneqq\dim A\ge2$, and set $\rho_{AB}\coloneqq(\id_A\otimes\cN)(\phi_{AA'})$. Use the coherent information $\Ic$ and variance $\Vc$ from \eqref{eq:Ic}--\eqref{eq:Vc}, evaluated on this state, and the standard normal distribution function $\Phi$ from \eqref{eq:normal}. Suppose $\Vc>0$, and fix $0<\err<1$. There is a finite constant $c_{\rho,\err}^{\mathrm{et}}$ such that, for all sufficiently large $n$, an unassisted entanglement transmission code for $\cN^{\otimes n}$ has infidelity at most $\err$ and target dimension $k_n$ satisfying
\begin{equation}\label{eq:second-transmission}
 \log_2 k_n\ge\left[
 n\Ic+\sqrt{n\Vc}\,\Phi^{-1}(\err)
          -\tfrac12\log_2 n-c_{\rho,\err}^{\mathrm{et}}\right]_+.
\end{equation}
The decoder consists of the completed Petz channel \eqref{eq:recovery-action} of an entanglement generation code from Theorem~\ref{thm:second}, followed by a local channel. The encoder may be required to be isometric without changing any of the displayed $n$, $\sqrt n$, or $\log n$ terms. No maximally mixed marginal assumption on $\phi_A$ is imposed.
\end{theorem}
\begin{proof}
Let $M_n$ denote the entanglement generation code dimension $m_n$ chosen in \eqref{eq:second-mn}. Its infidelity is at most $\err$ by Theorem~\ref{thm:second}. Apply Lemma~\ref{lem:eg-et-conversion} to this complete block code and set
\begin{equation}\label{eq:second-transmission-dimension}
 k_n\coloneqq\max\!\left\{1,
       \left\lfloor M_n\sqrt{1-\err}\right\rfloor\right\}.
\end{equation}
The new code uses the same $n$ instances of $\cN$ and has infidelity at most the same $\err$. Equation~\eqref{eq:conversion-log-loss} gives
\begin{equation}\label{eq:second-transmission-loss}
 \log_2 k_n\ge\log_2 M_n-1-\frac12\log_2\frac1{1-\err}.
\end{equation}
Combine \eqref{eq:second-transmission-loss} with \eqref{eq:second-distill} and $\log_2k_n\ge0$. This proves \eqref{eq:second-transmission} with
\begin{equation}\label{eq:second-transmission-constant}
 c_{\rho,\err}^{\mathrm{et}}
 =c_{\rho,\err}+1+\frac12\log_2\frac1{1-\err}.
\end{equation}
Here $c_{\rho,\err}$ is the constant from Theorem~\ref{thm:second}.

For the isometric version, set $a\coloneqq\dim A'$ and replace $k_n$ by $k_n^{\mathrm{iso}}\coloneqq\min\{k_n,a^n\}$. Lemma~\ref{lem:eg-et-conversion} provides an isometric encoder at this dimension and the same infidelity. Equation~\eqref{eq:conversion-isometric-log-loss}, applied to the block channel, proves \eqref{eq:second-transmission} with the possibly larger constant
\begin{equation}\label{eq:second-transmission-isometric-constant}
 c_{\rho,\err}+1+\log_2\frac1{1-\err}.
\end{equation}
If $d\le a$, then $M_n\le d^n\le a^n$ by \eqref{eq:integer-dimension}, so no cap is needed and the smaller constant \eqref{eq:second-transmission-constant} already permits an isometric encoder.
\end{proof}

The distinction between an auxiliary input and a logical code input is important here. The state $\phi_{AA'}$ determines the information quantities in \eqref{eq:second-transmission} and can have an arbitrary reference marginal. The actual entanglement transmission encoder nevertheless acts on one share of a prescribed $\Phi^{k_n}$, whose reference marginal is exactly $I/k_n$. These are different roles. The output processing in the theorem is trace preserving and is included in Bob's decoder; it is not postselection. We do not assert that the composite decoder is itself the canonical Petz decoder for the new encoder.

At fixed $\err$, the loss in \eqref{eq:second-transmission-loss} is independent of $n$. The coherent-information coefficient, the Gaussian term at $\Phi^{-1}(\err)$, and the coefficient $-1/2$ of $\log_2n$ are therefore all retained. When $M_n\to\infty$ and no cap is needed, \eqref{eq:conversion-asymptotic-loss} gives the more precise additional loss $\tfrac12\log_2[1/(1-\err)]+o(1)$. The same argument transfers a constant-remainder entanglement generation bound, such as Proposition~\ref{prop:commuting-second}, to entanglement transmission with a constant remainder. In contrast, the same-dimension fidelity guarantee $f^2$ in \eqref{eq:conversion-tradeoff} would require entanglement generation error at most $1-\sqrt{1-\err}$ to reach final error $\err$, changing the order-$\sqrt n$ term when $\Vc>0$.

\emph{Relation to earlier entanglement transmission results.}
The general inner-bound argument of Tomamichel, Berta, and Renes~\cite[arXiv version~2, Section~IV.B, Lemma~4 and Theorem~6]{TomamichelBertaRenes2016} invokes Proposition~20 of Morgan and Winter~\cite{MorganWinter2014}, which is stated for entanglement generation, in an entanglement transmission setting. The stated change from purified distance to infidelity does not by itself construct an entanglement transmission encoder. Lemma~\ref{lem:eg-et-conversion} supplies this missing task-conversion step: it retains the error tolerance and incurs only an $O_{\err}(1)$ loss in log-dimension, thereby establishing the second-order entanglement transmission conclusion of that argument. This does not assert that its finite one-shot dimension transfers unchanged.

Independently, Beigi, Datta, and Leditzky proved a second-order entanglement transmission lower bound with the coherent-information Gaussian term and an $O(\log n)$ remainder~\cite[Theorem~3.5]{BeigiDattaLeditzky2016}. Theorem~\ref{thm:second-transmission} does not claim the Gaussian rate as new. It gives a direct route through our entanglement generation bound and specifies the achievable logarithmic remainder as $-\tfrac12\log_2n-O(1)$ for each fixed auxiliary input. This is more explicit control of the third-order guarantee, not a claim that $-1/2$ is the optimal coefficient or that every finite-blocklength bound in earlier work is weaker. The exact flat-marginal result in Corollary~\ref{cor:flat} remains useful because it needs neither the dimension reduction nor the additional output processing.

\subsection{The zero-variance case}\label{subsec:zero-variance}
The positive-variance bounds in Theorems~\ref{thm:second} and~\ref{thm:second-transmission} use Lemma~\ref{lem:normal-testing}, whose finite-sample estimate divides by a power of the variance. Zero variance should therefore not be included by a formal limiting substitution. In this case the test itself can be solved exactly, giving a simpler result. The underlying reduction to commuting hypotheses is also discussed in~\cite[Section~5]{Li2014}. We include a direct argument for our unnormalized alternative and retain the finite-dimensional terms in the coding bound.

\begin{proposition}[Exact testing at zero variance]\label{prop:zero-variance}
Let $\rho_{AB}$ be a fixed state with $d=\dim A\ge2$, with coherent information and variance defined in \eqref{eq:Ic}--\eqref{eq:Vc}. If $\Vc=0$, then for every $n\ge1$ and $0\le\delta<1$,
\begin{equation}\label{eq:zero-DH}
 \Ih^\delta(A^n\rangle B^n)_{\rho^{\otimes n}}
 =n\Ic-\log_2(1-\delta).
\end{equation}
For each integer $2\le m\le d^n$, the full binary cost in Theorem~\ref{thm:main} is
\begin{equation}\label{eq:zero-exact-cost}
 e_{d^n,m}=\min\!\left\{1,
 \frac{d^nm-1}{d^{2n}-1}\left(d^n2^{-n\Ic}-1\right)\right\}.
\end{equation}
The theorem therefore certifies infidelity at most
$f_{d^n,m}(e_{d^n,m})$ for all its applicable tasks. In particular, for $0<\err<1$, the simpler explicit dimension
\begin{equation}\label{eq:zero-yield}
 m_n\coloneqq\max\{1,\lfloor\err\,2^{n\Ic}\rfloor\}
\end{equation}
is achievable with infidelity at most $\err$. If $\Ic=\log_2d$, the full cost \eqref{eq:zero-exact-cost} vanishes and certifies perfect recovery with $m=d^n$.
\end{proposition}
\begin{proof}
Work on $\supp\sigma$, where $\sigma=I_A\otimes\rho_B$ is positive definite, and let $P\coloneqq\Pi_\rho$. The equality $\Vc=0$ means
\begin{equation}\label{eq:zero-log}
 (\log_2\rho-\log_2\sigma-\Ic I)\sqrt\rho=0.
\end{equation}
The columns of $\sqrt\rho$ span $\supp\rho$. Since $\log_2\rho$ preserves that subspace, \eqref{eq:zero-log} shows that $\log_2\sigma$ also preserves it, with restriction $\log_2\rho-\Ic I$. Hermiticity makes its orthogonal complement invariant as well. Exponentiating on the two invariant subspaces gives
\begin{equation}\label{eq:zero-block}
 \sigma=2^{-\Ic}\rho\oplus\sigma_\perp,
 \qquad \sigma_\perp\ge0.
\end{equation}
The tensor-power alternative consequently obeys $\sigma^{\otimes n}\ge2^{-n\Ic}\rho^{\otimes n}$ and has restriction $2^{-n\Ic}\rho^{\otimes n}$ on $P^{\otimes n}$. Every feasible test has cost at least $2^{-n\Ic}(1-\delta)$, and the effect $T=(1-\delta)P^{\otimes n}$ attains it. This proves \eqref{eq:zero-DH}. On $\supp\rho^{\otimes n}$, the alternative $B_{d^n,m}$ is a scalar multiple of $\rho^{\otimes n}$, with scalar equal to the second argument of the minimum in \eqref{eq:zero-exact-cost}. On its orthogonal complement the null state is zero. Minimizing the binary cost separately on these two invariant subspaces proves \eqref{eq:zero-exact-cost}. For the explicit dimension, choose $\delta=0$ and $\nu=\err$ in \eqref{eq:integer-choice} and use \eqref{eq:zero-DH}. More generally, the relaxed dimension is proportional to $(\err-\delta)/(1-\delta)$, which decreases with $\delta\in[0,\err]$ because $\err<1$. Thus $\delta=0$ is the best tolerance for that relaxed zero-variance expression. Finally, if $\Ic=\log_2d$, the parenthesis in \eqref{eq:zero-exact-cost} is zero for every $n$.
\end{proof}

The noiseless maximally entangled example has $\Ic=\log_2d$ and $\Vc=0$. The full statement of the proposition now preserves the same exact cancellation as \eqref{eq:noiseless} and gives perfect entanglement generation of dimension $d^n$. Its simpler dimension \eqref{eq:zero-yield} is convenient but need not retain that cancellation. In the channel setting, Lemma~\ref{lem:eg-et-conversion} transfers either entanglement generation guarantee to entanglement transmission without a maximally mixed reference assumption, at the exact reduced dimension in \eqref{eq:conversion-dimension}.

\subsection{An erasure-channel illustration}\label{subsec:erasure-second}
The state \eqref{eq:erasure-state} gives a transparent interpretation of the variance. On its transmitted support, the log-likelihood ratio relative to $I_A\otimes\rho_B$ is $\log_2d$; on its erased support, it is $-\log_2d$. Their probabilities are $1-p$ and $p$. Therefore
\begin{equation}\label{eq:erasure-IV}
 \Ic=(1-2p)\log_2d,
 \qquad \Vc=4p(1-p)(\log_2d)^2.
\end{equation}
The hypotheses $\rho_{AB}$ and $I_A\otimes\rho_B$ commute: the transmitted part of the alternative is proportional to the identity, and the erasure flag is orthogonal to that part. For $0<p<1$, Proposition~\ref{prop:commuting-second} in Appendix~\ref{app:third-order} therefore improves the general logarithmic remainder to a constant:
\begin{equation}\label{eq:erasure-second}
 \log_2m_n\ge
 \left[n(1-2p)\log_2d+
    2\sqrt{np(1-p)}\log_2d\,\Phi^{-1}(\err)
     -c_{p,d,\err}\right]_+.
\end{equation}
The maximally entangled input permits entanglement transmission as well. At $p=1/2$ the first-order term vanishes, while the displayed $\sqrt n$ contribution is positive when $\err>1/2$. This finite-error phenomenon is also identified in~\cite{BeigiDattaLeditzky2016}; it is not a positive vanishing-error capacity assertion.

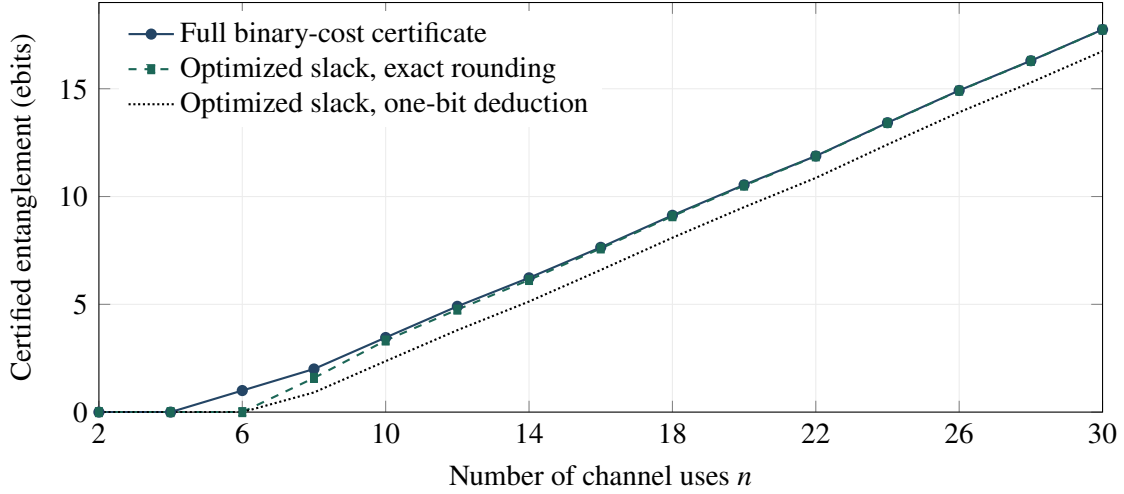
\begin{figure}[!t]
\centering
\begin{tikzpicture}
\begin{axis}[
 width=.90\linewidth,height=7cm,
 xlabel={Number of channel uses $n$},ylabel={Certified entanglement (ebits)},
 xmin=2,xmax=30,ymin=0,ymax=19,
 xtick={2,6,10,14,18,22,26,30},
 grid=major,grid style={gray!15},
 tick label style={font=\small},label style={font=\small},
 legend style={font=\small,at={(.02,.98)},anchor=north west,
               draw=none,fill=white,cells={anchor=west}},
 ]
\addplot[thick,inkblue,mark=*,mark size=1.7pt] coordinates {(2,0.0000000000) (4,0.0000000000) (6,1.0000000000) (8,2.0000000000) (10,3.4594316186) (12,4.9068905956) (14,6.2288186905) (16,7.6438561898) (18,9.1292830169) (20,10.5420645423) (22,11.8803488082) (24,13.4249535713) (26,14.9205389496) (28,16.2996730981) (30,17.7453001053)};
\addlegendentry{Full binary-cost certificate}
\addplot[thick,inkgreen,dashed,mark=square*,mark size=1.5pt] coordinates {(2,0.0000000000) (4,0.0000000000) (6,0.0000000000) (8,1.5849625007) (10,3.3219280949) (12,4.7548875022) (14,6.1292830169) (16,7.5924570373) (18,9.0874628413) (20,10.5077946402) (22,11.8649599151) (24,13.4123052050) (26,14.9099399481) (28,16.2905948881) (30,17.7414274845)};
\addlegendentry{Optimized slack, exact rounding}
\addplot[thick,black,densely dotted] coordinates {(2,0.0000000000) (4,0.0000000000) (6,0.0000000000) (8,0.9125717994) (10,2.3643455953) (12,3.8010730872) (14,5.1294120689) (16,6.5933871966) (18,8.0892086017) (20,9.5080826124) (22,10.8652185440) (24,12.4124075781) (26,13.9099679993) (28,15.2906085117) (30,16.7414334739)};
\addlegendentry{Optimized slack, one-bit deduction}
\end{axis}
\end{tikzpicture}
\caption{Keeping the integer choice and the finite-dimensional corrections. The example is a qubit erasure channel with $p=\err=1/10$ and a maximally entangled reference--input state. The upper curve uses \eqref{eq:erasure-full-cost} and the quadratic correction; the middle curve maximizes the dimension in \eqref{eq:integer-choice} over the slack. The dotted curve instead subtracts one bit from the corresponding unrounded yield and takes the positive part. All values are achievable lower bounds, not converses. For example, at $n=8$, the full certificate gives $m=4$, while the optimized relaxed choice gives $m=3$. Lines connect the displayed blocklengths; the script supplied with the source verifies the integer choices using rational arithmetic.}
\label{fig:integer-comparison}
\end{figure}

For completeness, the one-shot hypothesis test on $n$ copies can be evaluated without constructing exponentially large matrices. Fix $0\le\delta<1$ and let $J\sim\operatorname{Bin}(n,p)$. On the support of $\rho^{\otimes n}$, the likelihood ratio is $d^{n-2J}$. An optimal test retains the smallest erasure counts first. If $r$ is the smallest integer with $\Pr\{J\le r\}\ge1-\delta$ and $\gamma\coloneqq(1-\delta-\Pr\{J<r\})/\Pr\{J=r\}$, then
\begin{equation}\label{eq:erasure-beta}
 2^{-\Ih^\delta(A^n\rangle B^n)}
 =\sum_{j<r}\Pr\{J=j\}d^{2j-n}
       +\gamma\Pr\{J=r\}d^{2r-n}.
\end{equation}
The test is zero on the remaining, unused support of the alternative. This formula, together with \eqref{eq:integer-choice}, provides a finite-blocklength evaluation of the achievable bound. The full binary cost is equally explicit. On an erasure pattern with $j$ erasures, the ratio of the alternative $B_{d^n,m}$ to the null state is $c_{d^n,m}(d^{2j}-1)$, where $c_{d^n,m}$ is defined in \eqref{eq:testing-coefficients}. Therefore
\begin{equation}\label{eq:erasure-full-cost}
 e_{d^n,m}=\sum_{j=0}^n\Pr\{J=j\}
            \min\{1,c_{d^n,m}(d^{2j}-1)\},
 \qquad 2\le m\le d^n.
\end{equation}
The coordinates outside the support of the null state again contribute zero. Substitution into $f_{d^n,m}$ retains both improvements in the main theorem, whereas \eqref{eq:erasure-beta} with a chosen slack gives the simpler corollary. Figure~\ref{fig:integer-comparison} illustrates the difference. The accompanying script \texttt{anc/erasure\_bounds.py} includes a logarithmic-domain evaluation of the hypothesis test and evaluates both integer dimension bounds with exact rational arithmetic when $p$ and $\err$ are rational. It optimizes the slack over the binomial testing tradeoff and verifies the largest certified integer without floating-point comparisons at its boundary. It does not construct exponentially large state matrices. This numerical evaluation is separate from the proof of \thref{thm:second}.

\section{Conclusion}\label{sec:conclusion}
\subsection{Summary}
This paper connects a physical communication problem with a simpler statistical question. The communication problem is to share nearly ideal entanglement despite noise. The statistical question is to distinguish two specified quantum hypotheses. Our main theorem shows that the error of an explicit entanglement protocol can be bounded by the cost of this binary test. The same bound applies both to entanglement generation through a noisy channel and to one-way entanglement distillation from an imperfect shared state. The test is used to evaluate the protocol; Alice and Bob do not need to implement it jointly.

The decoder is based on the Petz recovery channel. The main simplification is to apply Cheng's trace inequality before averaging over possible codes. This order of steps avoids averaging the inverse operators appearing in the decoder and leaves a simple average over quantum subspaces. The resulting bounds apply to every integer target dimension, including dimensions that do not divide the original input dimension. They also retain the finite-dimensional terms needed to certify perfect recovery in the noiseless example.

A separate conversion connects entanglement generation with entanglement transmission, where the encoder must act on one share of a prescribed maximally entangled state. Rather than increase the allowed error, the conversion reduces the target dimension by a constant factor at fixed error. The reduction costs only a bounded number of ebits, regardless of how many times the channel is used. Consequently, arbitrary auxiliary reference--input states give entanglement transmission codes with the same coherent-information rate, Gaussian finite-size correction, and achievable logarithmic remainder as the entanglement generation codes. When the auxiliary reference marginal is already maximally mixed, the direct construction needs neither a dimension reduction nor extra processing after the Petz decoder.

Expressing the one-shot guarantee through hypothesis-testing coherent information makes the passage to many independent repetitions transparent. It gives a leading rate together with a quantitative correction for finite resources at the original error tolerance. We also obtain a constant remainder when the testing hypotheses commute. These results specify achievable performance, not generally matching limits on the best possible codes. They provide explicit recovery constructions and bounds from which more precise analyses for particular channels can begin.

\subsection{Future directions}
A first question is whether the positive phase terms discarded in the step from \eqref{eq:general-phase-average} to \eqref{eq:general-identity-phase} can yield sharper channel-specific guarantees without obscuring the proof. A second is to compare the full finite-dimensional binary cost with existing one-shot bounds under exactly matched tasks and error conventions. In such comparisons, the optimized certificate \eqref{eq:best-integer}, rather than the relaxed slack formula, retains the strongest guarantee proved here. A further question is to improve the third-order guarantee for genuinely noncommuting hypotheses. Even in commuting examples, a positive logarithmic coding term would require an improvement beyond the present binary-cost certificate. Finally, the Petz formula specifies a decoder but not an efficient implementation; structured codes may allow useful approximations with a provable fidelity loss. The simplicity of the final test bound makes these quantitative questions accessible within the direct recovery analysis.

\Needspace{9\baselineskip}
\section*{AI assistance}
ChatGPT assisted with proof formulation and exploration, exposition, literature checks, figure preparation, and numerical tests during the development of this manuscript. The checks performed during this assistance were internal reviews, not independent external peer review or formal verification. Responsibility for any submitted version and its claims rests with the authors.

\begingroup
\fontsize{9.5}{10.8}\selectfont
\setlength{\bibsep}{1.5pt}
\interlinepenalty=10000
\bibliographystyle{unsrtnat}
\bibliography{references}
\endgroup

\appendix
\section{Proof of the isometric-embedding trace inequality}\label{app:embedding}
We prove Lemma~\ref{lem:isometric-recovery-trace} without assuming familiarity with operator convexity. The main idea is elementary: restricting a quadratic maximization to vectors in the range of an isometry can only decrease its value. An integral representation then converts this inverse inequality into an inverse-square-root inequality. The final regularization is needed because that operator inequality cannot simply be asserted for support inverses.

\begin{proof}[Proof of Lemma~\ref{lem:isometric-recovery-trace}]
Retain the notation of the lemma: $J\colon\mathcal H\to\mathcal K$ is an isometry, $\Omega=J\rho J^\dagger$, $S\ge\Omega$, and $\Gamma=J^\dagger SJ\ge\rho$. First assume $S>0$, which also implies $\Gamma>0$.

For a positive definite operator $M$ and a vector $\ket v$, completing the square gives
\begin{align}
 2\operatorname{Re}\langle v|x\rangle-\langle x|M|x\rangle
 &=\langle v|M^{-1}|v\rangle
   -\norm{M^{1/2}\ket x-M^{-1/2}\ket v}^{2},
 \label{eq:inverse-variational-step-1}\\
 \langle v|M^{-1}|v\rangle
 &=\max_{\ket x}\{2\operatorname{Re}\langle v|x\rangle
                       -\langle x|M|x\rangle\}.
 \label{eq:inverse-variational}
\end{align}
The maximizing vector in \eqref{eq:inverse-variational} is $M^{-1}\ket v$. Apply this identity on $\mathcal K$ to $M=S+uI_{\mathcal K}$ and $J\ket v$, where $u\ge0$. Restricting the maximizing vectors to $\ket x=J\ket y$ gives
\begin{align}
 \langle v|J^\dagger(S+uI_{\mathcal K})^{-1}J|v\rangle
 &=\max_{\ket x\in\mathcal K}
   \{2\operatorname{Re}\langle Jv|x\rangle
                     -\langle x|(S+uI_{\mathcal K})|x\rangle\}
 \label{eq:resolvent-compression-step-1}\\
 &\ge\max_{\ket y\in\mathcal H}
   \{2\operatorname{Re}\langle v|y\rangle
                     -\langle y|(\Gamma+uI_{\mathcal H})|y\rangle\}
 \label{eq:resolvent-compression-step-2}\\
 &=\langle v|(\Gamma+uI_{\mathcal H})^{-1}|v\rangle.
 \label{eq:resolvent-compression}
\end{align}
We used $J^\dagger J=I_{\mathcal H}$ and $J^\dagger SJ=\Gamma$. Since this holds for every $\ket v$, it is an operator inequality.

For a positive scalar $x$, the substitution $u=xz^2$ shows that
\begin{align}
 \frac1\pi\int_0^\infty\frac{u^{-1/2}}{x+u}\,\mathrm du
 &=\frac{2}{\pi\sqrt x}\int_0^\infty\frac{\mathrm dz}{1+z^2}
 \label{eq:root-integral-step-1}\\
 &=x^{-1/2}.
 \label{eq:root-integral}
\end{align}
Apply this identity to each eigenvalue of $S$ and $\Gamma$, and integrate \eqref{eq:resolvent-compression} against the positive weight $u^{-1/2}/\pi$. The resulting operator inequality is
\begin{equation}\label{eq:root-compression}
 J^\dagger S^{-1/2}J\ge\Gamma^{-1/2}.
\end{equation}
All integrals converge because the operators are positive definite and the spaces are finite dimensional.

Congruence by $\sqrt\rho$ preserves the order. Define
\begin{equation}\label{eq:embedding-AB}
 A_0\coloneqq\sqrt\rho J^\dagger S^{-1/2}J\sqrt\rho,
 \qquad B_0\coloneqq\sqrt\rho\Gamma^{-1/2}\sqrt\rho.
\end{equation}
Then $A_0\ge B_0\ge0$, and
\begin{equation}\label{eq:embedding-square}
 \Tr A_0^2-\Tr B_0^2
 =\Tr[(A_0-B_0)(A_0+B_0)]\ge0.
\end{equation}
The inequality in \eqref{eq:embedding-square} uses only positivity: for positive semidefinite operators $X,Y$, $\Tr[XY]=\Tr[X^{1/2}YX^{1/2}]\ge0$. It does not require $A_0$ and $B_0$ to commute. Cyclicity of the trace and $\Omega=J\rho J^\dagger$ identify the two squared traces as
\begin{align}
 \Tr A_0^2&=\Tr[\Omega S^{-1/2}\Omega S^{-1/2}],
 \label{eq:embedding-identify-step-1}\\
 \Tr B_0^2&=\Tr[\rho\Gamma^{-1/2}\rho\Gamma^{-1/2}].
 \label{eq:embedding-identify}
\end{align}
This proves the lemma when $S>0$.

For singular $S$, set $S_\eta\coloneqq S+\eta I_{\mathcal K}$ and $\Gamma_\eta\coloneqq\Gamma+\eta I_{\mathcal H}$ for $\eta>0$. Since $J$ is an isometry, $J^\dagger S_\eta J=\Gamma_\eta$. The positive-definite argument gives
\begin{equation}\label{eq:embedding-regularized}
 \Tr[\Omega S_\eta^{-1/2}\Omega S_\eta^{-1/2}]
 \ge\Tr[\rho\Gamma_\eta^{-1/2}\rho\Gamma_\eta^{-1/2}].
\end{equation}
Since $0\le\Omega\le S$, the operator $\Omega$ annihilates $\ker S$; similarly, $\rho$ annihilates $\ker\Gamma$. To see the first assertion, a vector in $\ker S$ has zero expectation in $\Omega$, and a positive semidefinite operator annihilates every vector with zero expectation. Thus only strictly positive eigenvalues of $S$ and $\Gamma$ contribute to the traces in \eqref{eq:embedding-regularized}. Their inverse-square-root factors converge to the support inverses as $\eta\downarrow0$. Taking that limit proves \eqref{eq:isometric-recovery-trace}. The proof makes no order claim for generalized inverses on unused kernel directions.
\end{proof}

\section{An elementary proof of the two-copy unitary average}\label{app:haar}
Lemma~\ref{lem:two-copy-average} is often proved using representation theory. We give a self-contained argument using only invariance, diagonal phase changes, basis permutations, and one two-dimensional basis change. The first three steps determine the possible form of the average; two preserved traces then fix its coefficients.

\begin{proof}[Proof of Lemma~\ref{lem:two-copy-average}]
For an arbitrary operator $X$ on the two registers, set
$Y\coloneqq\mathbb E_G[(G\otimes G)X(G\otimes G)^\dagger]$.
Invariance of Haar measure under left multiplication implies that
\begin{equation}\label{eq:mean-invariance}
 (W\otimes W)Y(W\otimes W)^\dagger=Y
 \quad\text{for every unitary }W.
\end{equation}
We first use diagonal phase changes to determine which basis vectors $Y$ can mix. In a fixed orthonormal basis, let
$W_{\boldsymbol\theta}\coloneqq\sum_{k=0}^{d-1}e^{\mathrm i\theta_k}\proj k$,
where the phases $\theta_k$ can be chosen independently. Multiplying \eqref{eq:mean-invariance} by $W\otimes W$ on the right shows that $Y$ commutes with $W\otimes W$. Therefore
\begin{align}
 (W_{\boldsymbol\theta}\otimes W_{\boldsymbol\theta})Y\ket{ij}
 &=Y(W_{\boldsymbol\theta}\otimes W_{\boldsymbol\theta})\ket{ij}
 \label{eq:phase-eigenspace-step-1}\\
 &=e^{\mathrm i(\theta_i+\theta_j)}Y\ket{ij}.
 \label{eq:phase-eigenspace}
\end{align}
If the coefficient of $\ket{k\ell}$ in $Y\ket{ij}$ is nonzero, \eqref{eq:phase-eigenspace} requires
$e^{\mathrm i(\theta_k+\theta_\ell)}=e^{\mathrm i(\theta_i+\theta_j)}$
for \emph{every} choice of phases. Varying one phase at a time shows that every index must occur equally often in the two pairs. For $i=j$, the only possibility is $(k,\ell)=(i,i)$. For $i\ne j$, the only possibilities are $(k,\ell)=(i,j)$ and $(j,i)$. Thus
\begin{equation}\label{eq:phase-block-action}
 Y\ket{ii}=a_i\ket{ii},\qquad
 Y\ket{ij}=b_{ij}\ket{ij}+c_{ij}\ket{ji}\quad(i\ne j).
\end{equation}

We next use basis permutations to remove the dependence of these coefficients on the indices. For a permutation $\pi$, the unitary $W_\pi\ket i=\ket{\pi(i)}$ also satisfies \eqref{eq:mean-invariance}. Applying the commutation relation to \eqref{eq:phase-block-action} and comparing coefficients gives
\begin{equation}\label{eq:permutation-coefficients}
 a_{\pi(i)}=a_i,\qquad b_{\pi(i)\pi(j)}=b_{ij},\qquad
 c_{\pi(i)\pi(j)}=c_{ij}\quad(i\ne j).
\end{equation}
A permutation can send an arbitrary index to another, and an arbitrary ordered pair of distinct indices to another such pair. Consequently there are scalars $a,b,c$, independent of the indices, for which
\begin{equation}\label{eq:mean-basis-action}
 Y\ket{ii}=a\ket{ii},\qquad
 Y\ket{ij}=b\ket{ij}+c\ket{ji}\quad(i\ne j).
\end{equation}

One superposition determines the remaining relation among $a,b,c$. Define
$\ket\pm\coloneqq(\ket0\pm\ket1)/\sqrt2$, and take $W$ to map
$\ket0$ to $\ket+$ and $\ket1$ to $\ket-$, while acting as the identity on the other basis vectors. By \eqref{eq:mean-invariance} and \eqref{eq:mean-basis-action},
\begin{align}
 Y\ket{++}&=Y(W\otimes W)\ket{00}
 \label{eq:plus-eigenvector-step-1}\\
 &=(W\otimes W)Y\ket{00}
 \label{eq:plus-eigenvector-step-2}\\
 &=a\ket{++}.
 \label{eq:plus-eigenvector}
\end{align}
On the other hand, expanding $\ket{++}$ in the original basis gives
\begin{align}
 Y\ket{++}
 &=\tfrac12\bigl(Y\ket{00}+Y\ket{01}+Y\ket{10}+Y\ket{11}\bigr)
 \label{eq:mean-superposition-step-1}\\
 &=\tfrac12\bigl[a(\ket{00}+\ket{11})
                     +(b+c)(\ket{01}+\ket{10})\bigr].
 \label{eq:mean-superposition}
\end{align}
The coefficient of $\ket{01}$ in \eqref{eq:plus-eigenvector} is $a/2$, whereas in \eqref{eq:mean-superposition} it is $(b+c)/2$. Hence $a=b+c$. Comparing the action on every basis vector in \eqref{eq:mean-basis-action} now proves
\begin{align}
 Y&=bI+c\mathbb F
 \label{eq:mean-identity-swap-step-1}\\
 &=(b+c)\Pi_{\mathcal S}+(b-c)\Pi_{\mathcal A}.
 \label{eq:mean-identity-swap}
\end{align}

It remains only to determine two scalar coefficients by taking traces. The swap commutes with $G\otimes G$, as is seen by applying both orders of these operators to $\ket u\ket v$. Therefore the projectors in \eqref{eq:symmetry-projectors} commute with $G\otimes G$ as well. For either $\mathcal V=\mathcal S$ or $\mathcal V=\mathcal A$, cyclicity of the trace gives
\begin{align}
 \Tr[\Pi_{\mathcal V}Y]
 &=\mathbb E_G\Tr[\Pi_{\mathcal V}(G\otimes G)X(G\otimes G)^\dagger]
 \label{eq:preserved-sector-trace-step-1}\\
 &=\mathbb E_G\Tr[(G\otimes G)^\dagger\Pi_{\mathcal V}(G\otimes G)X]
 \label{eq:preserved-sector-trace-step-2}\\
 &=\Tr[\Pi_{\mathcal V}X].
 \label{eq:preserved-sector-trace}
\end{align}
Using $\Pi_{\mathcal S}\Pi_{\mathcal A}=0$ in \eqref{eq:mean-identity-swap}, these equalities become
\begin{equation}\label{eq:sector-coefficients}
 b+c=\frac{2\Tr[\Pi_{\mathcal S}X]}{d(d+1)},\qquad
 b-c=\frac{2\Tr[\Pi_{\mathcal A}X]}{d(d-1)}.
\end{equation}
Substitution into \eqref{eq:mean-identity-swap} proves the symmetric/antisymmetric expression in \eqref{eq:werner-twirl-step-1}. Equivalently, taking the trace and the trace against $\mathbb F$ in \eqref{eq:mean-identity-swap-step-1} gives the two linear equations
$d^2b+dc=\Tr X$ and $db+d^2c=\Tr[\mathbb F X]$.
Solving them gives the identity/swap expression in \eqref{eq:werner-twirl}.
This proves \eqref{eq:werner-twirl-step-1}--\eqref{eq:werner-twirl}. No positivity or Hermiticity assumption on $X$ was used.
\end{proof}

\section{Evaluating the refined integer dimension}\label{app:integer}
Corollary~\ref{cor:profile} gives the largest dimension certified by Theorem~\ref{thm:main}. We now make its fixed-tolerance condition explicit, retaining exact integer rounding and the quadratic correction. These formulas evaluate the stated bound, not the optimal performance of all physical codes.

Fix a state $\rho_{AB}$, let $d=\dim A\ge2$, and fix $0\le\delta<1$ and desired error $0<\err<1$. Define
\begin{equation}\label{eq:integer-h}
 h_\delta\coloneqq d\,2^{-\Ih^\delta(A\rangle B)_\rho}-1+\delta,
 \qquad u_\delta(m)\coloneqq\delta+\frac{dm-1}{d^2-1}h_\delta.
\end{equation}
Equation~\eqref{eq:dimension-IH} gives $h_\delta\ge0$, and
$v_{d,m}^{\delta}=\min\{1,u_\delta(m)\}$.
For each integer $2\le m\le d$, there are two cases. If
$\err\ge 1-1/(dm)$, then $f_{d,m}(v_{d,m}^{\delta})\le\err$ automatically. Otherwise $\err<f_{d,m}(1)$, and strict monotonicity of $f_{d,m}$ on $[0,1]$ gives the exact equivalence
\begin{align}
 f_{d,m}(v_{d,m}^{\delta})\le\err
 &\quad\Longleftrightarrow\quad u_\delta(m)\le r_{d,m}(\err),
 \label{eq:integer-inverse-step-1}\\
 r_{d,m}(\err)&\coloneqq
 \frac{2\err}{1+\sqrt{1-4\err/(dm)}}<1.
 \label{eq:integer-inverse}
\end{align}
The root $r_{d,m}$ in \eqref{eq:integer-inverse} solves $x-x^2/(dm)=\err$ and avoids subtracting nearly equal numbers. Under $\err<f_{d,m}(1)$, a capped value is infeasible because its error is $f_{d,m}(1)>\err$.

For a closed-form threshold in $m$, write
\begin{equation}\label{eq:integer-linear}
 a_\delta\coloneqq\frac{d h_\delta}{d^2-1},\qquad
 b_\delta\coloneqq\delta-\frac{h_\delta}{d^2-1},\qquad
 u_\delta(m)=a_\delta m+b_\delta.
\end{equation}
The feasible test $(1-\delta)I_{AB}$ gives
$2^{-\Ih^\delta}\le d(1-\delta)$, and hence
$0\le h_\delta\le(d^2-1)(1-\delta)$ and $0\le a_\delta\le d$.
For dimensions with $u_\delta(m)\le1$, direct expansion yields
\begin{align}
 f_{d,m}(u_\delta(m))\le\err
 &\quad\Longleftrightarrow\quad
 A_\delta m^2+(B_\delta-\err)m-C_\delta\le0,
 \label{eq:integer-quadratic-step-1}\\
 A_\delta&\coloneqq a_\delta(1-a_\delta/d)\ge0,
 \label{eq:integer-quadratic-step-2}\\
 B_\delta&\coloneqq b_\delta(1-2a_\delta/d),\qquad
 C_\delta\coloneqq b_\delta^2/d.
 \label{eq:integer-quadratic}
\end{align}
When $A_\delta>0$, the admissible positive real dimensions in this uncapped case satisfy
\begin{equation}\label{eq:integer-root}
 m\le\frac{\err-B_\delta+
       \sqrt{(\err-B_\delta)^2+4A_\delta C_\delta}}{2A_\delta},
 \qquad u_\delta(m)\le1.
\end{equation}
Intersect these constraints with $2\le m\le d$ and take the largest integer, retaining equality at either boundary. Dimensions with $u_\delta(m)>1$ must instead be checked using $1-1/(dm)\le\err$; extrapolating the quadratic past the cap would be incorrect. Equivalently, the equivalence in \eqref{eq:integer-inverse-step-1}, with the root in \eqref{eq:integer-inverse}, handles each dimension directly.

If $A_\delta=0$, no division by it is necessary: \eqref{eq:integer-quadratic-step-1} becomes the linear inequality
$(B_\delta-\err)m\le C_\delta$, together with the same cap condition.
In particular, $h_\delta=0$ gives $u_\delta(m)=\delta$ for all $m$. Every dimension is certified when $\delta\le\err$; when $\delta>\err$, the precise condition is
\begin{equation}\label{eq:integer-zero-h}
 m\le\frac{\delta^2}{d(\delta-\err)}.
\end{equation}
The case $h_0=0$ certifies zero error for the full target dimension $d$, as in the noiseless example.

Optimizing these fixed-tolerance conditions implements \eqref{eq:best-integer} and retains the full guarantee by \eqref{eq:profile-exact}.

\section{Error-preserving conversion from entanglement generation to entanglement transmission}\label{app:conversion}
This appendix proves Lemma~\ref{lem:eg-et-conversion}. We first concentrate the entanglement generation fidelity into one Kraus operator of the decoded channel. Encoding and decoding along its largest singular directions then gives an entanglement transmission code. An elementary inequality for these singular values shows that a constant factor in target dimension is enough to preserve the error. The argument changes the code, not the physical channel, and never conditions successful communication on a measurement outcome.

\begin{proof}[Proof of Lemma~\ref{lem:eg-et-conversion}]
Let $\psi_{RA'}$ be the normalized pure entanglement generation input, with $\dim R=M$, and set $\mathcal M\coloneqq\cD\circ\cN\colon A'\to B_0$. Its achieved fidelity is
\begin{equation}\label{eq:conversion-fidelity}
 f\coloneqq\Tr\!\left[\Phi^M_{RB_0}(\id_R\otimes\mathcal M)(\psi_{RA'})\right]
 \ge1-\err>0.
\end{equation}
We will prove the stronger statement that, for each integer $1\le k\le M$, an entanglement transmission code with decoder $\mathcal C_k\circ\cD$ attains
\begin{equation}\label{eq:conversion-tradeoff}
 F_{\mathrm{et}}^{(k)}\ge
 \min\!\left\{f,\left(\frac{Mf}{k}\right)^2\right\}.
\end{equation}
Here $F_{\mathrm{et}}^{(k)}$ denotes its overlap with $\Phi^k$, so its infidelity is $1-F_{\mathrm{et}}^{(k)}$. In particular, \eqref{eq:conversion-tradeoff} preserves the fidelity whenever $k\le M\sqrt f$.

\emph{Concentrating the overlap into one Kraus operator.}
Choose Kraus operators $(L_\alpha)_\alpha$ for $\mathcal M$, and define
\begin{equation}\label{eq:conversion-amplitudes}
 a_\alpha\coloneqq\bra{\Phi^M}_{RB_0}(I_R\otimes L_\alpha)\ket\psi_{RA'},
 \qquad f=\sum_\alpha|a_\alpha|^2.
\end{equation}
The equality follows by inserting the Kraus representation into \eqref{eq:conversion-fidelity}. Set
\begin{equation}\label{eq:conversion-K}
 K\coloneqq\frac1{\sqrt f}\sum_\alpha\overline{a_\alpha}L_\alpha
 \colon A'\to B_0.
\end{equation}
The coefficients $\overline{a_\alpha}/\sqrt f$ have squared absolute values summing to one. Complete them to the rows of a unitary matrix and apply this matrix to the list of Kraus operators. Unitary mixing preserves the channel and makes $K$ its first Kraus operator. Consequently,
\begin{equation}\label{eq:conversion-contraction}
 K^\dagger K\le I_{A'},\qquad
 \bra{\Phi^M}(I_R\otimes K)\ket\psi=\sqrt f.
\end{equation}
This is a choice of mathematical representation, not a physical operation on an environment. Cauchy--Schwarz and the maximally mixed marginal of $\Phi^M$ give
\begin{align}
 f&\le\left\|(I_R\otimes K^\dagger)\ket{\Phi^M}_{RB_0}\right\|^2
 \label{eq:conversion-square-sum-step-1}\\
 &=\bra{\Phi^M}(I_R\otimes KK^\dagger)\ket{\Phi^M}
 \label{eq:conversion-square-sum-step-2}\\
 &=\frac1M\Tr[K^\dagger K].
 \label{eq:conversion-square-sum}
\end{align}
Let $a\coloneqq\dim A'$. Write the singular values of $K$, padded with zeros to a list of length $M$, as
\begin{equation}\label{eq:conversion-singular-values}
 1\ge s_1\ge\cdots\ge s_M\ge0,
 \qquad \sum_{i=1}^M s_i^2=\Tr[K^\dagger K]\ge Mf.
\end{equation}
The upper bound on each singular value follows from \eqref{eq:conversion-contraction}. Combining that contraction bound with \eqref{eq:conversion-square-sum} also gives $Mf\le a$.

\emph{Encoding and decoding along singular vectors.}
First suppose $k\le a$. Choose orthonormal right and left singular vectors satisfying $K\ket{v_i}_{A'}=s_i\ket{u_i}_{B_0}$ for $1\le i\le k$, completing the families within the kernels if necessary. On a $k$-dimensional logical input $L_0$ and recovered output $\widehat L$, define
\begin{equation}\label{eq:conversion-VW}
 V_k\coloneqq\sum_{i=1}^k\ket{v_i}_{A'}\bra i_{L_0},
 \qquad
 W_k\coloneqq\sum_{i=1}^k\ket i_{\widehat L}\bra{u_i}_{B_0}.
\end{equation}
Alice uses the isometric encoding channel $\mathcal E_k(X)\coloneqq V_kXV_k^\dagger$ on one share of $\Phi^k_{R_kL_0}$, with $\dim R_k=k$. Thus the new input is $(\id_{R_k}\otimes\mathcal E_k)(\Phi^k_{R_kL_0})$; no operation is applied to its reference $R_k$. Bob follows the original decoder by
\begin{equation}\label{eq:conversion-output-channel}
 \mathcal C_k(Y)\coloneqq W_kYW_k^\dagger
   +\Tr[(I_{B_0}-W_k^\dagger W_k)Y]\xi_{\widehat L},
\end{equation}
where $\xi_{\widehat L}$ is a fixed state. Since $W_k^\dagger W_k$ is a projector, both terms in \eqref{eq:conversion-output-channel} are completely positive and their traces add to $\Tr Y$. The channel changes coordinates on the chosen output subspace and prepares a fixed state on its complement; it does not discard the complementary outcome.

One Kraus operator of the complete logical channel $\mathcal C_k\circ\mathcal M\circ\mathcal E_k$ is
\begin{equation}\label{eq:conversion-diagonal}
 W_kKV_k=\sum_{i=1}^k s_i\ket i_{\widehat L}\bra i_{L_0}.
\end{equation}
For Kraus operators $(A_\beta)_\beta$ of a channel between these matching $k$-dimensional spaces, expansion of the maximally entangled state gives
\begin{align}
 F_{\mathrm{et}}^{(k)}
 &=\sum_\beta\left|\bra{\Phi^k}(I\otimes A_\beta)\ket{\Phi^k}\right|^2
 \label{eq:conversion-kraus-fidelity-step-1}\\
 &=\frac1{k^2}\sum_\beta|\Tr A_\beta|^2.
 \label{eq:conversion-kraus-fidelity}
\end{align}
Retaining just the nonnegative contribution of \eqref{eq:conversion-diagonal} yields
\begin{equation}\label{eq:conversion-top-average}
 F_{\mathrm{et}}^{(k)}\ge\mu^2,
 \qquad \mu\coloneqq\frac1k\sum_{i=1}^k s_i.
\end{equation}

\emph{The scalar estimate.}
Set $t\coloneqq s_k$, so $0\le t\le\mu\le1$. For $i\le k$, the inequality $(s_i-t)(s_i-1)\le0$ gives $s_i^2\le(1+t)s_i-t$; for $i>k$, we have $s_i^2\le t^2$. Therefore
\begin{align}
 \sum_{i=1}^M s_i^2
 &\le k(1+t)\mu-kt+(M-k)t^2
   \label{eq:conversion-scalar-start}\\
 &\le k\mu+t(M\mu-k)
   \label{eq:conversion-scalar-linear}\\
 &\le\max\{k\mu,M\mu^2\}.
   \label{eq:conversion-scalar-end}
\end{align}
Equation~\eqref{eq:conversion-scalar-linear} uses $t^2\le\mu t$. The expression on that line is affine in $t\in[0,\mu]$, so its maximum is the larger of its two endpoint values, as written in \eqref{eq:conversion-scalar-end}. Together with \eqref{eq:conversion-singular-values}, this implies
\begin{equation}\label{eq:conversion-mu-bound}
 \mu\ge\min\!\left\{\frac{Mf}{k},\sqrt f\right\}.
\end{equation}
Indeed, a value of $\mu$ strictly below both quantities would make both arguments of the maximum in \eqref{eq:conversion-scalar-end} smaller than $Mf$. Equations~\eqref{eq:conversion-top-average} and \eqref{eq:conversion-mu-bound} prove \eqref{eq:conversion-tradeoff} when $k\le a$.

\emph{Completing the encoder when $k>a$.}
In this case $a<k\le M$. Use all $a$ right singular vectors in $V_k\coloneqq\sum_{i=1}^{a}\ket{v_i}\bra i$, and let $P_k\coloneqq V_k^\dagger V_k=\sum_{i=1}^{a}\proj i_{L_0}$. Replace the isometric encoder by the channel
\begin{equation}\label{eq:conversion-input-completion}
 \mathcal E_k(X)\coloneqq V_kXV_k^\dagger
       +\Tr[(I_{L_0}-P_k)X]\chi_{A'},
\end{equation}
for a fixed state $\chi_{A'}$. Its two terms are completely positive and their traces sum to $\Tr X$. Keep $W_k$ and \eqref{eq:conversion-output-channel}, using an orthonormal completion of the left singular vectors. Since $s_i=0$ for $i>a$, the operator $W_kKV_k$ is still precisely \eqref{eq:conversion-diagonal}. It is one Kraus operator of the completed logical channel, and all other Kraus contributions in \eqref{eq:conversion-kraus-fidelity} are nonnegative. Thus \eqref{eq:conversion-tradeoff} remains valid.

Finally, if the integer in \eqref{eq:conversion-dimension} is at least two, it satisfies $k\le M\sqrt{1-\err}\le M\sqrt f$. Equation~\eqref{eq:conversion-tradeoff} then gives $F_{\mathrm{et}}^{(k)}\ge f\ge1-\err$. If $k=1$, use the trivial zero-error protocol. The same reasoning applies to $k_{\mathrm{iso}}=\min\{k,a\}$, with an isometric encoder. Every operation just constructed is trace preserving, and no extra resource has been supplied.
\end{proof}

For completeness, the rounding and input-dimension qualifications also have explicit logarithmic bounds. For every $x>0$, the integer $\max\{1,\lfloor x\rfloor\}$ is at least $x/2$: this is immediate for $x<2$, and follows from $\lfloor x\rfloor>x-1\ge x/2$ for $x\ge2$. Apply this to $x=M\sqrt{1-\err}$ to obtain \eqref{eq:conversion-log-loss}. Moreover, \eqref{eq:conversion-square-sum} and $K^\dagger K\le I_{A'}$ give
\begin{equation}\label{eq:conversion-input-dimension}
 \dim A'\ge Mf\ge M(1-\err).
\end{equation}
Both $k$ and $\dim A'$ are at least $\max\{1,\lfloor M(1-\err)\rfloor\}$, so the isometric choice always obeys
\begin{equation}\label{eq:conversion-isometric-log-loss}
 \log_2 k_{\mathrm{iso}}\ge\log_2 M-1-\log_2\frac1{1-\err}.
\end{equation}
Equations~\eqref{eq:conversion-log-loss} and \eqref{eq:conversion-isometric-log-loss} are used only to bound the asymptotic cost. At finite dimension we retain the exact integers. At fixed $\err<1$, whenever $M\to\infty$ and no input-dimension cap is needed, the sharper relation is
\begin{equation}\label{eq:conversion-asymptotic-loss}
 \log_2 k=\log_2 M+\frac12\log_2(1-\err)+o(1).
\end{equation}
Thus the price of preserving the error is bounded in log-dimension, rather than an altered error argument in the Gaussian term.

\section{Logarithmic terms: testing versus coding}\label{app:third-order}\label{subsec:third-order}
This appendix separates the achievable logarithmic remainder in Theorem~\ref{thm:second} from the logarithmic term in classical testing. We then prove a constant-remainder bound for commuting hypotheses and identify the limitation of the present binary-cost certificate. Neither remainder is asserted to be optimal for coding.

\subsection{What the classical testing expansion does and does not imply}
For fixed finite-alphabet distributions $P,Q$ with the support and positive-variance assumptions of \cite[Proposition~2.3, Eq.~(2.46)]{Tan2014}, the classical testing divergence has expansion
\begin{equation}\label{eq:classical-third-order}
 D_H^\delta(P^{\otimes n}\Vert Q^{\otimes n})
 =nD(P\Vert Q)+\sqrt{nV(P\Vert Q)}\,\Phi^{-1}(\delta)
       +\tfrac12\log_2n+O(1).
\end{equation}
The convention in \cite{Tan2014} includes a factor $1/(1-\delta)$, changing \eqref{eq:classical-third-order} only by a constant at fixed $\delta$. This is a fixed-error testing expansion, not a coding yield. For noncommuting hypotheses, it cannot replace \lemref{lem:normal-testing}.

If \eqref{eq:classical-third-order} holds uniformly at $\delta_n=\err-n^{-1/2}$, its $+\tfrac12\log_2n$ cancels the slack $-\log_2(1/\nu_n)=-\tfrac12\log_2n$ in \eqref{eq:intro-rate}, leaving a constant coding remainder. A fixed-tolerance expansion alone does not justify this substitution. We avoid that issue by directly analyzing the binary cost, using the likelihood-tail estimate underlying refined classical testing~\cite[Theorems~1.1 and~1.3]{Tan2014}.

\subsection{A constant remainder for commuting hypotheses}
\begin{proposition}[Commuting-state refinement]\label{prop:commuting-second}
Let $\rho_{AB}$ be a fixed finite-dimensional state with $d\coloneqq\dim A\ge2$, set $\sigma_{AB}\coloneqq I_A\otimes\rho_B$, and suppose $[\rho_{AB},\sigma_{AB}]=0$ and $\Vc(A\rangle B)_\rho>0$. For every $0<\err<1$, there is a finite constant $c'_{\rho,\err}$ such that, for all sufficiently large $n$, the tasks in \thref{thm:second}, with $\Ic$ and $\Vc$ as in \eqref{eq:Ic}--\eqref{eq:Vc}, are achievable with infidelity at most $\err$ and
\begin{equation}\label{eq:commuting-yield}
 \log_2m_n\ge
 [n\Ic+\sqrt{n\Vc}\,\Phi^{-1}(\err)-c'_{\rho,\err}]_+.
\end{equation}
The channel statement uses a fixed pure reference--input state whose output is $\rho_{AB}$. For a maximally mixed reference marginal, Corollary~\ref{cor:flat} gives entanglement transmission directly; for an arbitrary reference marginal, Lemma~\ref{lem:eg-et-conversion} gives entanglement transmission with the same form of bound and a possibly larger constant $c'_{\rho,\err}$. No assumption on whether the log-likelihood values lie on a lattice is needed.
\end{proposition}
\begin{proof}
We prove the proposition by analyzing the full binary cost in a common eigenbasis. This retains the likelihood tail that is bounded more coarsely in the simple slack-based corollary. Diagonalize $\rho$ and $\sigma$ simultaneously. Let their eigenvalues be $p_x$ and $q_x$; on the set $p_x>0$, support inclusion implies $q_x>0$. Draw $X$ with probability $p_x$, define
\begin{equation}\label{eq:classical-likelihood}
 \ell(X)\coloneqq\log_2(p_X/q_X),
 \qquad S_n\coloneqq\sum_{i=1}^n\ell(X_i),
\end{equation}
and take independent copies $X_i$. Then $\ee S_n=n\Ic$ and $\Var(S_n)=n\Vc$. The alternative eigenvalues need not sum to one: $\Tr\sigma=d$. They still define this likelihood variable under the normalized null distribution.

For a real $R$, the coordinatewise optimal binary decision gives
\begin{align}
 g_n(R)&\coloneqq\cost(\rho^{\otimes n},2^R\sigma^{\otimes n})
 \label{eq:commuting-cost-step-1}\\
 &=\ee\min\{1,2^{R-S_n}\}
 \label{eq:commuting-cost-step-2}\\
 &=\Pr\{S_n\le R\}+\ee[2^{R-S_n}\mathbf1_{\{S_n>R\}}].
 \label{eq:commuting-cost}
\end{align}
Coordinates outside $\supp\rho^{\otimes n}$ contribute zero to this cost.
The Berry--Esseen theorem~\cite[Theorem~1.1]{Tan2014} supplies a constant $b\ge0$, depending only on the finite likelihood distribution, for which
\begin{equation}\label{eq:commuting-BE}
 \sup_{R\in\mathbb R}\left|
 \Pr\{S_n\le R\}-\Phi\!\left(\frac{R-n\Ic}{\sqrt{n\Vc}}\right)
 \right|\le\frac b{\sqrt n}.
\end{equation}
Equation~\eqref{eq:commuting-BE} also bounds the probability of every unit-length interval:
\begin{equation}\label{eq:commuting-interval}
 \Pr\{R+j<S_n\le R+j+1\}
 \le\frac a{\sqrt n},
 \qquad a\coloneqq\frac1{\sqrt{2\pi\Vc}}+2b.
\end{equation}
Indeed, subtract the two distribution functions, use \eqref{eq:commuting-BE} at both endpoints, and bound the increment of $\Phi$ by its maximum density times the interval length divided by $\sqrt{n\Vc}$. Splitting the expectation in \eqref{eq:commuting-cost} over these intervals gives
\begin{align}
 \ee[2^{R-S_n}\mathbf1_{\{S_n>R\}}]
 &\le\sum_{j=0}^\infty2^{-j}
           \Pr\{R+j<S_n\le R+j+1\}
 \label{eq:commuting-tail-bound1-step-1}\\
 &\le\frac{2a}{\sqrt n}.
 \label{eq:commuting-tail-bound1}
\end{align}
Thus, uniformly in $R$,
\begin{equation}\label{eq:commuting-cost-normal}
 \Phi\!\left(\frac{R-n\Ic}{\sqrt{n\Vc}}\right)-\frac b{\sqrt n}
 \le g_n(R)\le
 \Phi\!\left(\frac{R-n\Ic}{\sqrt{n\Vc}}\right)+\frac{b+2a}{\sqrt n}.
\end{equation}

Set $z_\err\coloneqq\Phi^{-1}(\err)$ and choose a fixed $C>0$ so large that
\begin{equation}\label{eq:commuting-C}
 \frac{C}{\sqrt{\Vc}}\min_{u\in[z_\err-1,z_\err]}\varphi(u)>b+2a.
\end{equation}
For sufficiently large $n$, the mean value theorem then shows that
$g_n(R_n)\le\err$ for $R_n\coloneqq n\Ic+\sqrt{n\Vc}\,z_\err-C$.
If $2^{R_n}<2$, choose $m_n=1$ and use the zero-error trivial protocol. Since $R_n<1$, the positive-part yield with constant $C+1$ is then zero. For the rest of the proof assume $2^{R_n}\ge2$ and choose $m_n=\lfloor2^{R_n}\rfloor\ge2$.
The dimension inequality \eqref{eq:dimension-bound} implies
$\ell(X)\le\log_2d$ almost surely. Positive variance makes this inequality strict in expectation, so $\Ic<\log_2d$ and $m_n\le d^n$ for all sufficiently large $n$.

Finally, the alternative in \thref{thm:main}, applied with left dimension $d^n$, obeys
\begin{equation}\label{eq:commuting-alternative-domination}
 B_{d^n,m_n}\le m_n\sigma^{\otimes n}.
\end{equation}
This follows from $\lambda_{d^n,m_n}\le m_n$ in \eqref{eq:testing-coefficients} and the nonnegative subtracted multiple of $\rho^{\otimes n}$. The testing cost is nondecreasing in its second argument. Therefore the theorem certifies error at most
$\cost(\rho^{\otimes n},B_{d^n,m_n})\le g_n(\log_2m_n)\le g_n(R_n)\le\err$.
The exact rounding loss is less than $\log_2(1+1/m_n)\le1$ by \eqref{eq:rounding-loss}. The trivial target covers the other case, giving \eqref{eq:commuting-yield} with $c'_{\rho,\err}=C+1$. In the general channel-input case, apply Lemma~\ref{lem:eg-et-conversion} to the entanglement generation code. Equation~\eqref{eq:conversion-log-loss}, or \eqref{eq:conversion-isometric-log-loss} when an input-dimension cap is required, adds only a constant to this remainder and preserves the error $\err$.
\end{proof}

Commutativity here concerns the \emph{analysis hypotheses}, not the operational state being unentangled. For example, a qubit dephasing channel
$\mathcal Z_p(X)\coloneqq(1-p)X+pZXZ$, with $Z\coloneqq\proj0-\proj1$ and $0<p<1/2$, has a maximally entangled input producing
\begin{equation}\label{eq:dephasing-example}
 \rho_{AB}=(1-p)\Phi^2_{AB}+p\Phi^-_{AB},
 \qquad \sigma_{AB}=I_{AB}/2.
\end{equation}
Here $\Phi^-_{AB}$ denotes the projector onto $(\ket{00}-\ket{11})/\sqrt2$. The hypotheses commute, while $\rho_{AB}$ is entangled. With
$h_2(p)\coloneqq-p\log_2p-(1-p)\log_2(1-p)$,
\begin{equation}\label{eq:dephasing-IV}
 \Ic=1-h_2(p)>0,
 \qquad \Vc=p(1-p)\left(\log_2\frac{1-p}{p}\right)^2>0.
\end{equation}
Thus \prref{prop:commuting-second} gives a nontrivial constant-remainder quantum coding statement. This is not the sharpest known dephasing-channel result: Tomamichel, Berta, and Renes established a positive $\tfrac12\log_2n$ term for that channel by exploiting its relation to classical coding~\cite{TomamichelBertaRenes2016}. The proposition instead gives a common argument for all states satisfying its commutativity assumption.

\subsection{Why this does not prove a positive logarithmic coding term}
For commuting hypotheses with $0<\Ic<\log_2d$ and $\Vc>0$, fix $0<\err<1$, set $z_\err\coloneqq\Phi^{-1}(\err)$, and consider integer target dimensions satisfying
$\log_2m_n=n\Ic+\sqrt{n\Vc}\,z_\err+\theta\log_2n+O(1)$, where $\theta>0$.
Equation~\eqref{eq:commuting-cost-normal} and a Taylor expansion give
\begin{equation}\label{eq:commuting-positive-log}
 g_n(\log_2m_n)=\err+
    \frac{\theta\,\varphi(z_\err)}{\sqrt{\Vc}}
           \frac{\log_2n}{\sqrt n}+O(n^{-1/2}).
\end{equation}
The positive logarithmic contribution dominates the remainder. To show that the finite-dimensional subtraction and Cheng's quadratic correction cannot cancel it, set
$d_n\coloneqq d^n$, $c_n\coloneqq(d_nm_n-1)/(d_n^2-1)$, and
$\eta_n\coloneqq(d_n-m_n)/(m_n(d_n^2-1))$.
On a common $n$-copy eigenvector, let $p>0$ and $q$ be the null and alternative eigenvalues. Set $x=m_nq/p$. The alternative $B_{d_n,m_n}$ has ratio $(1-\eta_n)x-c_n\ge0$ by positivity.
For $x\le1$, both minima are uncapped and differ by $\eta_nx+c_n\le\eta_n+c_n$. For $x\ge1$, the difference is zero or $1-(1-\eta_n)x+c_n\le\eta_n+c_n$. It is nonnegative because $(1-\eta_n)x-c_n\le x$. Hence
$0\le\min\{1,x\}-\min\{1,(1-\eta_n)x-c_n\}\le\eta_n+c_n$.
Averaging over the null eigenvalues, and bounding the quadratic correction by $1/(d_nm_n)$, yields
\begin{equation}\label{eq:certificate-difference}
 0\le g_n(\log_2m_n)-f_{d_n,m_n}(e_{d_n,m_n})
 \le\eta_n+c_n+\frac1{d_nm_n}.
\end{equation}
Since $0<\Ic<\log_2d$, the right side of \eqref{eq:certificate-difference} decays exponentially and cannot cancel the positive correction in \eqref{eq:commuting-positive-log}. This limits the stated \emph{error bound}, not the optimal code or the actual Petz-decoder performance.

\end{document}